\documentclass[aps,twocolumn,prb,preprintnumbers,amsmath,amssymb]{revtex4}
\usepackage{graphicx, bm}
\usepackage{dcolumn}
\usepackage{float}
\usepackage{latexsym}
\usepackage{amsmath}
\usepackage{graphics}
\usepackage{amssymb}
\usepackage{layout}
\usepackage{verbatim}
\usepackage{amsfonts,epsfig}
\usepackage{color}
\usepackage{multirow}
\usepackage{setspace}

\newcommand{\beqnar}{\begin{eqnarray}}
\newcommand{\eeqnar}{\end{eqnarray}}

\newcommand{\beq}{\begin{equation}}
\newcommand{\eeq}{\end{equation}}

\begin{document}
\title{Intrinsic plasmons in 2D Dirac materials}
\author{S. Das Sarma and Qiuzi Li}
\affiliation{Condensed Matter Theory Center, Department of Physics, University of Maryland, College Park, Maryland 20742}
\date{\today}
\begin{abstract}
We consider theoretically, using the random phase approximation (RPA), low-energy {\it intrinsic} plasmons for two-dimensional (2D) systems obeying Dirac-like linear chiral dispersion with the chemical potential set precisely at the charge neutral Dirac point. The ``intrinsic Dirac plasmon" energy has the characteristic $\sqrt{q}$ dispersion in the 2D wave-vector $q$, but vanishes as $\sqrt{T}$ in temperature for both monolayer and bilayer graphene. The intrinsic plasmon becomes overdamped for a fixed $q$ as $T \rightarrow 0$ since the level broadening (i.e. the decay of the plasmon into electron-hole pairs due to Landau damping) increases as $1/\sqrt{T}$ as temperature decreases, however, the plasmon mode remains well-defined at any fixed $T$ (no matter how small) as $q \rightarrow 0$. We find the intrinsic plasmon to be well-defined as long as $q < \frac{k_B T}{e^2}$. We give analytical results for low and high temperatures, and numerical RPA results for arbitrary temperatures, and consider both single-layer and double-layer intrinsic Dirac plasmons. We provide extensive comparison and contrast between intrinsic and extrinsic graphene plasmons, and critically discuss the prospects for experimentally observing intrinsic Dirac point graphene plasmons.
\end{abstract}


\maketitle

\section{Introduction}

Collective plasma oscillations of free carriers in doped or gated graphene\cite{HwangDas_PRB07} (we would refer to this situation as ``extrinsic" graphene where the chemical potential or the Fermi level is doped away from the Dirac point) have attracted considerable interest both from fundamental and technological perspectives\cite{vafek_PRL06,ryzhii2007,gangadharaiah_PRL08,polini_PRB08,yuliu_PRB08,kramberger_PRL08,lujiong_PRB09,jablanmarinko_PRB09,dassarma2009,stauberperes_PRB10,
langer_njp10,liuyu_PRB10,
tudorovskiy_PRB10,hwangsensardas_PRB10,mishchenko_prl10,muniz_PRB10,koch_PRB10,schutt_PRB11,politano_PRB11,
Cesar_diamond11,abedinpur_PRB11,walter_PRB11,tegenkamp_jph11,
ganchoon_PRB12,gomezsanto_EPL12,Dong20121889,krstaji_RPB12,krstajipee_PRB12,thomas_NJP12,politano_PRB12,gorbach_PRA13,zhujj_PRB13}. The fundamental interest arises from the fact that graphene plasmons are apriori quantum-mechanical entities with no classical analogs whatsoever\cite{dassarma2009} (since in classical physics energy is always proportional to the square of the momentum and never has a linear dispersion as in graphene). This is manifested in the fact that the long wave-length plasma dispersion relation in extrinsic graphene goes as $\omega_p = \sqrt{2 r_s \hbar v_F q E_F}$, where $E_F = \hbar v_F k_F= \hbar v_F \sqrt{\pi n}$ is the Fermi energy (i.e. the chemical potential at $T=0$) associated with a doping carrier density of $n$ and $v_F$ is the (constant) graphene velocity defining the linear energy dispersion, and $r_s = e^2 /(\kappa \hbar v_F )$ is the so-called graphene fine-structure constant (defining the dimensionless strength of Coulomb interaction with $\kappa$ being the background lattice dielectric constant), with an ``$\hbar$" appearing
explicitly in the definition of the plasma frequency ($\omega_p = \sqrt{2 e^2 q v_F \hbar \sqrt{\pi n}/\kappa}$ in terms of the experimentally controlled variables $q$, $n$, and $\kappa$). This is in sharp contrast to the corresponding parabolic dispersion systems\cite{dashwang_PRL98,hwangdas_PRB01} with an effective mass $m$ where $\omega_p$ ($\propto \sqrt{\dfrac{n e^2}{\kappa m}}$) is the same classically or quantum-mechanically in the long wavelength ($q \rightarrow 0$) limit in any dimensionality. The technological interest arises from the considerable recent progress in graphene nanoplasmonics\cite{koppens_nano11,nikitin_PRB11,jufeng_NN11,Echtermeyer_natcommun11,feizhe_nanoletter11,
yanhugen_nanaotech12,grigorenko_natpho12,feizhe_nature12,jianing_nature12,thongra_ACSnano12,davoyan_PRL12,
zhantr_PRB12,carbotte_PRB12,Ilic_opletter12,rast_PRB13} for prospective optoelectronic applications\cite{fengnianxia_natnano09,xiaodongxu_Nanoletter10,mueller_natphoto10,farhan_natnano11,
Farhan_PRB11,liu_Nature11,sun_ACSnano10,bonaccorso_natpho10,ashkan_science11,qiaobao_ACSNano12}.

In contrast to the extensively studied extrinsic graphene plasmons, there has been little interest in the collective modes of {\it intrinsic} or undoped graphene, where the chemical potential sits right at the Dirac point with a completely filled valence band a completely empty conduction band at $T=0$. (Our interest here is in low-energy $\sim$ meV two-dimensional collective modes and not very high-energy $\sim 10$ eV band- or so-called $\pi$- plasmons where the whole valence band charge response is involved\cite{eberlein_PRB08,yuan_PRB11,shin_prb11,yanjunFP_prl11,kinyanjui_EPL12,despoja_PRB13}.) By definition, the doping carrier density vanishes at the Dirac point with $n=0$ ($\propto k_F \propto E_F$), and the extrinsic graphene dispersion relation, $\omega_p \propto n^{1/4}$, implies that no intrinsic graphene plasmons (or more generally, Dirac plasmons) are possible.

The above is certainly true strictly at $T=0$ where there can be no free carriers for $E_F =0$. But, for non-zero temperatures, $T \neq 0$, the gapless nature of graphene leads to a thermal population of electrons (in the conduction band) and holes (in the valence band) with equal density ($n_e = n_h = n$). This thermal electron-hole excitation process is known to be a power law (in fact $n \propto T^2$)  due to the gaplessness of graphene\cite{HwangScreen_PRB09}. Putting $n \propto T^2$ in the formula for the graphene extrinsic plasma frequency, we conclude that there should be a finite-temperature intrinsic graphene plasmon with a long-wavelength plasma frequency going as $\omega_p \sim \sqrt{q T}$. But, finite temperature implies that the collective mode will decay into electron-hole pairs even at long wavelength, and therefore, such an intrinsic Dirac plasmon may be ill-defined even for $q \rightarrow 0$ since its decay (i.e. damping) rate (or level broadening) $\gamma$ could exceed the mode frequency $\omega_p$ making the mode an overdamped excitation of little interest.

In the current work, we theoretically study intrinsic Dirac plasmons for both monolayer and bilayer graphene and for single- and double-layer systems. We obtain both asymptotic theoretical analytical results at long wavelengths and low/high temperatures, and quantitative numerical results for arbitrary wavevectors and temperatures. We use the RPA approach which should be well-valid for graphene plasmons at arbitrary wavevectors by virtue of its relatively small value of $r_s$ ($\lesssim 1$ typically). We compare and contrast the temperature dependence of intrinsic and extrinsic graphene plasmon frequency (and their Landau damping) in order to comment on the feasibility of the experimental observation of our theoretical predictions. The possible existence of high-temperature {\it intrinsic} graphene plasmons (with $E_F = 0$) is a qualitative difference between graphene and gapped 2D semiconductor-based electron/hole systems.

\section{Theory and results}
\label{sec2:theory}

Within the RPA, the collective plasmon modes of an electron system are given by the zeros of the complex dielectric function $\varepsilon (q, \omega)$:
\beq
\varepsilon(q,\omega) = 1 - V(q) \Pi(q,\omega)=0
\label{eq:vareps}
\eeq
where $V(q)$ is the relevant bare electron-electron (i.e. Coulomb) interaction and $\Pi (q,\omega)$ is the noninteracting polarizability of the system. (As an aside we note that $\varepsilon (q, \omega) = 1 - V(q) \Pi(q,\omega)$) is the {\it exact} expression for the microscopic dielectric function if $\Pi (q, \omega)$ is the exact {\it interacting} irreducible polarizability function which, of course, is unknown --- RPA consists of replacing the exact $\Pi (q, \omega)$ by the corresponding non-interacting or bare polarizability function.) Equation~\eqref{eq:vareps}, as it stands, applies for a single-component (i.e. single-layer in our case) system --- for the double-layer case $\varepsilon (q,\omega)$ should be interpreted as a matrix in the layer index with Eq.~\eqref{eq:vareps} being interpreted as a determinantal equation $|1- V \Pi| =0$.

In general, Eq.~\eqref{eq:vareps} will have complex solutions in frequency, $\omega = \omega_p(q) + i \gamma(q)$, with $\omega_p$ and $\gamma$ being respectively the collective mode (i.e. plasma) frequency (which we will often refer to as the plasmon) and its damping. If $\omega_p \gg \gamma$, the plasmon collective mode is well-defined, and by contrast for $\gamma \gtrsim \omega_p$, the plasmon is heavily damped (or even overdamped) and is not particularly relevant experimentally as a self-sustaining normal mode of the system.

Before proceeding with the theoretical details for intrinsic graphene collective modes, we write down the full formal expression for the graphene non-interacting polarizability\cite{HwangDas_PRB07,wunsch2006} to be used in Eq.~\eqref{eq:vareps}:
\beq
\Pi(q,\omega)=-\frac{4}{A}\underset{\mathbf{k},s,s'}{\sum}\frac{f_{s,\mathbf{k}}
-f_{s',\mathbf{k'}}}{\omega+\epsilon_{s,\mathbf{k}}
-\epsilon_{s',\mathbf{k'}}+i\eta}F_{s,s'}(\mathbf{k},\mathbf{k'})\label{eq:pola}
\eeq
where $A$ is the area of the 2D layers and the factor of $4$ arises from the valley/spin degeneracy (two each) of graphene. In Eq.~\eqref{eq:pola}, $\mathbf{k' \equiv k+q}$ and $s, s' = \pm 1$ with $F_{s,s'}(\mathbf{k, k'})=(1+\cos\theta)/2$ arising from the matrix element effect associated with the chiral nature of Dirac fermions. The functions $\epsilon_{s,\mathbf{k}}$, $\epsilon_{s',\mathbf{k'}}$ are single-particle energies for wavevector  $\mathbf{k}$, $\mathbf{k'}$ respectively, and $f_{s,\mathbf{k}}$, $f_{s',\mathbf{k'}}$ are the corresponding non-interacting Fermi distribution functions. We mention that explicit forms for the polarizability function in monolayer and bilayer graphene were derived in Refs.~[\onlinecite{HwangDas_PRB07}] and [\onlinecite{sensarma2010}]
 respectively at the zero temperature. The finite-temperature polarizability, which cannot be obtained in a closed analytic form for arbitrary $T$, can be directly obtained from Eq.~\eqref{eq:pola} using finite-temperature Fermi distribution functions or (numerically more conveniently) by using the following integral identity to obtain the finite-$T$ polarizability from its known analytic form\cite{HwangDas_PRB07} at $T=0$
 \beq
 \Pi(q,\omega,\mu; T) = \int_0^{\infty} d \mu' \dfrac{\Pi(q, \omega, \mu';T=0)}{4 k_B T \text{cosh}^2[(\mu-\mu')/2k_B T]}
 \eeq
We also note that the 2D Coulomb interaction is given by:
\beq
V(q) = \dfrac{2\pi e^{2}}{\kappa q} e^{-q|z|}
\eeq
where $z=0$ for single-layer systems and $|z|=d$, where $d$ is the interlayer separation for double layer systems.

\subsection{Monolayer graphene}

\begin{figure}[htb]
\begin{center}
\includegraphics[width=0.99\columnwidth]{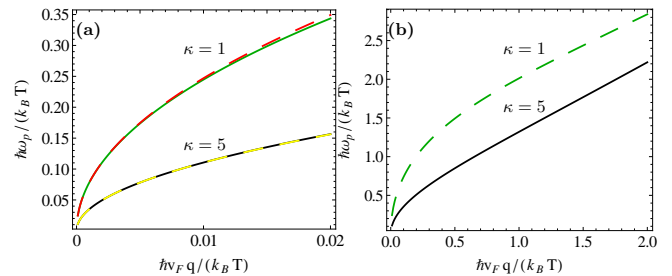}
  \caption{(Color online). Plasmon dispersion of intrinsic MLG. (a) presents the plasmon dispersion in the high temperature  regime ($k_B T \gg \hbar v_F q$). The dashed and solid lines correspond to the analytical results (given in Eq.~\eqref{eq:iomega}) and the numerical results, respectively. (b) Numerical results for the plasmon  dispersion as a function of $\hbar v_F q/(k_B T)$. The dashed  and solid lines correspond to $\kappa= 1$ and $\kappa =5$, respectively. }
\label{fig1:imdis}
\end{center}
\end{figure}

\begin{figure}[htb]
\begin{center}
\includegraphics[width=0.99\columnwidth]{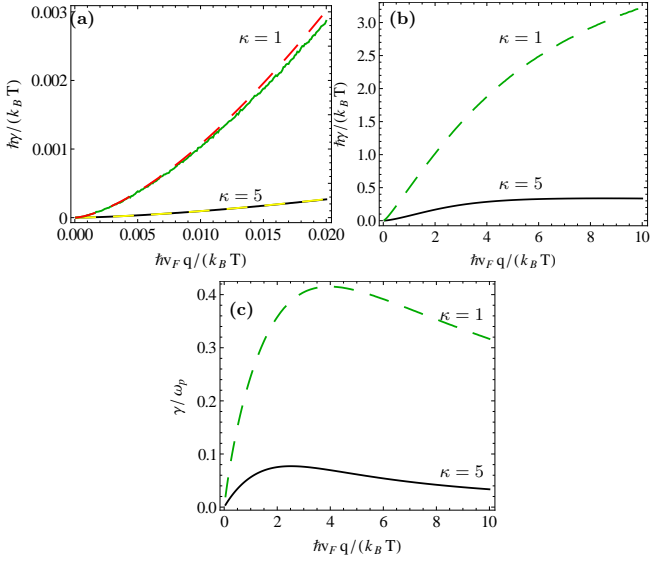}
  \caption{(Color online).  Plasmon damping rate of intrinsic MLG. (a) presents the plasmon damping rate in the high temperature  regime ($k_B T \gg \hbar v_F q$). The dashed and solid lines correspond to the analytical results (given in Eq.~\eqref{eq:igamma}) and the numerical results, respectively. (b) Numerical results for $\hbar \gamma/(k_B T)$ as a function of $\hbar v_F q/(k_B T)$. (c) $\gamma_p/\omega$  as a function of $\hbar v_F q/(k_B T)$. The dashed and solid lines correspond to $\kappa= 1$ and $\kappa =5$, respectively.}
\label{fig2:imdis}
\end{center}
\end{figure}

We first consider the intrinsic plasmon modes in monolayer graphene (MLG). We use units such that $\hbar =1$ throughout so that frequency/energy and wavevector/momentum have the same units in our notations.

For intrinsic graphene, there are no free carriers at zero temperature and the chemical potential $\mu$ is precisely at the Dirac point: $\mu=0$. We note that this is true even for $T \neq 0$, i.e. $\mu(T) = \mu(0) =E_F =0$ for intrinsic graphene, by definition.  The zero-temperature polarizability $\Pi(q,\omega)$ for intrinsic graphene is $-\frac{q^2}{4 \sqrt{v_F^2 q^2 - (\omega+i 0)^2}}$ as given in Ref.~[\onlinecite{HwangDas_PRB07}]. Since the real part of $V(q) \Pi(q,\omega)$ is pure negative, there can be no 2D plasmon modes in intrinsic graphene within RPA at $T=0$ according to Eq.~\eqref{eq:vareps}. Thus our work on intrinsic Dirac plasmons, using RPA, focuses entirely on finite temperature collective modes in undoped intrinsic graphene.

\begin{figure}[htb]
\begin{center}
\includegraphics[width=0.99\columnwidth]{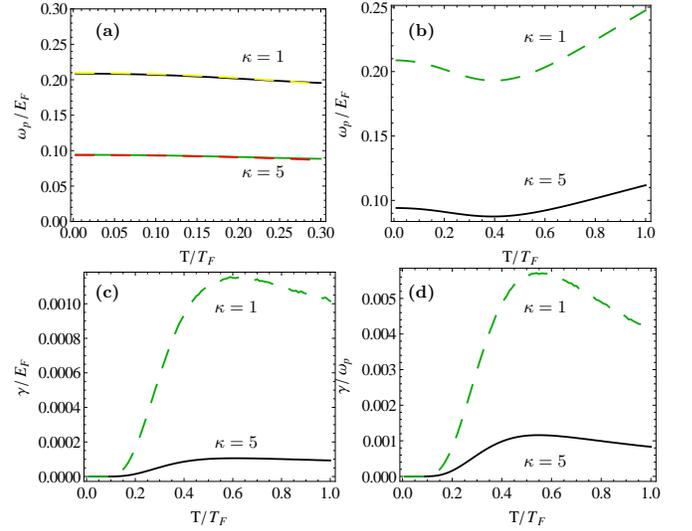}
  \caption{(Color online). Results for extrinsic MLG with $q/k_F = 0.01$. (a) presents the plasmon dispersion in the low temperature  regime. The dashed and solid lines correspond to the analytical results (given in Eq.~\eqref{eq:exomega}) and the numerical results, respectively.  (b) Numerical results for $\omega_p/E_F$ as a function of $T/T_F$.  (c) Numerical results for $\gamma/E_F$ as a function of $T/T_F$. (d) Numerical results for $\gamma/\omega_p$ as a function of $T/T_F$. The dashed and solid lines correspond to $\kappa= 1$ and $\kappa =5$, respectively. }
\label{fig3:exdist}
\end{center}
\end{figure}

\begin{figure}[htb]
\begin{center}
\includegraphics[width=0.99\columnwidth]{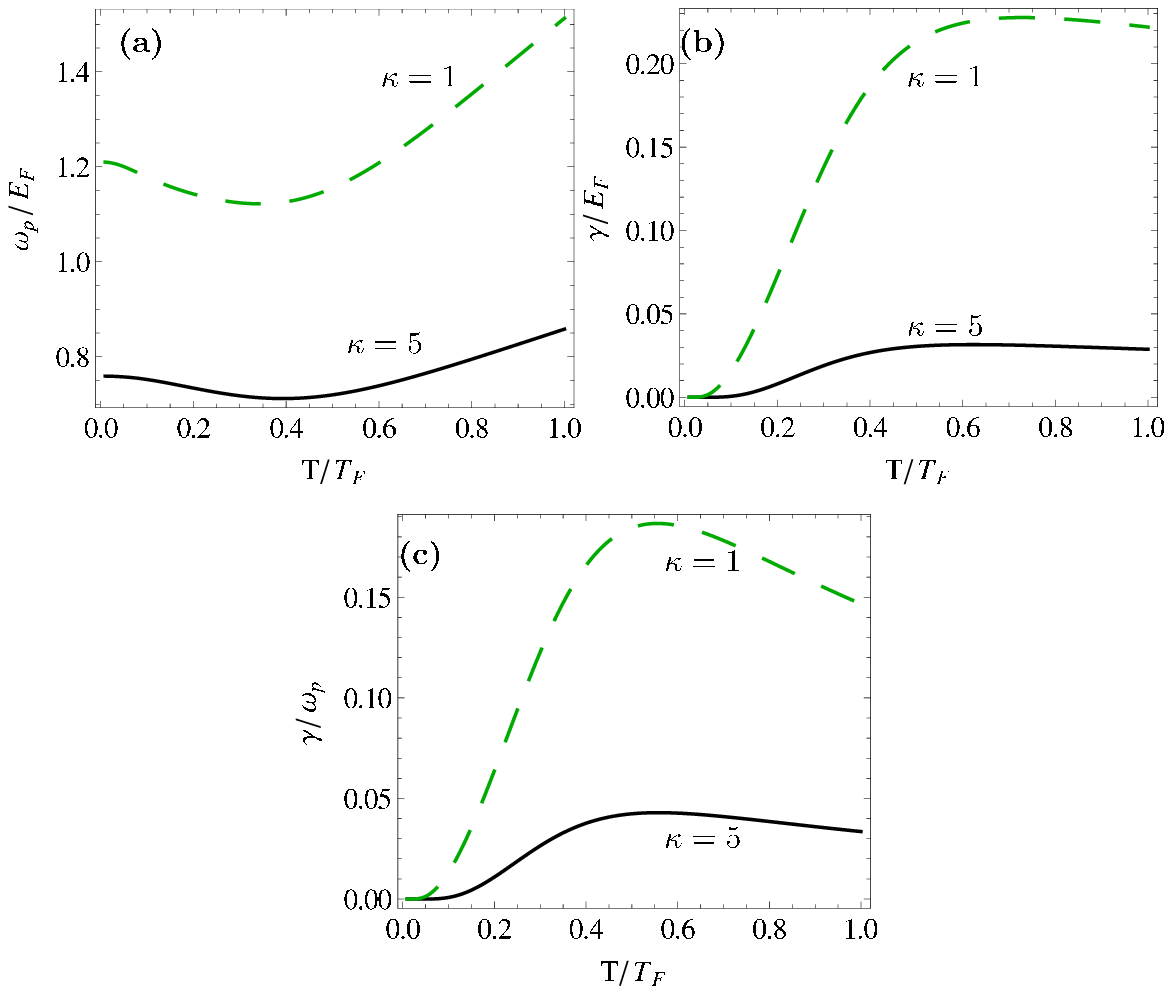}
  \caption{(Color online). Results for extrinsic  MLG with $q/k_F = 0.5$.  (a) Numerical results for $\omega_p/E_F$ as a function of $T/T_F$.  (b) Numerical results for $\gamma/E_F$ as a function of $T/T_F$. (c) Numerical results for $\gamma/\omega_p$ as a function of $T/T_F$. The dashed and solid lines correspond to $\kappa= 1$ and $\kappa =5$, respectively.}
\label{fig4:exdistq0.5}
\end{center}
\end{figure}

\begin{figure}[htb]
\begin{center}
\includegraphics[width=0.99\columnwidth]{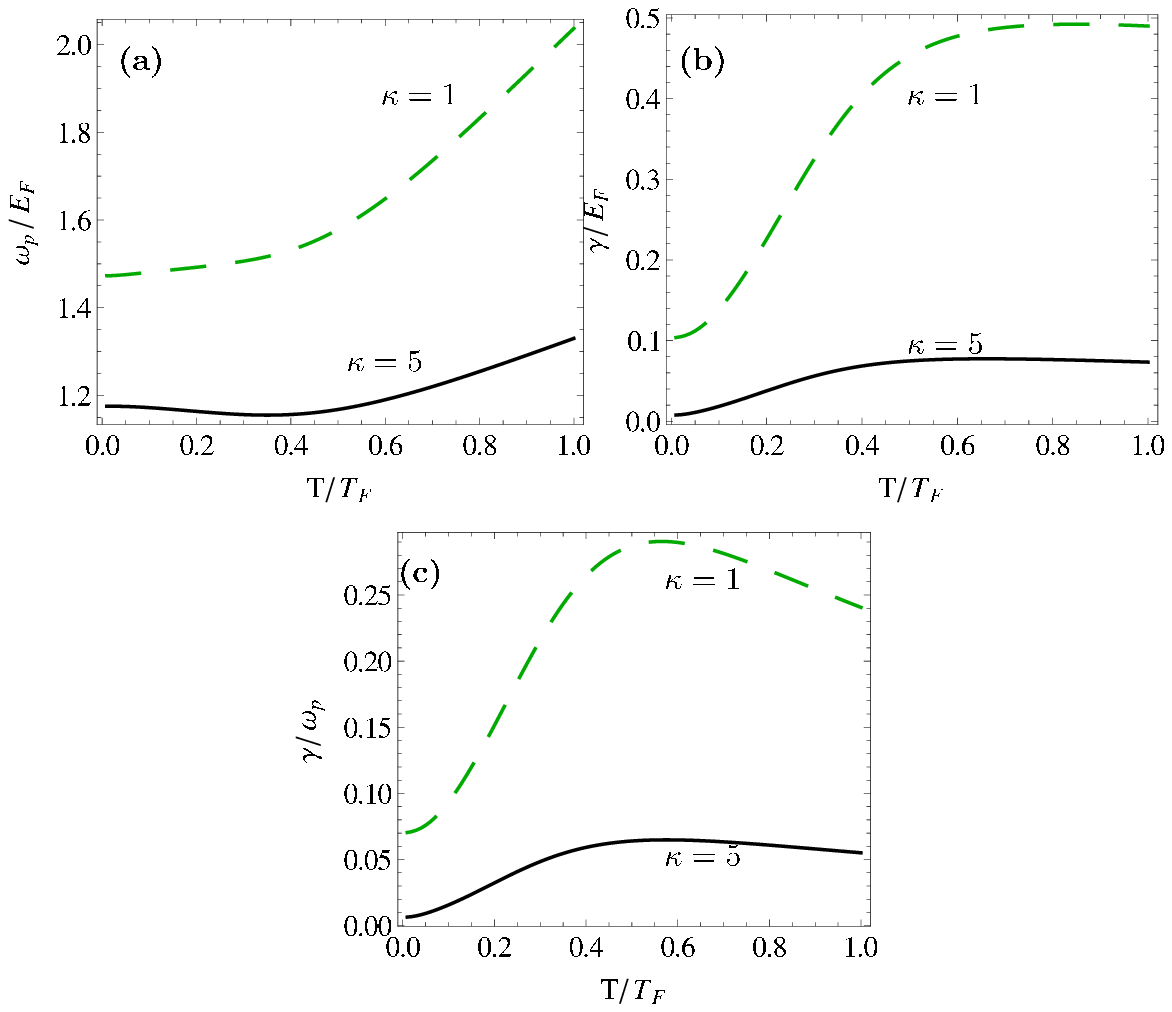}
  \caption{(Color online).  Results for extrinsic  MLG with $q/k_F = 1.0$.  (a) Numerical results for $\omega_p/E_F$ as a function of $T/T_F$.  (b) Numerical results for $\gamma/E_F$ as a function of $T/T_F$. (c) Numerical results for $\gamma/\omega_p$ as a function of $T/T_F$. The dashed and solid lines correspond to $\kappa= 1$ and $\kappa =5$, respectively. }
\label{fig5:exdistq1}
\end{center}
\end{figure}

\begin{figure}[htb]
\begin{center}
\includegraphics[width=0.99\columnwidth]{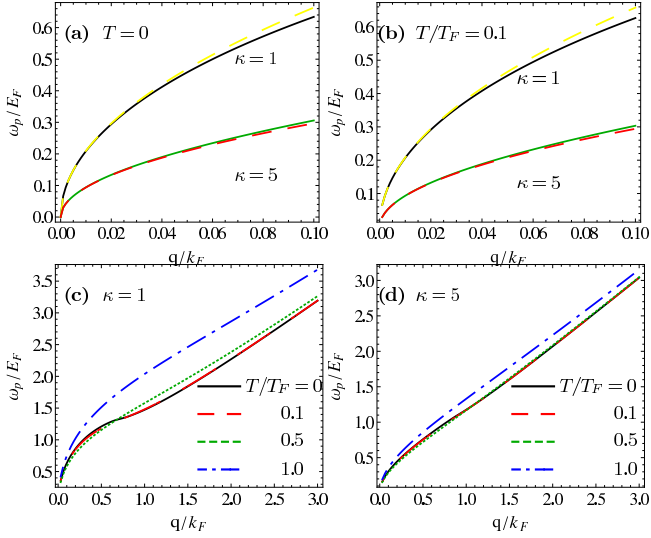}
  \caption{(Color online). Plasmon dispersion as a function of $q/k_F$ for extrinsic  MLG. (a) and (b) present $\omega_p/E_F$ in the long wavelength limit. The dashed and solid lines correspond to the analytical results (given in Eq.~\eqref{eq12:exlow}) and the numerical results, respectively. (a) is for $T=0$ and (b) is for $T/T_F = 0.1$. (c) and (d) present $\omega_p/E_F$ for different temperatures. (c) is for $\kappa = 1$ and (d) is for $\kappa = 5$.  }
\label{fig6:exdisq}
\end{center}
\end{figure}

\begin{figure}[htb]
\begin{center}
\includegraphics[width=0.99\columnwidth]{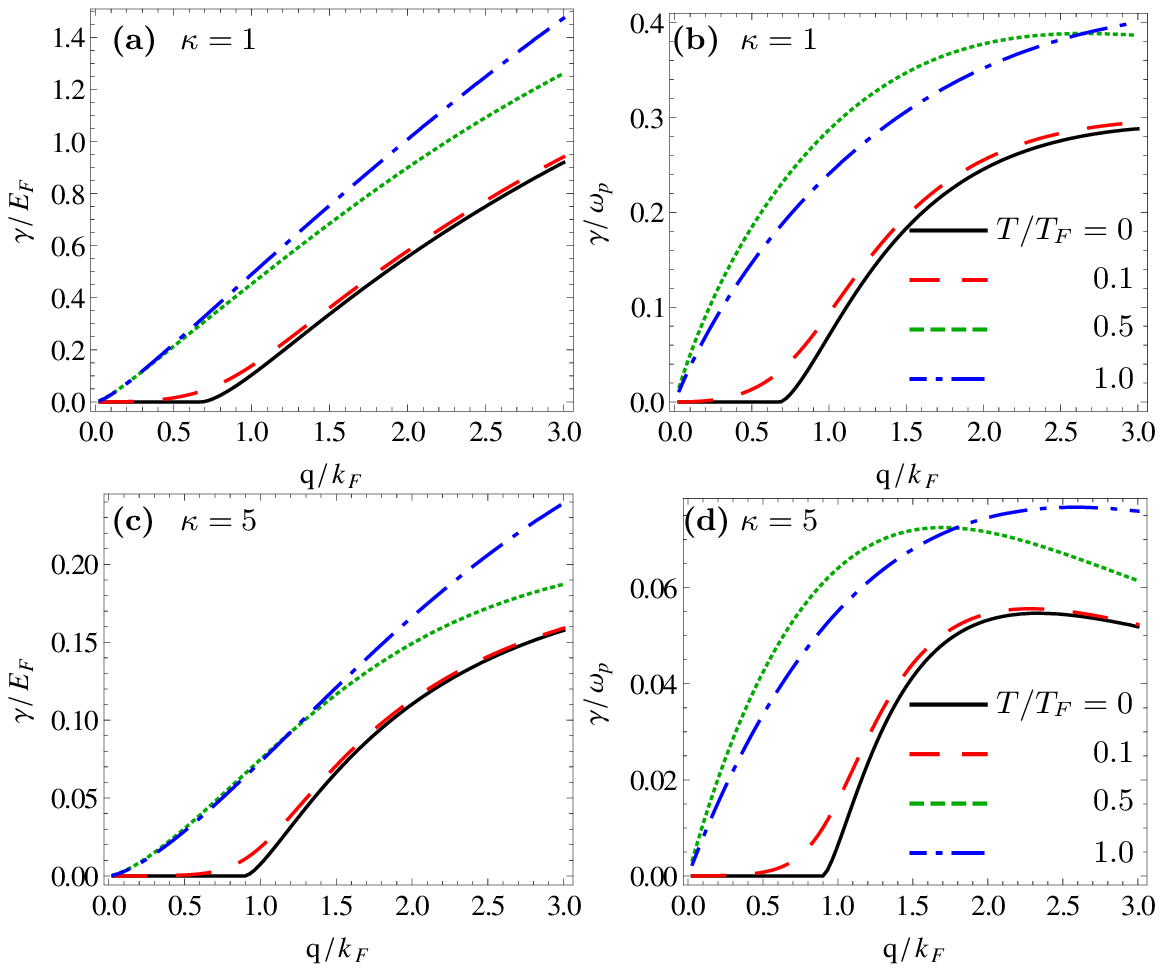}
  \caption{(Color online). Plasmon damping rate as a function of $q/k_F$ for extrinsic  MLG. (a) and (c) $\gamma/E_F$ as a function of $q/k_F$ for different temperatures. (b) and (d) present $\gamma/\omega_p$  as a function of $q/k_F$. (a) and (b) are for $\kappa = 1$; (c) and (d) are for $\kappa = 5$.  Note that the legend applies to all sub-figures.}
\label{fig7:exdamq}
\end{center}
\end{figure}

Putting $\mu=0$ and taking the long-wavelength limit $v_F q/k_B T \rightarrow 0$, we get from Eqs.~\eqref{eq:vareps} and \eqref{eq:pola} the following expression for the finite-temperature ($k_B T/v_F q \gg 1$) intrinsic graphene polarizability function (We mention that the complete analytical $\Pi(q,\omega)$ at $T=0$ for both intrinsic and extrinsic case, i.e. intraband and interband, can be found in Ref.~[\onlinecite{HwangDas_PRB07}]):
\beq
\Pi(q,\omega) \approx \frac{2\ln2}{\pi}\frac{q^{2}}{\omega^{2}}k_{B}T+\frac{i}{16}\frac{q^{2}}{\sqrt{\omega^2 - q^2 v_F^2}}\frac{\omega}{k_B T}\label{eq5:ianapo}
\eeq
which reduces to (in the limit $\omega \gg q v_F$ as $q \rightarrow 0$):
\beq
\Pi(q,\omega) \approx \frac{2\ln2}{\pi}\frac{q^{2}}{\omega^{2}}k_{B}T+\frac{i}{16}\frac{q^{2}}{k_{B}T}\label{eq6:ianapo}
\eeq
Putting Eq.~\eqref{eq6:ianapo} in Eq.~\eqref{eq:vareps}, and solving for the complex frequency defining the intrinsic plasmon, we get:
\begin{eqnarray}
\omega_{p} & = & \sqrt{(4\ln2) r_{s}\hbar v_{F}qk_{B}T} \label{eq:iomega}\\
\gamma &  = &\frac{\pi\hbar v_{F}qr_{s}}{8\sqrt{k_{B}T}}\sqrt{(\ln2) r_{s}\hbar v_{F}q} \label{eq:igamma}
\end{eqnarray}
We have restored $\hbar$ in Eqs.~\eqref{eq:iomega} and \eqref{eq:igamma} for the sake of clarity and usefulness (and $r_s = e^2 /\kappa \hbar v_F$ is the graphene fine-structure constant).

Equations \eqref{eq:iomega} and \eqref{eq:igamma} define the long-wavelength (and necessarily finite temperature) intrinsic MLG plasmon, which is determined by the variables temperature and wavevector (and not by any carrier density $n$ as in all ordinary plasmon modes). We note that $\omega_p, \gamma \rightarrow 0$ as $q \rightarrow 0$, but obeying different power laws: $\omega_p \sim \sqrt{q}$ consistent with the 2D plasmon behavior and $\gamma \sim q^{3/2}$. We also note that $\omega_p \sim \sqrt{T}$ and $\gamma \sim 1/\sqrt{T}$ whereas $\omega_p \sim \sqrt{r_s}$ and $\gamma \sim r_s^{3/2}$ with $r_s \sim \kappa^{-1}$ giving the dependence on the background dielectric constant.

We note that $\omega_p/\gamma=16 \kappa k_B T /(\pi e^2 q)$ from Eqs.~\eqref{eq:iomega} and \eqref{eq:igamma}, and therefore the long-wavelength intrinsic plasmon is well-defined as long as
\beq
q<\frac{16 \kappa k_B T}{ \pi e^2}
\eeq
which defines the condition for the existence of a well-defined long-wavelength intrinsic MLG plasmon. Thus, the intrinsic plasmon remains well-defined at long-wavelength for arbitrarily low temperature as long as one is probing wavevectors shorter than the critical wavevector $q_c$ defined by
\beq
q_c = \frac{16 \kappa k_B T}{ \pi e^2}
\eeq
For $q < q_c$, the MLG intrinsic plasmon exists as a well-defined long wavelength collective mode, and for $q  > q_c$, it is overdamped (i.e. $\gamma > \omega_p$).

Before providing our full numerical results for the MLG intrinsic plasmon for arbitrary $q$ and $T$, we briefly compare the analytical results for intrinsic and extrinsic MLG plasmons, which were earlier considered in Refs.~[\onlinecite{HwangDas_PRB07}] and [\onlinecite{dassarma2009}]. The $T=0$ plasmon dispersion for extrinsic (i.e. doped) graphene is given in the long wavelength limit by:
\beq
\omega_p = \left(2 r_s \hbar v_F q E_F\right)^{1/2} = \left(\frac{2 e^2 \hbar v_F q}{\kappa} \sqrt{\pi n}\right)^{1/2}
\label{eq:exomega}
\eeq
where $n$ is the carrier density (with a Fermi level $E_F = \hbar v_F \sqrt{\pi n} \neq 0$). It is easy to obtain the low-temperature analytical result for the MLG plasmon dispersion by using the finite-temperature expansion for the chemical potential: $\mu (T) \approx E_F \left[1- \frac{\pi^2}{6} \left(\frac{T}{T_F}\right)^2\right]$ for $T \ll T_F = E_F/k_B$. We get for $T \ll T_F$:
\beq
\omega_p = \left(2 r_s \hbar v_F q E_F \left[1-\frac{\pi^2}{6}\left(\frac{T}{T_F}^2\right)\right]\right)^{1/2}
\label{eq12:exlow}
\eeq
Thus, $\omega_p (T)=\omega_p(T=0) \left[1- \frac{\pi^2}{12} \left(\dfrac{T}{T_F}\right)^2\right]$, which is a small correction to the $T=0$ result. We note that the Landau damping for extrinsic plasmons at long wavelengths and low temperatures is exponentially suppressed, going as $e^{-T/T_F}$. We remark here that the reason that the finite-$T$ extrinsic plasmon is exponentially weak by Landau damped (as $T \rightarrow 0$) whereas the corresponding intrinsic plasmon has power law ($\gamma \sim 1/\sqrt{T}$) divergent Landau damping as $T \rightarrow 0$ is that, by definition, the intrinsic plasmon is always in the high-temperature regime for any temperatures since $E_F = k_B T_F =0$ for intrinsic plasmons. (Below we will discuss the high-temperature limit for the extrinsic plasmon.)

\begin{figure}[htb]
\begin{center}
\includegraphics[width=0.99\columnwidth]{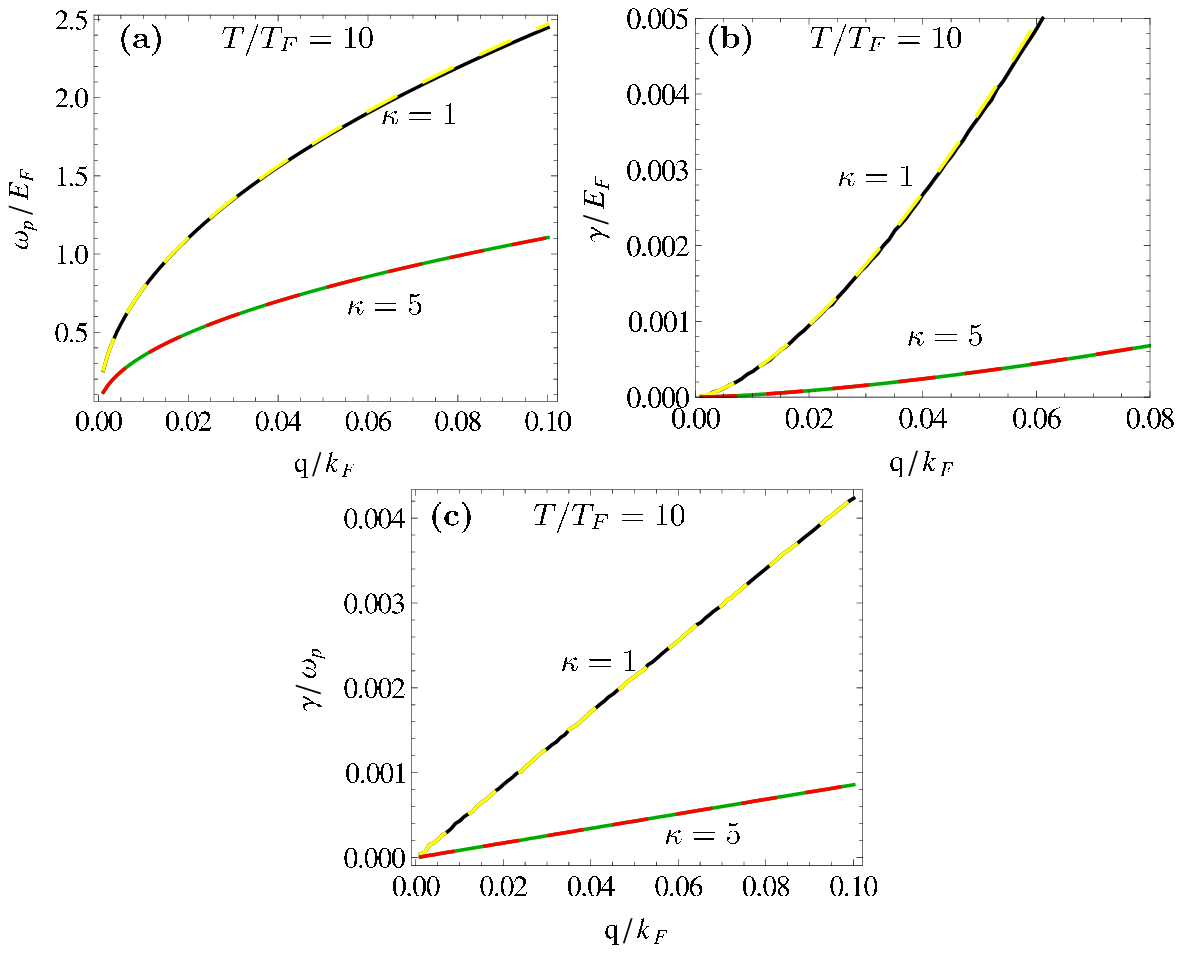}
  \caption{(Color online). Results for extrinsic  MLG with $T/T_F = 10.0$.  (a) presents plasmon dispersion in the long wavelength limit. The dashed and solid lines correspond to the analytical results (given in Eq.~\eqref{eq17:exomehigh}) and the numerical results, respectively. (b) Plasmon damping rate versus $q/k_F$. The dashed and solid lines correspond to the analytical results (given in Eq.~\eqref{eq18:exgamhigh}) and the numerical results, respectively. (c) $\gamma/\omega_p$ versus $q/k_F$. The dashed and solid lines correspond to the analytical results (given in Eq.~\eqref{eq17:exomehigh}) and the numerical results, respectively.}
\label{fig8:exdisqhight}
\end{center}
\end{figure}

\begin{figure}[htb]
\begin{center}
\includegraphics[width=0.99\columnwidth]{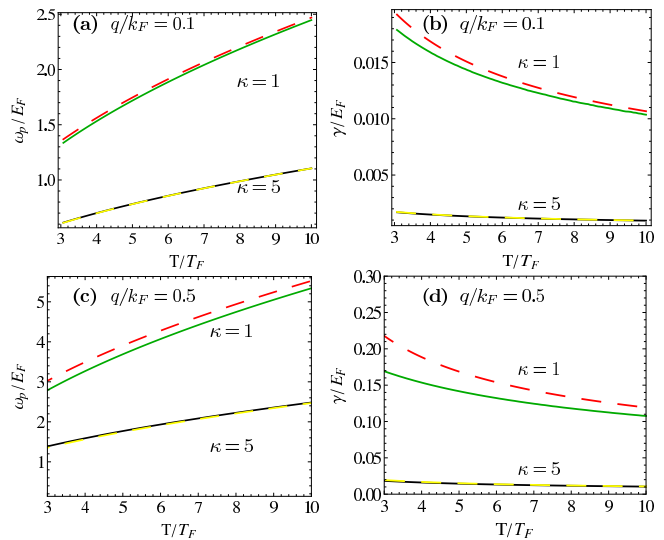}
  \caption{(Color online). Results  for extrinsic  MLG.  (a) and (c) present plasmon dispersion versus $T/T_F$. The dashed and solid lines correspond to the analytical results (given in Eq.~\eqref{eq17:exomehigh}) and the numerical results, respectively. (b) and (d) present plasmon damping rate versus $T/T_F$. (a) and (b) are for $q/k_F = 0.1$. (c) and (d) are for $q/k_F = 0.5$. The dashed and solid lines correspond to the analytical results (given in Eq.~\eqref{eq18:exgamhigh}) and the numerical results, respectively.  }
\label{fig9:exhtcom}
\end{center}
\end{figure}

\begin{figure}[htb]
\begin{center}
\includegraphics[width=0.99\columnwidth]{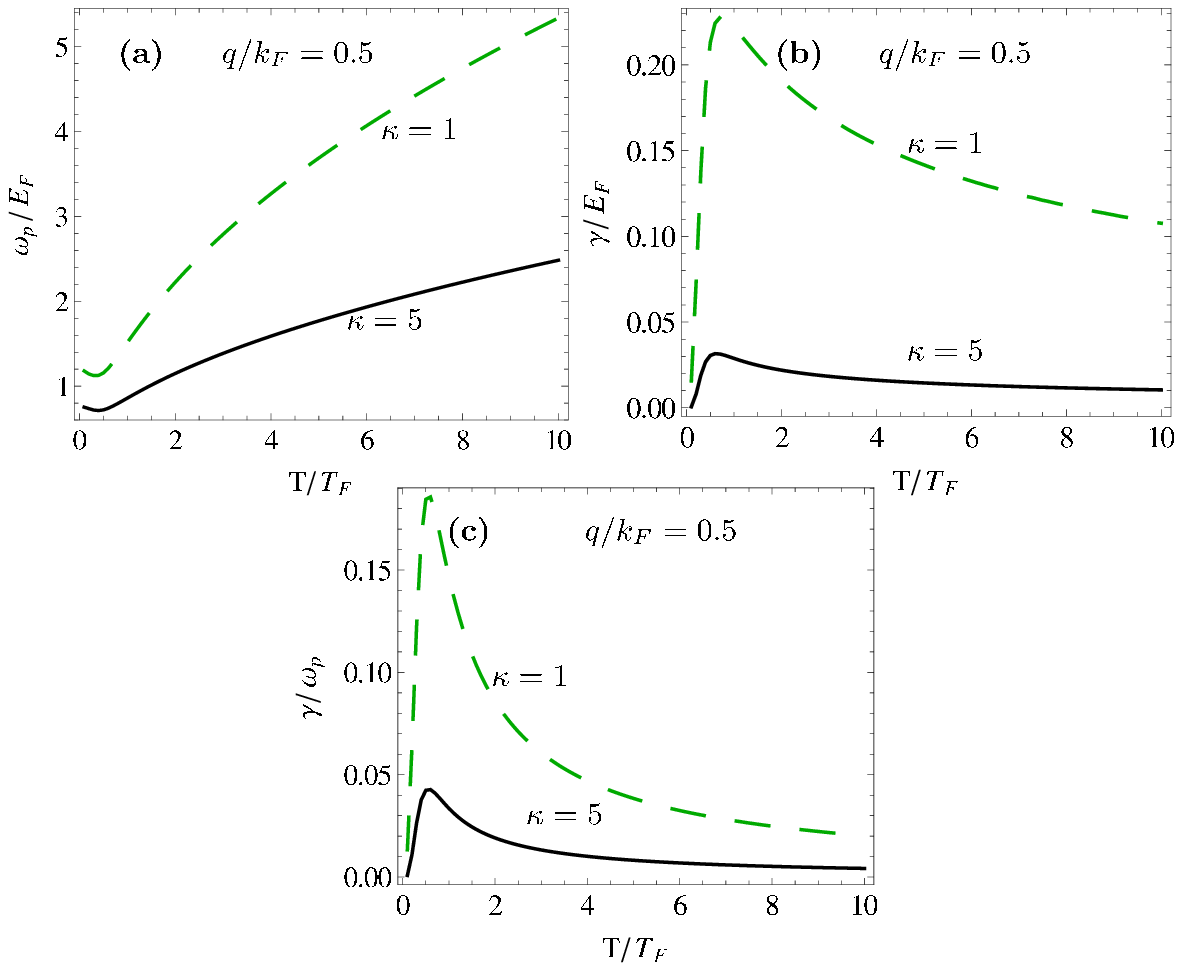}
  \caption{(Color online). Results  for extrinsic  MLG  with $q/k_F = 0.5$ over a wide $T/T_F =0 - 10$ range.  (a) Numerical results for $\omega_p/E_F$ as a function of $T/T_F$.  (b) Numerical results for $\gamma/E_F$ as a function of $T/T_F$. (c) Numerical results for $\gamma/\omega_p$ as a function of $T/T_F$. The dashed and solid lines correspond to $\kappa= 1$ and $\kappa =5$, respectively.  }
\label{fig10:exhttq0.5}
\end{center}
\end{figure}

An interesting exercise (alluded to in the Introduction of this paper) is to ask whether the long-wavelength intrinsic plasmon dispersion (i.e. Eq.~\eqref{eq:iomega}) can be obtained from the corresponding extrinsic plasmon dispersion (i.e. Eq.~\eqref{eq:exomega}) by reinterpreting the doping carrier density $n$ in Eq.~\eqref{eq:exomega} as the thermally excited carrier density $n(T)$ for the intrinsic case. The thermally excited carrier density $n(T)$ for intrinsic graphene with the Fermi level at the Dirac point ($E_F = 0$) is easily calculated to be:
\beq
n(T) = \int D(E)f(E)dE
\label{eq13:exden}
\eeq
where $D(E) = 4 E / (2 \pi v_F^2)$ is the graphene density of states. Integrating over the Fermi distribution function $f(E)$ at temperature $T$ with $\mu =0$ we get:
\beq
n(T) = \frac{\pi}{6}\frac{(k_B T )^2}{\hbar^2 v_F^2}
\label{eq14:exden}
\eeq
Inserting Eq.~\eqref{eq13:exden} for $n$ in Eq.~\eqref{eq:exomega} we get:
\beq
\omega_p = \left[\left(2 \sqrt{\dfrac{\pi^2}{6}}\right)r_s (\hbar v_F q)(k_B T)\right]^{1/2}
\label{eq15:exomega}
\eeq
Eq.~\eqref{eq15:exomega} has the same parameter dependence $\sqrt{r_s \hbar v_F q k_B T}$ as in the correct intrinsic plasmon dispersion given by Eq.~\eqref{eq:iomega} with the only difference is that the prefactor in Eq.~\eqref{eq15:exomega} is $2 \sqrt{\pi^2 /6} \approx 2.6$ versus the prefactor in Eq.~\eqref{eq:iomega} is $4 \ln 2 \approx 2.8$. Thus, the plasma frequency is given by $1.67\sqrt{r_s \hbar v_F q (k_B T)^{1/2}}$ and by $1.60 \sqrt{r_s \hbar v_F q (k_B T)^{1/2}}$ in Eq.~\eqref{eq15:exomega}.

We now consider the high-temperature limit of the extrinsic plasmon dispersion for gated or doped graphene taking $T \gg T_F (= E_F/k_B)$. The asymptotic high-temperature expression for $\Pi(q,\omega)$ in doped graphene is given by (again for $q v_F \ll \omega$):
\begin{eqnarray}
\Pi(q,\omega)\approx\frac{2\ln2}{\pi}\frac{q^{2}}{\omega^{2}}k_{B}T \left(1+\frac{{T_F}^4}{128 (\ln 2)^3 T^4} \right) \nonumber \\
+ \frac{i}{16}\frac{q^{2}}{\sqrt{\omega^{2}-q^{2}}}\frac{\omega}{k_{B}T} \left(1-\frac{\omega^2}{48 k_B^2 T^2}\right)
\label{eq16:exomhigh}
\end{eqnarray}
We emphasize that Eq.~\eqref{eq16:exomhigh} is valid for extrinsic graphene in the limit of $T \gg T_F$ and $q v_F \ll \omega$. Using Eq.~\eqref{eq16:exomhigh} to solve for the plasmon modes in Eq.~\eqref{eq:vareps}, we get:
\begin{eqnarray}
\omega_{p} & = & \sqrt{4\ln2 \ r_{s}\hbar v_{F}qk_{B}T \left(1+\frac{{T_F}^4}{128 (\ln 2)^3 T^4} \right)} \label{eq17:exomehigh}\\
\gamma&= &\frac{\pi\hbar v_{F}qr_{s}}{8\sqrt{k_{B}T}}\sqrt{\ln 2 \ r_{s}\hbar v_{F}q} \left(1-\frac{\hbar v_F q r_s \ln2}{12 k_B T}\right) \label{eq18:exgamhigh}
\end{eqnarray}
A direct comparison between Eqs.~\eqref{eq17:exomehigh}, \eqref{eq18:exgamhigh} and Eqs.~\eqref{eq:iomega}, \eqref{eq:igamma} show that the $T / T_F \rightarrow \infty$ limit of the extrinsic plasmon dispersion and broadening indeed agree with the corresponding intrinsic plasmon results in the leading order, as indeed it must. We mention, however, that this agreement is only in the $T/T_F \rightarrow \infty $ limit (i.e. the $T_F =0$ limit of the extrinsic situation). Thus, there is a correction to the leading-order extrinsic plasmon dispersion in Eq.~\eqref{eq17:exomehigh} going as $O(T_F^4/T^4)$ which, by definition, cannot exist in the intrinsic plasmon dispersion where the leading-order dispersion comes entirely as $\sqrt{q T}$ with no correction term in temperature. All higher-order temperature corrections to the intrinsic plasmon dispersion occur in higher-order terms in the wavevector $q$.

\begin{figure}[htb]
\begin{center}
\includegraphics[width=0.99\columnwidth]{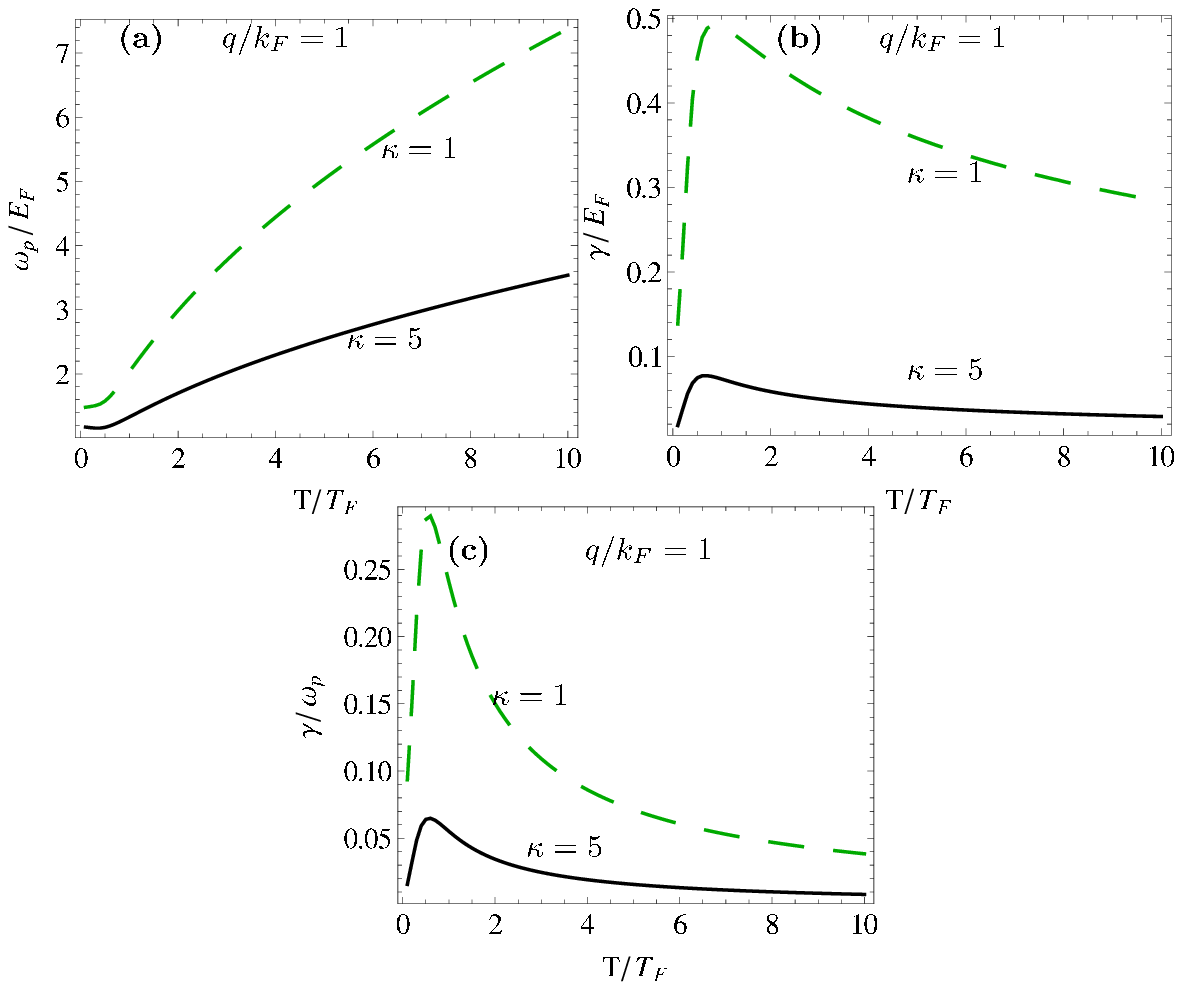}
  \caption{(Color online). Results  for extrinsic  MLG  with $q/k_F = 1.0$  over a wide $T/T_F =0 - 10$ range.  (a) Numerical results for $\omega_p/E_F$ as a function of $T/T_F$.  (b) Numerical results for $\gamma/E_F$ as a function of $T/T_F$. (c) Numerical results for $\gamma/\omega_p$ as a function of $T/T_F$. The dashed and solid lines correspond to $\kappa= 1$ and $\kappa =5$, respectively.  }
\label{fig11:exhttq1}
\end{center}
\end{figure}

\begin{figure}[htb]
\begin{center}
\includegraphics[width=0.99\columnwidth]{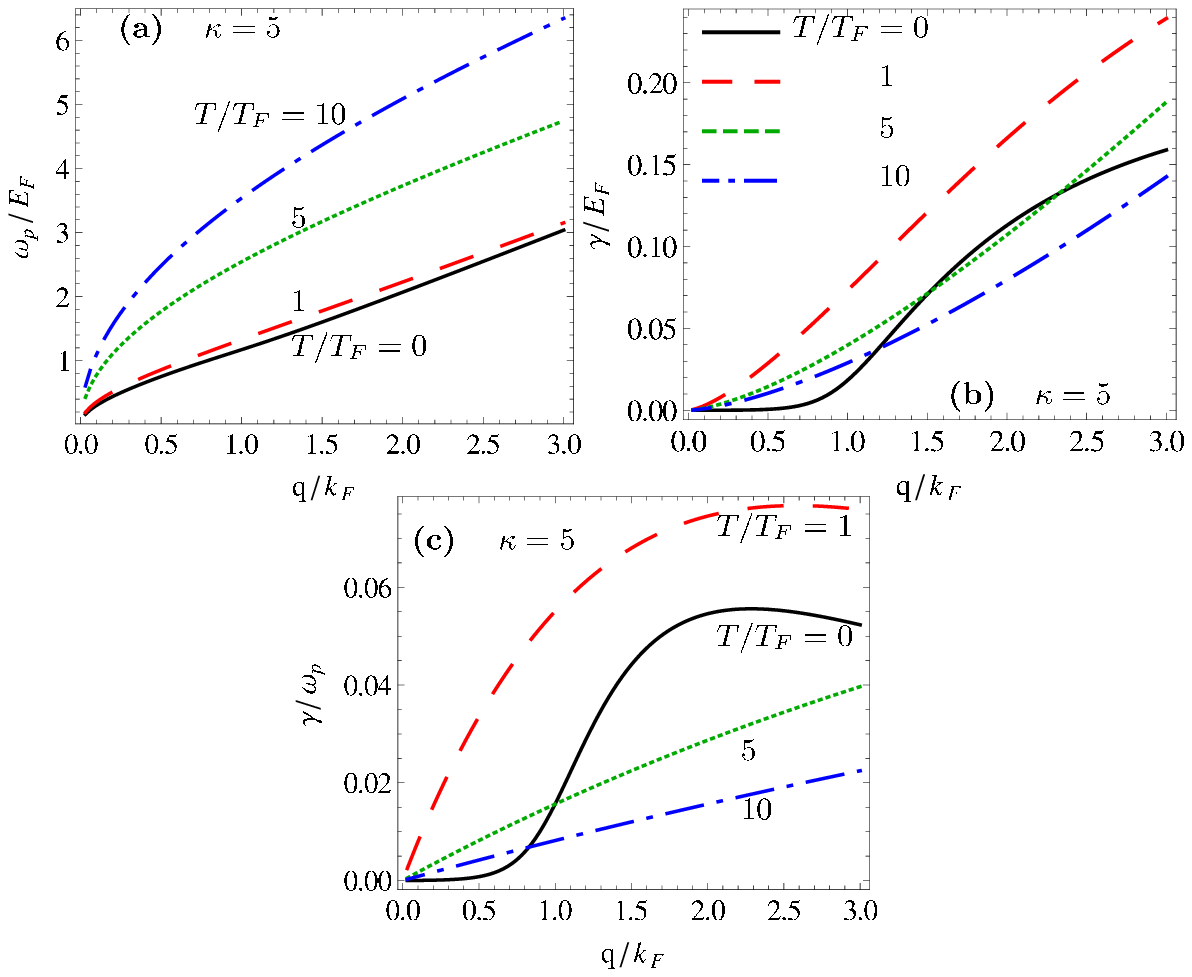}
  \caption{(Color online). Results  for extrinsic  MLG  with $\kappa = 5$.  (a) Numerical results for $\omega_p/E_F$ as a function of $q/k_F$.  (b) Numerical results for $\gamma/E_F$ as a function of $q/k_F$. (c) Numerical results for $\gamma/\omega_p$ as a function of $q/k_F$. The solid, dashed, dotted and dot-dashed lines correspond to $T/T_F =0,\ 1, \ 5$, and $10$, respectively.}
\label{fig12:exhtk5}
\end{center}
\end{figure}

\begin{figure}[htb]
\begin{center}
\includegraphics[width=0.99\columnwidth]{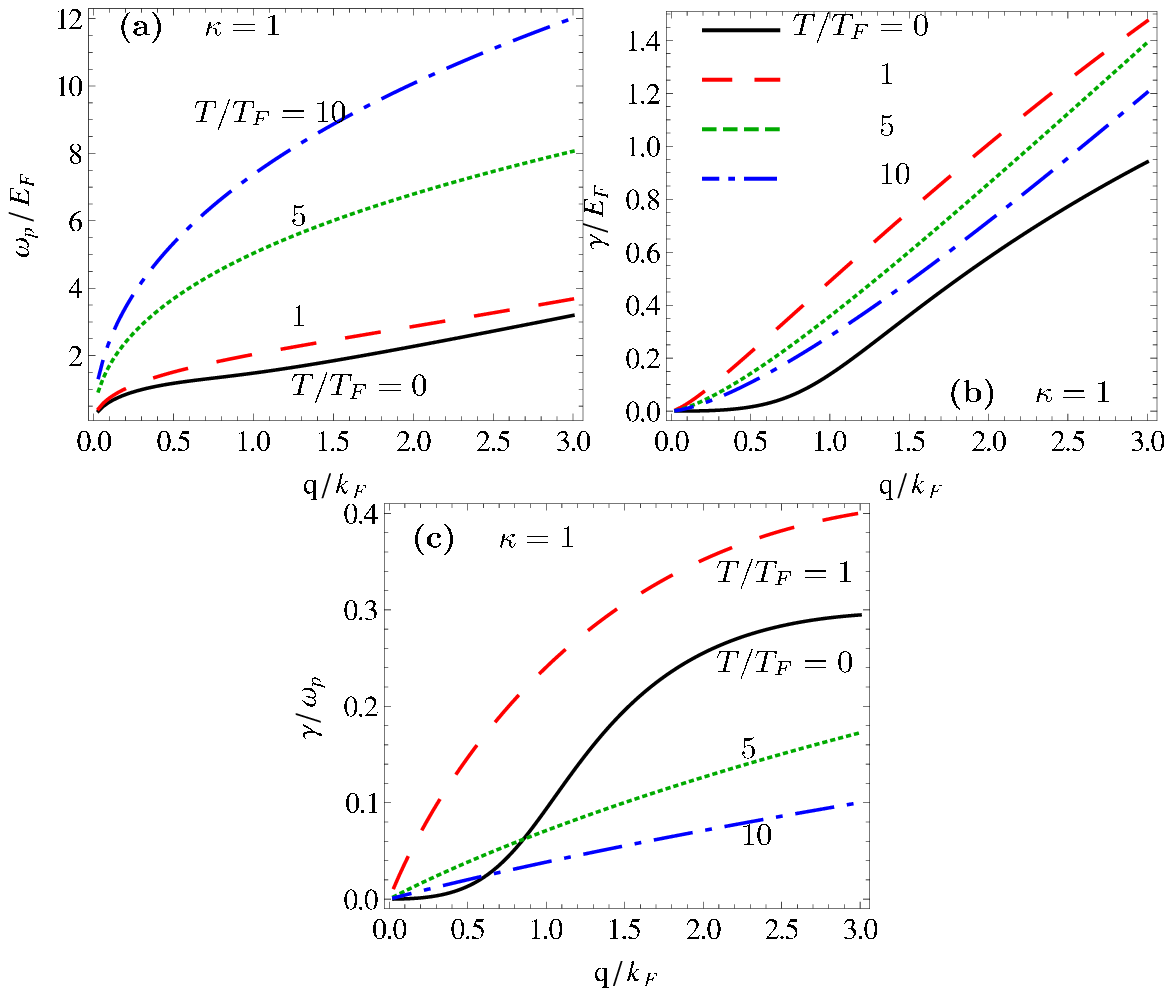}
  \caption{(Color online). Results  for extrinsic  MLG  with $\kappa = 1$.  (a) Numerical results for $\omega_p/E_F$ as a function of $q/k_F$.  (b) Numerical results for $\gamma/E_F$ as a function of $q/k_F$. (c) Numerical results for $\gamma/\omega_p$ as a function of $q/k_F$. The solid, dashed, dotted and dot-dashed lines correspond to $T/T_F =0,\ 1, \ 5$, and $10$, respectively.}
\label{fig13:exhtk1}
\end{center}
\end{figure}

We now present in Figs.~\ref{fig1:imdis} and \ref{fig2:imdis} our calculated numerical results for the intrinsic graphene energy dispersion and level broadening for arbitrary values of wavevector $q$ and temperature $T$. (Our analytical results presented above in Eqs.~\eqref{eq:iomega} and \eqref{eq:igamma} are necessarily restricted to the $\hbar v_F q \ll k_B T$ regime.) In presenting our results, we find that there are only two dimensionless (independent) variables that completely characterize the intrinsic graphene plasmon properties: $\kappa$, and $\hbar v_F q/k_B T$. In Figs.~\ref{fig1:imdis} and \ref{fig2:imdis}, we use two values of $\kappa =1$ (suspended graphene) and $5$ (graphene on boron nitride (BN) substrate) as representative examples of strongly ($\kappa =1$, i.e. $r_s \approx 2.2$) and weakly ($\kappa =5$, i.e. $r_s \approx 0.4$) interacting systems to show our numerical results for the numerically calculated plasmon energy $\hbar \omega_p$ and level broadening $\hbar \gamma$ in units of $k_B T$ as functions of the dimensionless 2D wavevector $\hbar v_F q/(k_B T)$. Our numerical results solve Eq.~\eqref{eq:vareps} to obtain the complex solution with the real part being the plasma frequency and the imaginary part the broadening. We obtain $\Pi(q,\omega)$ at arbitrary temperatures numerically in order to solve for the plasmon modes at arbitrary temperatures and wavevectors. In Fig.~\ref{fig1:imdis}, the plasma frequency is shown as a function of wavevector, both for the small-$q$ regime and over a large range of $q$. The small-$q$ results serve to verify the accuracy of our asymptotic analytic result given in Eq.~\eqref{eq:iomega}. In Fig.~\ref{fig2:imdis} we depict our calculated plasmon broadening as a function of wavevector again for small-$q$ (Fig.~\ref{fig2:imdis}(a)) and extended-$q$ (Fig.~\ref{fig2:imdis}(b)) regions whereas in Fig.~\ref{fig2:imdis}(c) we depict the dimensionless ratio $\gamma/\omega_p$ as a function of the dimensionless variable $\hbar v_F q/k_B T$.

The most notable, and perhaps somewhat unexpected, feature of our numerical results in Figs.~\ref{fig1:imdis} and \ref{fig2:imdis} is that the intrinsic plasmon mode remains well-defined, i.e. $\omega_p > \gamma$, for all values of $\hbar v_F q / k_B T$ with a shallow maximum around $\hbar v_F q \gtrsim k_B T$ manifesting a surprising nonmonotonic behavior for both values of $\kappa$ in Fig.~\ref{fig2:imdis}(c). For $\kappa=1$ ($r_s = 2.2$), suspended graphene, the maximum value of $\omega_p/\gamma$ reaches $0.4$, but for $\kappa = 5$ ($r_s = 0.4$), graphene on BN, the maximum value of $\omega_p/\gamma$ is below $0.1$. Thus, the intrinsic MLG plasmon should, in principle, be experimentally observable, particularly on substrates with large dielectric constant where $\gamma/\omega_p \ll 1$ making the Landau damping problem fairly irrelevant. Our results in Figs.~\ref{fig1:imdis} and \ref{fig2:imdis} also indicate that the leading order formula of Eqs.~\eqref{eq:iomega} and \eqref{eq:igamma} remain reasonably well-valid for arbitrary values of $\hbar v_F q/k_B T$.

Although our focus in the current work is the intrinsic Dirac point plasmon mode for undoped graphene, it is useful to compare the temperature dependence of the extrinsic plasmon in doped graphene with that of intrinsic graphene, particularly since the temperature dependence of extrinsic graphene plasmon has not much been studied in the literature. Understanding temperature dependent plasmon dispersion and damping of doped graphene is also relevant here since extrinsic and intrinsic graphene plasmons become the same at very high temperatures ($T \gg T_F$). We therefore provide a large set of finite-temperature results for extrinsic plasmon dispersion and damping, comparing them with our derived analytical low-and high-temperature results and with the corresponding intrinsic plasmon results. Our motivation for such a detailed finite-temperature RPA study of extrinsic graphene plasmons comes partially from the fact that temperature could in principle be used (in addition to wavevector and/or carrier density) to tune the plasmon energy in graphene (particularly at lower carrier densities and higher temperatures where $T/T_F$ is not necessarily extremely small), a fact which has not been much appreciated in the literature.

In showing our full numerical solutions for $\omega_p(q)$ and $\gamma(q)$ using Eq.~\eqref{eq:vareps} [and the full finite-temperature $\Pi(q,\omega)$] for extrinsic graphene, the first problem we face is that there are far too many independent variables (i.e. $q, T, n, \kappa$) which determine the plasmon properties. Since three of these variables are independent continuously tunable experimental variables (i.e. $q, T, n$), we need four-dimensional plots (for several values of $\kappa$, i.e. for different substrates) or perhaps even five-dimensional plots showing $\omega_p$ and $\gamma$ as functions of $q, T, n,$ and $\kappa$. A significant simplification arises from using $k_F (=\sqrt{\pi n})$ and $E_F(= \hbar v_F k_F = \hbar v_F \sqrt{\pi n})$ as the unit of wavevector and energy, respectively so that the carrier density shows up implicitly as a scaling variable rather than explicitly, eliminating one variable. We also show results only for two values of the background dielectric constant $\kappa=1$ (suspended graphene) and $5$ (graphene on h-BN substrates) corresponding to $r_s = e^2/(\hbar v_F \kappa) = 2.2$  and $0.4$, respectively. Thus, we present all our numerical results for $\omega_p/E_F$ and $\gamma/E_F$ as functions of $q/k_F$ and $T/T_F$ for $\kappa =1$ and $5$ in the following. A particular goal of the presented numerical results for arbitrary $q$ and $T$ is comparison with our analytical low- and high-temperature results in Eqs.~\eqref{eq12:exlow} and \eqref{eq17:exomehigh}/\eqref{eq18:exgamhigh}, respectively. Since we will be presenting a very large number of figures, we do not discuss all the figures individually in the text below, instead only highlighting the important salient features. We provide detailed figure captions in the figures themselves which should be self-explanatory.

In Figs.~\ref{fig3:exdist}-\ref{fig5:exdistq1} we present results as a function of $T/T_F$ for fixed $q/k_F = 0.01; 0.5; 1$. The extrinsic plasmon energy for fixed $q$ generically shows a non-monotonic dependence on temperature with a shallow minimum around $T \sim 0.4 T_F$. This arises from the fact that the degenerate system ($T \ll T_F$) has a plasma frequency decreasing with increasing temperature (according to Eq.~\eqref{eq12:exlow}) whereas the nondegenerate system has the plasma frequency increasing (as $\sim T$ for $T \gg T_F$)  with temperature. The broadening $\gamma$ is suppressed exponentially for small $T/T_F$ except for large $q$($\gtrsim k_F$) where intraband Landau damping starts playing a role, particularly for larger values of $r_s$ (smaller $\kappa$). The analytic formula, Eq.~\eqref{eq12:exlow}, for the low-temperature plasmon dispersion (and damping) seems to work very well (somewhat surprisingly) all the way to $T/T_F \sim 0.4$, i.e. all the way to the shallow minimum in $\omega_p(T)$. According to Figs.~\ref{fig3:exdist}-\ref{fig5:exdistq1}, the plasmon damping manifests a shallow maximum around the same value of $T/T_F$ ($\sim 0.4$) where the plasmon energy is a maximum, and thus $\gamma/\omega_p$ shows a generic peak for $T/T_F \sim 0.4$ (which is much sharper for larger $r_s$ values). In general, we find $\gamma/\omega_p <1$ for smaller $q$ values in the $T/T_F <1$ regime --- thus the plasmon is well-defined for $T < T_F$.

In Figs.~\ref{fig6:exdisq}-\ref{fig8:exdisqhight}, we show the plasmon energy and broadening at various finite $T$ values as a function of wavevector --- thus, Figs.~\ref{fig6:exdisq}-\ref{fig8:exdisqhight} are effectively complementary to Figs.~\ref{fig3:exdist}-\ref{fig5:exdistq1}. First, we note that the analytic results are essentially in exact agreement with the full numerical results upto $q \sim 0.5 k_F$. In Fig.~\ref{fig7:exdamq}, we show that $\gamma < \omega_p$ is well-satisfied for all $q$ values well up to $T \sim T_F$. In general, the broadening is exponentially suppressed at low $q$ and low $T$, but increasing either $q$ or $T$ eventually leads to intraband and interband Landau damping. In Fig.~\ref{fig7:exdamq}, the onset of the intraband Landau  broadening, where the extrinsic plasmon dispersion enters the intraband electron-hole single particle excitation continuum even at $T=0$, can clearly be seen around $q \sim 0.4- 0.7 k_F$, whereas for higher $T$ values, the extrinsic plasmon can decay even at long wavelength due to interband electron-hole excitation process.

Whereas in Figs.~\ref{fig3:exdist}-\ref{fig7:exdamq} we focus on low-temperature ($T \lesssim T_F$) extrinsic plasmon dispersion and damping, we now present in Figs.~\ref{fig8:exdisqhight}-\ref{fig13:exhtk1} higher-$T(> T_F)$ results. The higher-temperature ($T>T_F$) extrinsic plasmon results are relevant for understanding intrinsic plasmon behavior in graphene since, as emphasized in our analytical theory (see Eqs.~\eqref{eq17:exomehigh}, \eqref{eq18:exgamhigh}, and the discussions following their derivations), the leading-order (in $T_F/T$) results for both plasmon energy and damping for intrinsic and extrinsic plasmons are the same for $T \gg T_F$. Physically, the reason for this is obvious: For $T \gg T_F$, the thermal inter-band electron-hole excitations dominate the collective behavior over the contribution by the doped carriers even for extrinsic doped graphene (of course, the actual temperature scale needed to satisfy the $T \gg T_F$ condition increases as $\sqrt{n}$ with increasing the doping density).

One of the most important as well as interesting aspects of the numerical results shown in Figs.~\ref{fig8:exdisqhight}-\ref{fig13:exhtk1} is the great quantitative accuracy of our leading order analytic high-temperature ($T \gg T_F$) results (i.e. Eqs.\eqref{eq17:exomehigh} and \eqref{eq18:exgamhigh}) for extrinsic plasmon dispersion and damping as compared with the full RPA finite-temperature numerical solutions for $\omega_p(q)$ and $\gamma(T)$. In particular, our analytical theory seems to hold very well all the way down to $T \gtrsim T_F$ although the analytic theory represents an asymptotic expansion in $T_F/T$. This remarkable reliability of the high-temperature analytic theory (for $T \geqslant T_F$) as well as that of the low-temperature ($T \leqslant T_F$) analytic theory (derived as an asymptotic expansion in $T/T_F$), which we discussed in the context of Figs.~\ref{fig3:exdist}-\ref{fig7:exdamq} above, implies that the analytical finite-temperature theory developed in the current work could be extremely useful for experimental works in graphene plasmonics with no need for the full numerical solution of the RPA theory which is, in fact, quite complex and demanding at finite temperatures since the finite-temperature graphene polarizability $\Pi(q,\omega; T)$ does not have any simple analytical form and must be carefully calculated through a numerical integration at each value of $T$. If necessary, one could easily develop a numerical interpolation scheme between our low-temperature and high-temperature analytical theories (e.g. using a suitable Pad\'{e}  approximation) which should provide a reasonable and quantitatively accurate theory at arbitrary temperatures. Since the high-temperature analytical extrinsic plasmon theory (i.e. Eqs.~\eqref{eq17:exomehigh} and \eqref{eq18:exgamhigh}) essentially agree with the intrinsic plasmon results, the $T \gg T_F$ results provided in Figs.~\ref{fig8:exdisqhight}-\ref{fig13:exhtk1} could be construed as numerical results for the intrinsic plasmon as well.

\subsection{Double layer graphene}

Collective modes of two (i.e. ``double") parallel 2D graphene layers (along the $x-y$ plane) separated by a distance ``d" in the third direction ($z-$direction) were first theoretically considered by Hwang and Das Sarma\cite{hwang2009b} and later by other authors\cite{varifeh_PLA10,yuan_PRB11,Bahrami20123518,staubersantos_PRB12,stauber_NJP12,tuan_commphys12,profumo_PRB12,badalyan_PRB12,
badalyanpeeter_PRB12,Triolacenrico_PRB12}. In the current work, we focus on the double-layer system considering their plasmon modes at finite temperatures, when one (or both of the layers) is (are) intrinsic or undoped. Thus, our work is the finite-temperature generalization of the Hwang-Das Sarma work, concentrating on intrinsic plasmons in undoped double layers.

The collective modes of a double-layer system is obtained by diagonalizing the $2 \times 2$ determinantal equation \cite{dasmadhukar_PRB82} defined in Eq.~\eqref{eq:vareps}, which gives:
\beq
(1-V_{1111}\Pi_{11})(1-V_{2222}\Pi_{22})-V_{1212}^2\Pi_{11}\Pi_{22}=0
\label{eq19:dbeps}
\eeq
where $\Pi_{ll}$ (with $l=1,2$) is the polarizability of the $l$th layer, $V_{llll}$ is the Coulomb interaction (i.e. $2 \pi e^2/(\kappa_l q)$) in the $l$th layer and $V_{ll'll'}$ is the Coulomb interaction between electrons in the $l$ and $l'$ layers. We ignore here (and throughout this paper) the possibility of electron hopping between the two layers which is an excellent approximation for graphene. For our double-layer system we have:
\beq
V_{1111} = V_{2222} = \frac{2 \pi e^2}{\kappa q}
\label{eq20:v}
\eeq
and
\beq
V_{1212} = \frac{2 \pi e^2}{\kappa q} e^{-q d}
\label{eq21:v2}
\eeq
where we have assumed both layers to be submerged in the same background dielectric with $\kappa$ as the common background lattice dielectric constant. (A generalization\cite{stauber_NJP12} to the situation with $\kappa_1$, $\kappa_2$ is straightforward, but will not be considered in our work since it will add more parameters to a problem which already has far too many variables.)

Using the known analytical expressions for $\Pi=\Pi_{11} = \Pi_{22}$ in the long-wavelength limit for intrinsic graphene (Eqs.~\eqref{eq5:ianapo} and \eqref{eq6:ianapo}), we can solve Eq.~\eqref{eq19:dbeps} to get the following two coupled long-wavelength collective modes ($\omega_{op}$ and $\omega_{ap}$) for the double-layer system when both layers are intrinsic graphene:
\begin{eqnarray}
\omega_{op} &=&  \sqrt{(8 \ln 2)  (\hbar v_F q) ( k_B T) r_s} \label{eq22:dint}\\
\omega_{ap} &=& \sqrt{(4 \ln 2) (r_s d q) (\hbar v_F q) k_B T} \label{eq23:dint}\\ \nonumber \\
\gamma_{op}  &=&  \frac{\pi r_{s} \hbar v_{F}q}{\sqrt{ 8 k_{B}T}}\sqrt{(\ln 2) r_{s}\hbar v_{F}q} \label{eq24:dint} \\
\gamma_{ap}  &=& \frac{\pi r_{s} \hbar v_{F}q^2 d}{8\sqrt{k_{B}T}}\sqrt{(\ln2) r_{s}\hbar v_{F}q^2 d} \label{eq25:dint}
\end{eqnarray}
Equations \eqref{eq22:dint}-\eqref{eq25:dint} define the long wavelength collective modes and their damping for an intrinsic double-layer graphene system where both layers are undoped (and the Fermi level in both layers is sitting at the Dirac point).

We first discuss the implications of our derived (Eqs.~\eqref{eq22:dint}-\eqref{eq25:dint}) analytical results for double-layer graphene intrinsic plasmons. It sounds crazy that two undoped graphene layers (i.e. no free carriers whatsoever) can have two low-energy collective modes ($\omega_{op}$ and $\omega_{ap}$ above) when they are proximate to each other, but it is apparently true. One of these modes, $\omega_{op}$, is nothing other than the combined collective intrinsic plasmon mode of each independent layer with $\omega_{op}^2 = 2 \omega_p^2$, where $\omega_p$ is the intrinsic plasmon frequency of a single graphene layer as given in Eq.~\eqref{eq:iomega}. Thus, $\omega_{op}$ is simply the in-phase intrinsic plasma oscillation of the two intrinsic plasmons in the two layers. The $\omega_{op}$ mode is sometimes referred to as the ``optical plasmon" mode\cite{dasmadhukar_PRB82} of the $2-$component (i.e. two layers) double-layer system since it involves the in-phase collective charge density oscillation of the two layers (analogous to an optical phonon mode in a lattice). The second mode, $\omega_{ap}$, which has no analog in the single layer system, is the acoustic plasmon mode\cite{dasmadhukar_PRB82} where the charge density oscillates out of phase between the two layers.

The $\omega_{op} (\propto \sqrt{q})$ obviously has the same dispersion as the single layer plasmon whereas the $\omega_{ap} \propto q$ has an acoustic dispersion linear in $q$ at long wavelengths. Both modes have the basic intrinsic plasmon property of $\omega_p \propto \sqrt{T}$ as expected, and vanishes in the $T \rightarrow 0$ limit. The broadening (Eqs.~\eqref{eq24:dint} and \eqref{eq25:dint}) has a higher-order $q-$dependence, $q^{3/2}(q^3)$ for $\omega_{op} (\omega_{ap})$, respectively, ensuring that both modes are well-defined collective modes in the long-wavelength limit. In particular, we have:
\begin{eqnarray}
\omega_{op}/\gamma_{op} &=&  \frac{8 k_B T \kappa}{\pi e^2 q}\label{eq26:dint}\\
\omega_{ap}/\gamma_{ap} &=& \frac{16 k_B T \kappa}{\pi e^2 q^2 d}  \label{eq27:dint}
\end{eqnarray}
Thus, the optical plasmon mode of the double layer system has very similar behavior for $\omega_p/\gamma$ as in the corresponding single layer intrinsic plasmon case except that the ratio of $\omega_p/\gamma$ is a factor of 2 smaller for the double layer case than the single-layer case. The $\omega_{op}$ mode therefore remains well-defined (i.e. $\omega_{op} > \gamma_{op}$) down to a cut-off wavevector
\beq
q_{oc} = \frac{8 \kappa k_B T }{\pi e^2}
\label{eq28:dinqc}
\eeq
with $\omega_{op} > \gamma_{op}$ for all $q < q_{oc}$.

The situation for the intrinsic acoustic plasmon $\omega_{ap}$ is, however, qualitatively different since $\omega_{ap} \propto q$ (rather than $\sqrt{q}$ as for $\omega_{op}$) and thus it decreases fast as $q \rightarrow 0$. The cut-off wavevector for a well-defined $\omega_{ap}$ mode is:
\beq
q_{ac} = 4 \sqrt{\frac{\kappa k_B T}{d \pi e^2}}
\label{eq29:dinac}
\eeq
with $\omega_{ap} > \gamma_{ap}$ for all $q < q_{ac}$.

We point out that the analytical results for the acoustic plasmon given above (Eqs.~\eqref{eq23:dint} and \eqref{eq25:dint}) apply only when $k_B T > \hbar v_F/(4 \ln 2\ r_s d)$ in order to satisfy the criterion $\omega > \hbar v_F q$ used in the expansion of $\Pi(q,\omega)$ to derive the long-wavelength plasmon dispersion relation.

\begin{figure}[htb]
\begin{center}
\includegraphics[width=0.99\columnwidth]{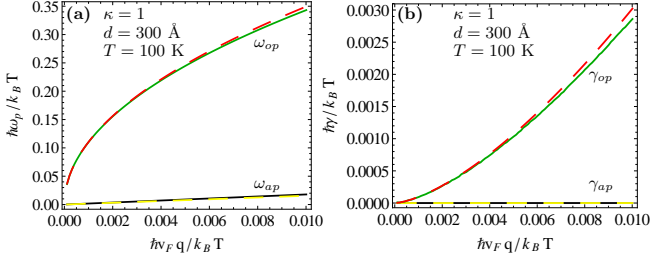}
  \caption{(Color online). Results for an intrinsic double-layer graphene system in the small-$q$ regime with $\kappa =1$, $d=300$ \AA \ and $T=100$ K. (a) Plasmon  dispersion versus $\hbar v_F q/(k_B T)$.  (b) Plasmon damping rate versus $\hbar v_F q/(k_B T)$.  The dashed and solid lines correspond to the analytical results (given in Eqs.~\eqref{eq22:dint}-\eqref{eq25:dint}) and the numerical results, respectively. }
\label{fig14:dint_com}
\end{center}
\end{figure}

\begin{figure}[htb]
\begin{center}
\includegraphics[width=0.99\columnwidth]{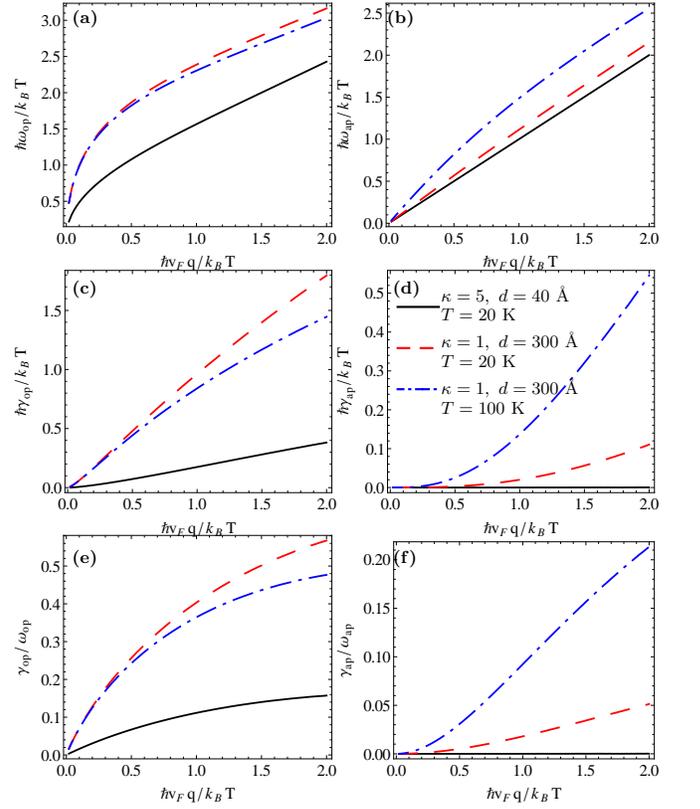}
  \caption{(Color online). Numerical results for an intrinsic double-layer graphene system. (a) and (b) Plasmon dispersion versus $\hbar v_F q/(k_B T)$; (c) and (d) Plasmon damping rate versus $\hbar v_F q/(k_B T)$; (e) and (f) $\gamma/\omega_p$ versus $\hbar v_F q/(k_B T)$.  (a), (c) and (e) are for the optical plasmon mode. (b), (d) and (f) are for the acoustic plasmon mode. For $\kappa = 5$, $d=40$ \AA, and $T=20$ K, the acoustic plasmon mode $\omega_{ap}$ is degenerate with the
boundary of the single particle excitation region. Note that the legend applies to all sub-figures. }
\label{fig15:dint_com}
\end{center}
\end{figure}

In Figs.~\ref{fig14:dint_com} and \ref{fig15:dint_com} we provide our full RPA numerical results for the double-layer intrinsic plasmon modes and compare them with our analytical results (Eqs.~\eqref{eq22:dint}-\eqref{eq25:dint}) obtained above. To keep the number of presented figures tractable we only change $q$ and $T$ for showing our results. In general, the theory no longer scales with $\hbar v_F q/k_B T$ as the corresponding single-layer intrinsic problem does because of the presence of the layer separation $d$ in the problem. For a fixed ``$d$" and ``$T$", however, we can still show results using $k_B T$ as the energy unit remembering that these double-layer results of Figs.~\ref{fig14:dint_com} and \ref{fig15:dint_com} apply only for fixed results of $\kappa, d, T$ as shown in the figure (but for varying $q$).

In Fig.~\ref{fig14:dint_com} we show the coupled intrinsic plasmon modes of double-layer graphene for small values of $\hbar v_F q/k_B T$ where our analytical expressions derived in Eqs.~\eqref{eq22:dint}-\eqref{eq25:dint} are essentially exact. The $\omega_{op}$ ($\omega_{ap}$) modes show the expected $\sqrt{q}$ ($q$) dispersion, and typically $\omega_{op} > \gamma_{op}$ ($\omega_{ap} > \gamma_{ap}$). In Fig.~\ref{fig15:dint_com} we depict the typical plasmon dispersion and damping for the coupled modes for three values of $\kappa, d, T$ over a broader range of $\hbar v_F q / k_B T$, finding that for layer values of $d (= 300$ \AA) and smaller $\kappa (=1)$ the damping could be quite large.

Next, we consider the double-layer graphene plasmons when one layer is doped (``extrinsic") and one undoped (``intrinsic"), i.e. one layer has carrier density $n \neq 0$ and the other has $n=0$ at $T=0$. The long wavelength and low temperature RPA collective modes of such a mixed intrinsic-extrinsic graphene double-layer system  are easily derived to be given by:

\begin{footnotesize}
\begin{eqnarray}
\omega_{op} &=& \left[2 r_s \hbar^2 v_F^2 q k_{F}  \left(1+ (2 \ln2)\frac{  T}{T_{F}}-\frac{\pi^2}{6}\frac{T^2}{T^2_{F}} \right) \right]^{1/2}   \label{eq30:dexinana}\\
\omega_{ap} &=& \sqrt{(8\ln 2) r_s \hbar v_F q^2 d k_B T \Big[ 1 - \frac{ (2 \ln2)   T}{T_{F}} + \frac{ 4 (\ln2)^2 T^2}{T^2_{F}}\Big]} \label{eq31:dexinana}\\
\gamma_{op}  &=&  \left[\frac{\pi^2 r_s^3 \hbar^3 v_F^3 q^3 E_{F}}{128 k_B^2 T^2}\left(1+ (2 \ln 2) \frac{T}{T_{F}}-\frac{\pi^2}{6}\frac{T^2}{T^2_{F}} \right)  \right]^{1/2}  \label{eq32:dexinana} \\
\gamma_{ap}  &=& \sqrt{\frac{\pi^2 \ln 2 r_s^3 \hbar^3 v_F^3 q^6 d^3}{8k_B T}  \Big [1-(10 \ln2) \frac{T}{T_{F}}+ 60 (\ln2)^2\frac{T^2}{T^2_{F}} \Big] } \label{eq33:dexinana}
\end{eqnarray}
\end{footnotesize}

Here $r_s = e^2/(\kappa \hbar v_F)$ refers to both layers, and $k_F = (\pi n)^{1/2}$ and $T_F = E_F/k_B = \hbar v_F k_F/k_B$ refer to the doped extrinsic layer.

While the above results are valid in the $T/T_F \ll 1$ (as well as leading order in $q$) limit, we can also obtain the high-temperature ($T/T_F \gg 1$) asymptotic analytical results to be exactly the same as those given in Eqs.~\eqref{eq22:dint}-\eqref{eq25:dint} for double-layer intrinsic graphene. This is, of course, expected since in the $T \gg T_F$ limit, there is no difference in the leading order between intrinsic and extrinsic graphene.

\begin{figure}[htb]
\begin{center}
\includegraphics[width=0.99\columnwidth]{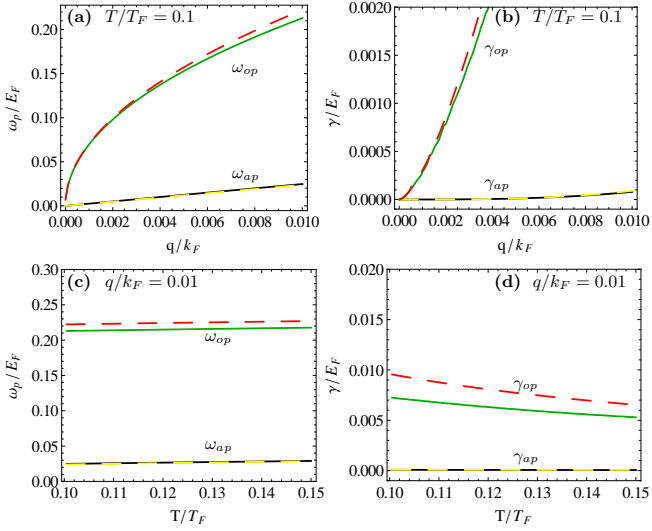}
  \caption{(Color online). Results for a mixed double-layer intrinsic-extrinsic graphene system with $n_1 = 10^{12}$ cm$^{-2}$, $n_2/n_1 = 0.0$, $d=300$ \AA \ and $\kappa=1$. (a) $\omega_p/E_F$ versus $q/k_F$ for $T/T_F = 0.1$. (b)$\gamma/E_F$ versus $q/k_F$ for $T/T_F = 0.1$. (c) $\omega_p/E_F$ versus $T/T_F$ for $q/k_F = 0.01$. (d)$\gamma/E_F$ versus $T/T_F$ for $q/k_F = 0.01$. The dashed and solid lines correspond to the analytical results (given in Eqs.~\eqref{eq30:dexinana}-\eqref{eq33:dexinana}) and the numerical results, respectively. }
\label{fig16:dex_com}
\end{center}
\end{figure}

\begin{figure}[htb]
\begin{center}
\includegraphics[width=0.99\columnwidth]{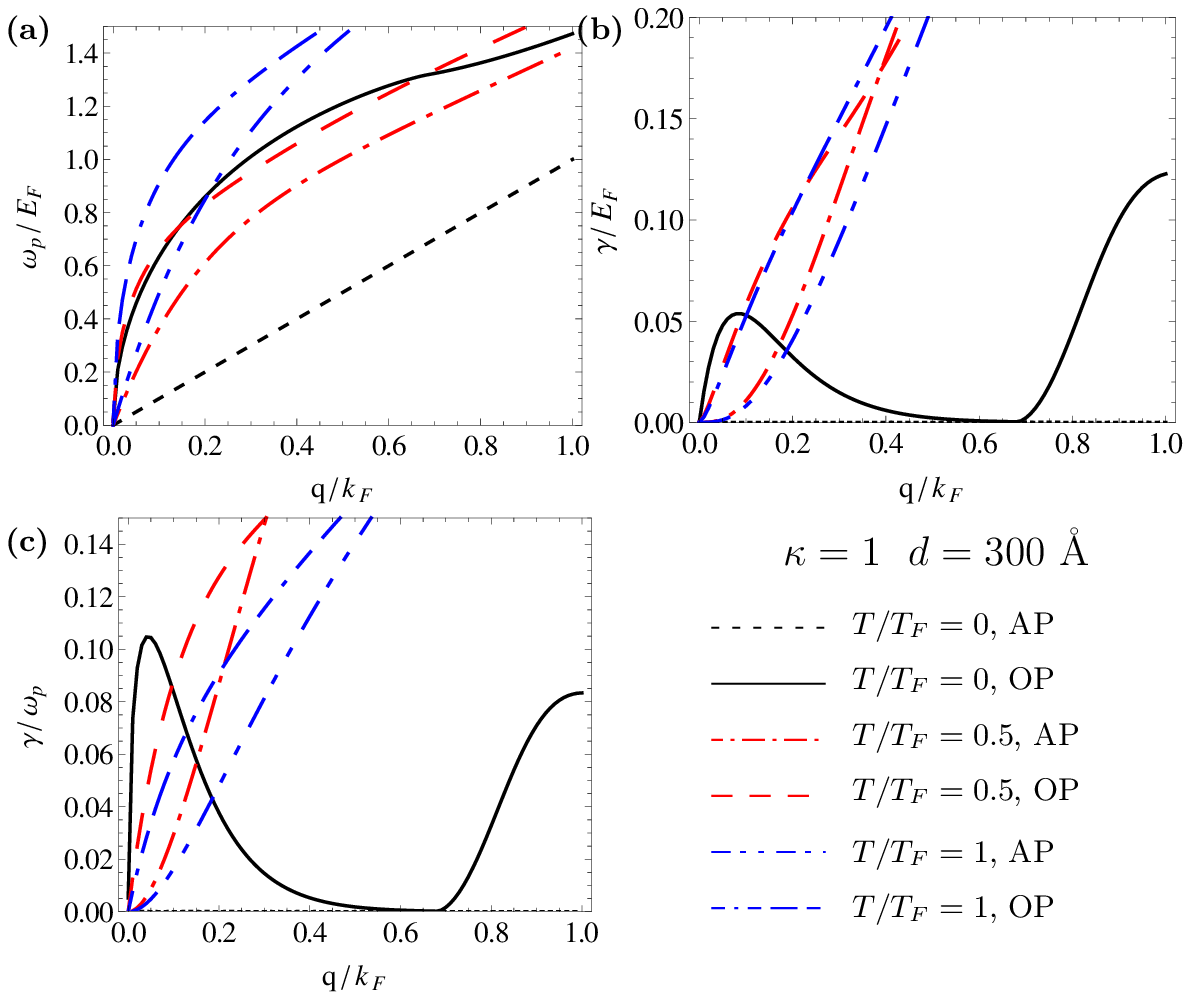}
  \caption{(Color online). Numerical results for a mixed double-layer intrinsic-extrinsic graphene system over a wide $q/k_F = 0-1$ range with $n_1 = 10^{12}$ cm$^{-2}$, $n_2/n_1 = 0.0$, $d=300$ \AA \ and $\kappa=1$.  (a) $\omega_p/E_F$ versus $q/k_F$ for different temperatures. (b) $\gamma/E_F$ versus $q/k_F$ for different temperatures. (c) $\gamma/\omega_p$ versus $q/k_F$ for different temperatures. For $T/T_F = 0$, the acoustic plasmon mode $\omega_{ap}$ is degenerate with the
boundary of the single particle excitation region. Note that the legend applies to all sub-figures.}
\label{fig17:dexin_3}
\end{center}
\end{figure}

\begin{figure}[htb]
\begin{center}
\includegraphics[width=0.99\columnwidth]{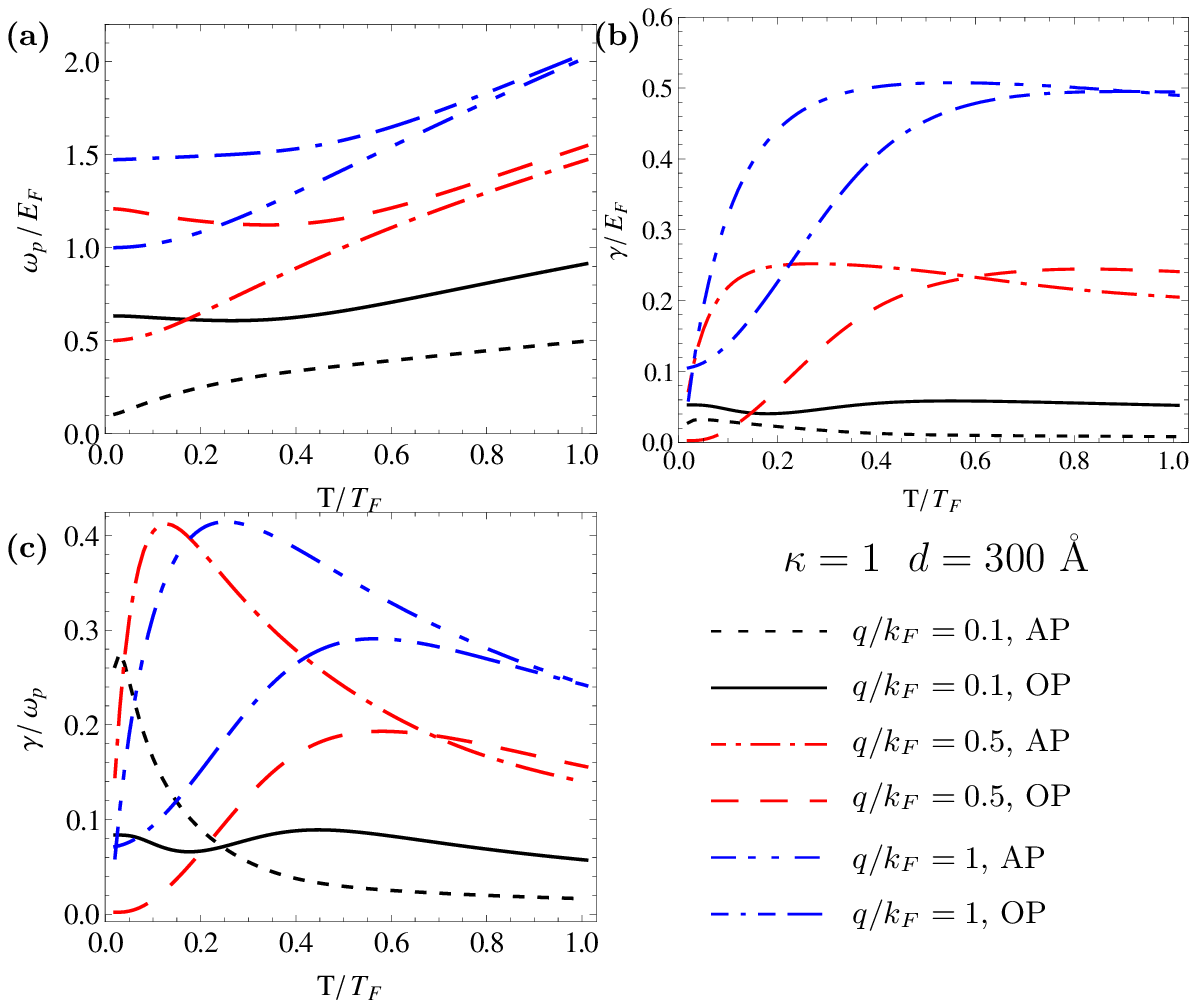}
  \caption{(Color online). Numerical results for a mixed double-layer intrinsic-extrinsic graphene system over a wide $T/T_F = 0-1$ range with $n_1 = 10^{12}$ cm$^{-2}$, $n_2/n_1 = 0.0$, $d=300$ \AA \ and $\kappa=1$.  (a) $\omega_p/E_F$ versus $T/T_F$ for different $q/k_F$. (b) $\gamma/E_F$ versus $T/T_F$ for different $q/k_F$. (c) $\gamma/\omega_p$ versus $T/T_F$ for different $q/k_F$. Note that the legend applies to all sub-figures. }
\label{fig18:dexin_3a}
\end{center}
\end{figure}

\begin{figure}[htb]
\begin{center}
\includegraphics[width=0.99\columnwidth]{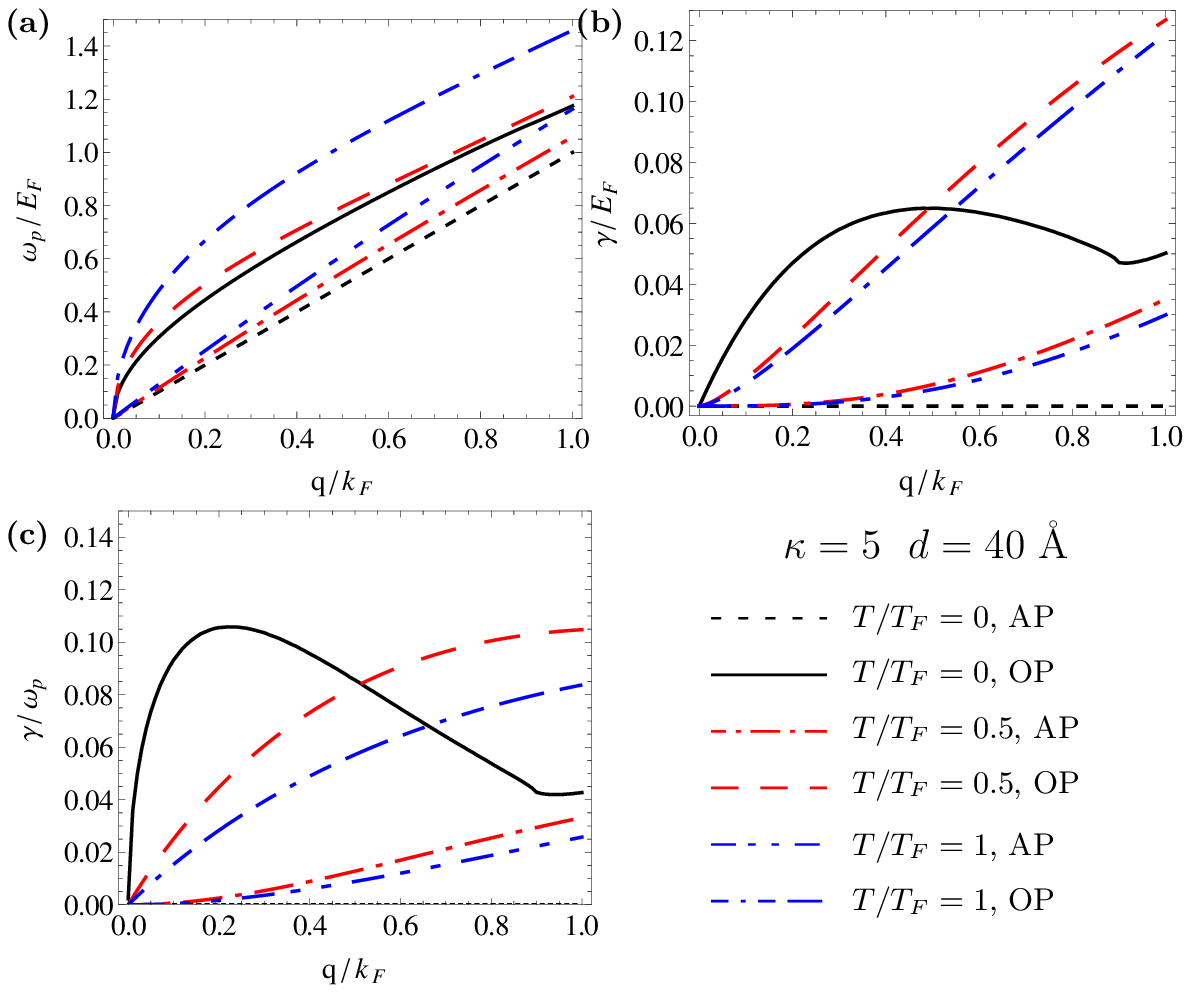}
  \caption{(Color online). Numerical results for a mixed double-layer intrinsic-extrinsic graphene system over a wide $q/k_F = 0-1$ range with $n_1 = 10^{12}$ cm$^{-2}$, $n_2/n_1 = 0.0$, $d=40$ \AA \ and $\kappa=5$.  (a) $\omega_p/E_F$ versus $q/k_F$ for different temperatures. (b) $\gamma/E_F$ versus $q/k_F$ for different temperatures. (c) $\gamma/\omega_p$ versus $q/k_F$ for different temperatures. For $T/T_F = 0$, the acoustic plasmon mode $\omega_{ap}$ is degenerate with the
boundary of the single particle excitation region. Note that the legend applies to all sub-figures. }
\label{fig19:dexin_4}
\end{center}
\end{figure}

\begin{figure}[htb]
\begin{center}
\includegraphics[width=0.99\columnwidth]{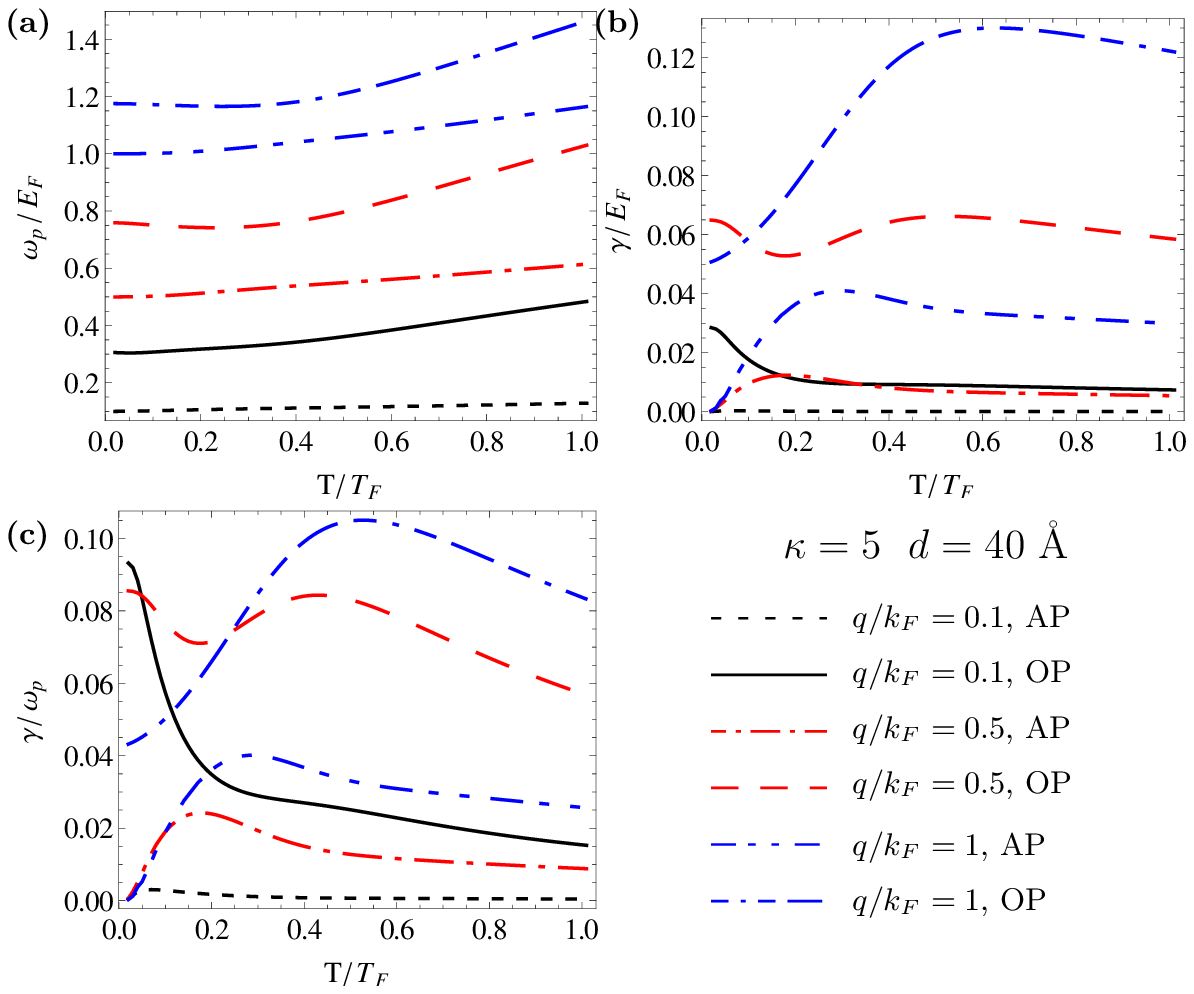}
  \caption{(Color online). Numerical results for a mixed double-layer intrinsic-extrinsic graphene system over a wide $T/T_F = 0-1$ range with $n_1 = 10^{12}$ cm$^{-2}$, $n_2/n_1 = 0.0$, $d=40$ \AA \ and $\kappa=5$.  (a) $\omega_p/E_F$ versus $T/T_F$ for different $q/k_F$. (b) $\gamma/E_F$ versus $T/T_F$ for different $q/k_F$. (c) $\gamma/\omega_p$ versus $T/T_F$ for different $q/k_F$.  Note that the legend applies to all sub-figures.  }
\label{fig20:dexin_4a}
\end{center}
\end{figure}

The mixed double-layer intrinsic-extrinsic graphene system depends in a complicated manner on a large number of independent parameters: $q, n, T, d, \kappa$. The plasmon modes now really depend on all five of these parameters (plus the value of $v_F$ which in principle is also a free parameter of the theory). We refrain from overloading the readers with a large number of results varying all five parameters freely. We provide some representative plasmon dispersion and damping results in Figs.~\ref{fig16:dex_com}-\ref{fig20:dexin_4a} showing the full numerical RPA solutions for the plasmon energy and damping for the mixed double-layer system, emphasizing that our low-temperature (high temperature) analytical results seem to work very well in the $T < T_F$ ($T > T_F$) regimes, providing an easy and effective way for quick calculations of the plasmon dispersion and damping in double-layer intrinsic-extrinsic plasmon systems. The captions in each figure (in Figs.~\ref{fig16:dex_com}-\ref{fig20:dexin_4a}) clearly describe in details the parameter values and the numerical results for the mixed intrinsic-extrinsic double layer system.

Finally, we conclude this subsection with a brief discussion of the finite-temperature double-layer extrinsic plasmon system when both layers are doped. Since the $T=0$ case for this system was considered in details by Hwang and Das Sarma\cite{hwang2009b} with some follow-up $T \neq 0$ double-layer plasmon calculations in the literature\cite{varifeh_PLA10,yuan_PRB11,Bahrami20123518,staubersantos_PRB12,stauber_NJP12,tuan_commphys12,profumo_PRB12,badalyan_PRB12,
badalyanpeeter_PRB12,Triolacenrico_PRB12}, we only provide some brief analytical results for the sake of completeness (and comparison with our intrinsic plasmon double layer results). We mention that the extrinsic double-layer plasmons are in principle determined by six independent parameters ($n_1, n_2, T, q, d, \kappa$), and providing complete numerical results here will simply make our paper far too long.

Using the asymptotic forms for the graphene polarizability in the long wavelength limit, we get the following analytical formula for the low-temperature ($T \ll T_F$) plasmon dispersion and damping in the leading order:
\begin{eqnarray}
\omega_{op} &=& \sqrt{2 r_s \hbar^2 v_F^2 q (k_{F1} + k_{F2}) \left(1-\frac{\pi^2}{6}\frac{T^2}{T_{F1} T_{F2}} \right)}   \label{eq34:dexana} \\
\omega_{ap} &=& \sqrt{\frac{4 r_s \hbar^2 v_F^2 d q^2 k_{F1} k_{F2}}{k_{F1}+ k_{F2}}  \frac{(1-\frac{\pi^2}{6}\frac{T^2}{T^2_{F1}})(1-\frac{\pi^2}{6}\frac{T^2}{T^2_{F2}})}{1-\frac{\pi^2}{6}\frac{T^2}{T_{F1} T_{F2}}}} \nonumber \\
\label{eq35:dexana}
\end{eqnarray}
where $k_{F1,2} = \sqrt{\pi n_{1,2}}$, $k_B T_{F1,2} = E_{F1,2}=\hbar v_F k_{F1,2}$ depend on the doping carrier density $n_{1,2}$ in the two layers, and $T \ll T_{F1,2}$ is assumed in obtaining Eqs.~\eqref{eq34:dexana} and \eqref{eq35:dexana}. The corresponding low-temperature long-wavelength damping $\gamma$ for both plasmon modes is exponentially suppressed as in the corresponding single-layer extrinsic plasmon case.

We can also carry out the high-temperature $T \gg T_{F1,2}$ asymptotic expansion of $\Pi_{1,2}$ to derive the corresponding high-temperature results and we get precisely Eqs.~\eqref{eq22:dint}-\eqref{eq25:dint} in the leading order in $T_{F1,2}/T$, i.e. $T_F$ drops out in the leading order, leaving us precisely the intrinsic double-layer plasmon results, as expected. The precise agreement between the extrinsic and the intrinsic results arises in the leading order in $T_F/T$ because the finite-$T$ chemical potential in the extrinsic case goes as $\mu \approx \frac{E_f}{ 4 \ln 2} \frac{T_F}{T}$ for $T_F \ll T$.

\begin{figure}[htb]
\begin{center}
\includegraphics[width=0.99\columnwidth]{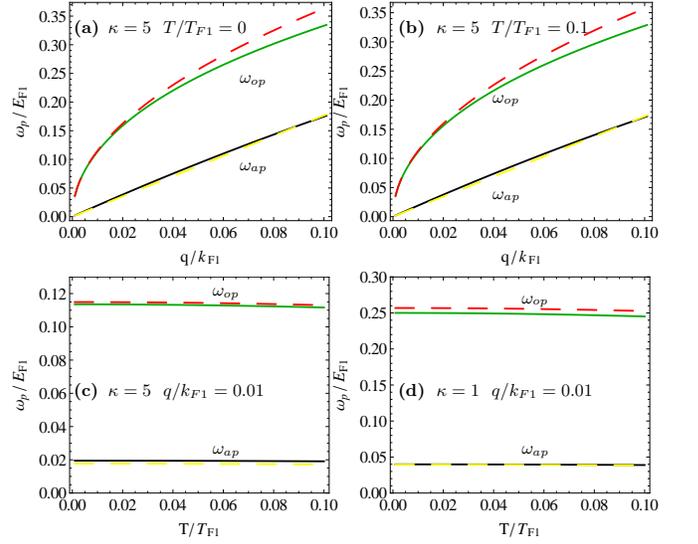}
  \caption{(Color online). Results for a double-layer extrinsic graphene system  with $n_1 = 10^{12}$ cm$^{-2}$, $n_2/n_1 = 0.25$, $d=300$ \AA. (a) and (b) $\omega_p/E_{F1}$ versus $q/k_{F1}$. (c) and (d) $\omega_p/E_{F1}$ versus $T/T_{F1}$. The dashed and solid lines correspond to the analytical results (given in Eqs.~\eqref{eq34:dexana}-\eqref{eq35:dexana}) and the numerical results, respectively. }
\label{fig21:dex_com}
\end{center}
\end{figure}

\begin{figure}[htb]
\begin{center}
\includegraphics[width=0.99\columnwidth]{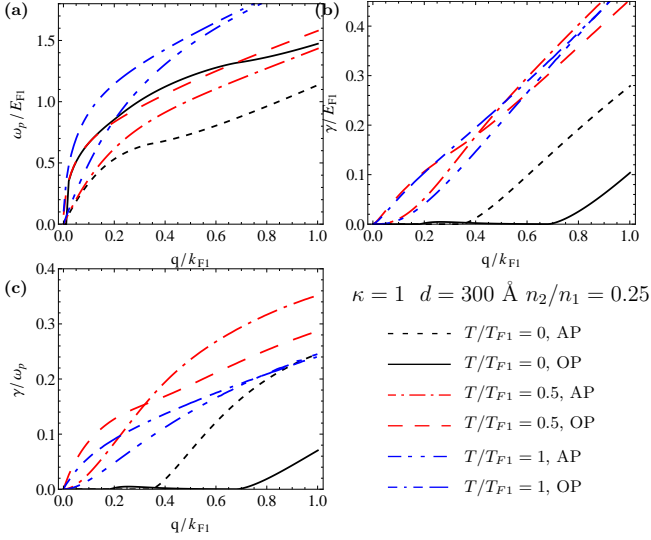}
  \caption{(Color online). Results for a double-layer extrinsic graphene system  with $n_1 = 10^{12}$ cm$^{-2}$, $n_2/n_1 = 0.25$, $d=300$ \AA \ and $\kappa=1$. (a) $\omega_p/E_{F1}$ versus $q/k_{F1}$ for different temperatures. (b) $\gamma/E_{F1}$ versus $q/k_{F1}$ for different temperatures. (c) $\gamma/\omega_p$ versus $q/k_{F1}$ for different temperatures. Note that the legend applies to all sub-figures. }
\label{fig22:dex_1}
\end{center}
\end{figure}

\begin{figure}[htb]
\begin{center}
\includegraphics[width=0.99\columnwidth]{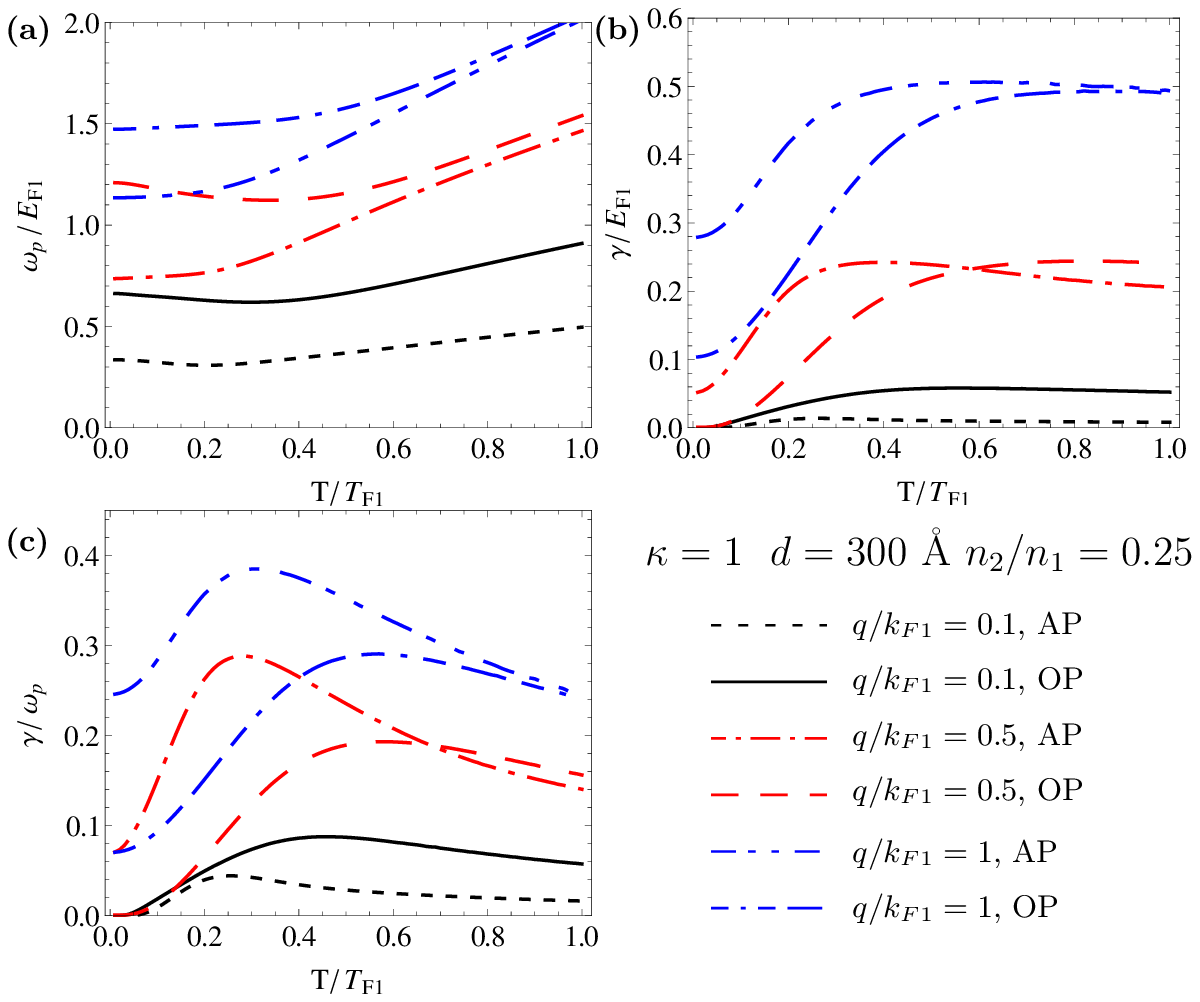}
  \caption{(Color online). Results for a double-layer extrinsic graphene system  with $n_1 = 10^{12}$ cm$^{-2}$, $n_2/n_1 = 0.25$, $d=300$ \AA \ and $\kappa=1$. (a) $\omega_p/E_{F1}$ versus $T/T_{F1}$ for different $q/k_{F1}$. (b) $\gamma/E_{F1}$ versus $T/T_{F1}$ for different $q/k_{F1}$. (c) $\gamma/\omega_p$ versus $T/T_{F1}$ for different $q/k_{F1}$. Note that the legend applies to all sub-figures. }
\label{fig23:dex_b}
\end{center}
\end{figure}

We just show three representative sets of numerical results for double-layer extrinsic graphene in Figs.~\ref{fig21:dex_com}-\ref{fig23:dex_b}. In Fig.~\ref{fig21:dex_com}, the small-$q$ and small-$T$ numerical results are shown manifesting excellent agreement with our $T \ll T_F$ analytical results. In Fig.~\ref{fig22:dex_1} we show the results for the same parameters in a much more expanded scale of $q/k_{F1}$. In Fig.~\ref{fig23:dex_b} we show the temperature dependence at fixed $q$.

\subsection{Bilayer graphene}

Finally, we very briefly consider the theory for intrinsic plasmons in undoped bilayer graphene (BLG) for the sake of completeness. For our purpose, we assume the bilayer graphene to have parabolic chiral band structure\cite{wang2007,borghi2009,sensarma2010,wang2010,gamayun2011} (characterized by an effective mass $m$ rather than a velocity $v_F$) in contrast to the linear chiral band structure of MLG (which is what we have considered so far in this work).

The BLG 2D polarizability function is given by\cite{sensarma2010}:
\begin{equation}
\Pi(q,\omega)=-\frac{4}{L^{2}}\underset{\mathbf{k},s,s'}{\sum}
\frac{f_{s,\mathbf{k}}-f_{s',\mathbf{k'}}}{\hbar \omega+\epsilon_{s,\mathbf{k}}-\epsilon_{s',\mathbf{k'}}+i\eta}
F_{s,s'}(\mathbf{k},\mathbf{k'})\label{eq34:bpola}\end{equation}
where $\mathbf{k'=k+q}$, $s,s'=\pm1$, $\epsilon_{s,\mathbf{k}} = s \hbar^2 \mathbf{k}^2/(2m) $ and $F_{s,s'}=(1+\cos2\theta)/2$ is the chiral wavefunction overlap. The functions ``$f$" in Eq.~\eqref{eq34:bpola} are the Fermi distribution functions at temperature $T$. The bilayer dynamical polarizability first calculated by Sensarma {\it et al.}\cite{sensarma2010} who also first obtained the bilayer graphene plasmon dispersion for the extrinsic (doped) system. Blow we give the analytical results for intrinsic BLG plasmons in the absence of any doping.

In the long wavelength limit ($q \rightarrow 0$) and at high temperatures $\hbar^2 q^2 /(2m) \ll k_B T$, we have the following asymptotic formula for BLG $\Pi (q,\omega)$ in the intrinsic ($n=0$) regime (from a direct expansion of Eq.~\eqref{eq34:bpola} above):
\begin{equation}
\Pi(q,\omega)\approx\frac{4\ln2}{\pi}\frac{q^{2}}{\omega^{2}}k_{B}T+i\frac{q^{2}}{8 k_{B}T}\label{eq35:ibanapo}\end{equation}
Using Eq.~\eqref{eq:vareps} for the plasmon oscillation, we get the mode dispersion and damping for intrinsic BLG to be:
\begin{eqnarray}
\omega_{p} & = & \sqrt{\frac{8 e^2 (\ln2)  (k_{B}T) q}{\kappa}} \label{eq36:ibananplas}  \\
\gamma & =& \frac{\pi e^2 q }{2 \kappa \sqrt{2 \kappa  k_{B}T}}\sqrt{(\ln2)  e^2 q} \label{eq37:ibananplas}
\end{eqnarray}
This gives for the BLG intrinsic plasmon:
\begin{equation}
\omega_p / \gamma = \frac{8 k_B T \kappa}{\pi e^2 q}
\label{eq38:inbicri}
\end{equation}
For well-defined modes, we must have $\omega_p / \gamma > 1$, leading to the condition that:
\begin{equation}
k_B T > \frac{\pi e^2 q}{8 \kappa}
\label{eq39:inbicri}
\end{equation}
Thus, similar to the MLG intrinsic plasmon although $\gamma \propto 1/\sqrt{T}$, there is a well-defined intrinsic plasma mode at long wavelength for arbitrarily low temperatures.

Comparing the BLG intrinsic plasmon dispersion given in Eq.~\eqref{eq36:ibananplas} with the corresponding MLG expression given in Eq.~\eqref{eq:iomega}, we see that the MLG and the BLG results are identical except for an extra factor of $2$ ($8 \ln2$ in Eq.~\eqref{eq36:ibananplas} and $4 \ln 2$ in Eq.~\eqref{eq:iomega}) in the BLG case. (Remember that $r_s = e^2/(\kappa \hbar v_F)$ making Eq.~\eqref{eq:iomega} equivalent to $\omega_p = \sqrt{(4 \ln 2) e^2 k_B T q/\kappa}.$) Thus the intrinsic plasmon frequency, in sharp contrast to the extrinsic plasmon frequency, is independent of the Fermi velocity (MLG) or the effective mass (BLG) and depends only on a single material constant $\kappa$, the background dielectric constant. We note that the expressions for the intrinsic plasmon damping is also very similar for MLG (Eq.~\eqref{eq:igamma}) and BLG (Eq.~\eqref{eq38:inbicri})---again except for a factor of $\sqrt{8}$, the two expressions are identical (and independent of $v_F$ or $m$).

We can also obtain the analytical results for the intrinsic plasmons in the double-layer BLG system composed of two BLG layers separated by a distance ``$d$". Solving the $2\times 2$ determinantal equation (Eq.~\eqref{eq18:exgamhigh}) for the double-layer BLG system, we get the following analytic leading-order results for the intrinsic plasmons of the double-layer BLG system:
\begin{eqnarray}
\omega_{op} &=&  \sqrt{(16 \ln 2) q k_B T \frac{e^2}{\kappa}} \label{eq40:dbint} \\
\omega_{ap} &=& \sqrt{(8 \ln 2) q^2 d k_B T \frac{e^2}{\kappa} } \label{eq41:dbint}  \\
\gamma_{op}  &=&  \frac{\pi q e^2}{\kappa \sqrt{ k_{B}T}}\sqrt{(\ln2) \frac{e^2}{\kappa} q } \label{eq42:dbint} \\
\gamma_{ap}  &=& \frac{\pi q^2 d e^2}{2 \kappa \sqrt{2 k_{B}T}}\sqrt{(\ln2) q^2 d \frac{e^2}{\kappa} }  \label{eq43:dbint}
\end{eqnarray}
These formula for the intrinsic double-layer BLG plasmons are essentially identical to the corresponding double-layer MLG intrinsic plasmons (Eqs.\eqref{eq22:dint}-\eqref{eq25:dint}) except for numerical factors. Again, these plasmon mode dispersion and damping do not depend on the effective mass or the Fermi velocity and are universal properties of the intrinsic system.

\section{Discussion}

Given the very large number of systems considered in this work (MLG, BLG, intrinsic, extrinsic, single layer, double-layer, $\cdots$) and the numerous analytical (as well as numerical) results presented in Sec.\ref{sec2:theory}, it is useful to summarize all the analytical results in Tables~\ref{tab1} and \ref{tab2} so that the similarities/differences/connections among the various derived results for plasmon modes (and their damping) become apparent. Some of the results in Tables~\ref{tab1} and \ref{tab2} were obtained in the literature before, but our emphasis in this work is on intrinsic plasmon modes and their temperature dependence. Results in the last row of Tables~\ref{tab1} and \ref{tab2} are for the ordinary (non-chiral) 2D electron gas systems (as occurring, for example, in semiconductor quantum wells) , which are by definition extrinsic systems since the large band gap (between conduction and valence bands) in the semiconductor ensures that for $n=0$, there is no plasmon mode in the system.

The key issue to be discussed in this section is the feasibility of an experimental observation of the graphene intrinsic Dirac point plasmon, which will be the manifestation of a qualitatively new phenomenon since such intrinsic plasmons at zero doping density in a charge neutral system is essentially impossible in any 2D (or 3D) semiconductors where the existence of free carriers necessitates doping by some means. (As an aside we mention that even in a very narrow gap semiconductor, e.g. In As or InSb, the band gap is $\gtrsim 100$ meV, which implies that $T > 2000$ K is necessary for any appreciable thermal population of free carriers making it impossible to study any kind of ``intrinsic plasmons" in undoped semiconductors.)

There has been a great deal of recent experimental interest in studying the properties of graphene Dirac point (i.e. the charge neutrality point) in both MLG and BLG systems\cite{weitz2010,Ponomarenko_arX11,velasco_natnano12,Bao03072012}. Various exotic quantum phases \cite{min2008b,yafis_PRB09,zhang2010,lemonik2010,nandkishore2010,nandkishore2010b,nadkishorelevi_PRB10,vafek2010,jjmac_PRB11} and non-Fermi liquid behavior\cite{dassarma2007b} are theoretically predicted at the graphene Dirac point, and thus the possibility of understanding the Dirac point behavior through the observation of intrinsic plasmon properties is both interesting and intriguing. For example, our theory of intrinsic Dirac point plasmon modes developed in this work is based entirely on the random phase approximation assuming a generic Fermi liquid ground state, and consequently, any observed qualitative departure in the experimentally observed Dirac point plasmon behavior (e.g. a completely different temperature dependence compared with that given in Tables.~\ref{tab1} and \ref{tab2}) would imply a failure of RPA indicating the fundamental and qualitative importance of interaction effects or perhaps even the emergence of a new spontaneously symmetry-broken ground state as has been predicted theoretically in the literature\cite{igor_PRL06,nomura2006,yang2006,drut_criticalPRB09,neto2009,lemonik_PRB12,kotov_RMP12}. Since collective modes (e.g. Goldstone modes, Higgs bosons, zero sound modes, etc.) typically tell us a great deal about the fundamental nature of the field theoretic vacuum (i.e. the ground state of the system), the study of graphene intrinsic plasmons could turn out to be a very useful route to understanding the nature of the Dirac point. A good example of the possible usefulness of intrinsic plasmons could be in the determination of whether the nonperturbative aspects of electron-electron interaction induce a chiral anomaly in graphene leading to the spontaneous formation of an energy gap at the Dirac point as has been predicted in some lattice quantum Monte Carlo simulation\cite{drut_PRL09,drut_latticePRB09,drut_criticalPRB09}. Such an energy gap at the Dirac point, if it exists in the experimental samples, would show up as an exponential (going as $e^{-\Delta/T}$ where $\Delta$ is the induced gap) suppression of the intrinsic plasmon energy (i.e. the simple power laws in temperature derived in our RPA theory will fail qualitatively), which could be directly experimentally observed thus validating (or not) the existence of a Dirac point chiral anomaly.

\begin{table*}[!ht]
\begin{center}
\renewcommand{\arraystretch}{1.28}
\begin{footnotesize}
    \begin{tabular}{|c|c|c||c|c||c|}
      \cline{1-6} \multirow{2}{*}{Material}&\multicolumn{2}{c||}{\multirow{2}{*}{System}}&\multicolumn{2}{c||}{{High Temperature $T \rightarrow \infty$}}&\\
      \cline{4-5} &\multicolumn{2}{c||}{}&{$\omega_p$}&{$\gamma$}& \raisebox{3.1ex}[0pt]{Comments}\\
        \cline{1-6}
      \cline{2-6} \multirow{3}{36pt}{MLG}
      \rule{0pt}{16pt}
      &\multicolumn{2}{c||}{\raisebox{1.6ex}[0pt]{Intrinsic}}
      &\raisebox{1.6ex}[0pt]{$\omega_{pi}$}
      &\raisebox{1.6ex}[0pt]{$\gamma_{pi}$}
      & \raisebox{2.6ex}[0pt]{\multirow{2}{68pt}{}}\\
       \cline{2-6}
       \cline{2-6}\rule{0pt}{20pt}
       &\multicolumn{2}{c||}{\raisebox{1.4ex}[]{Extrinsic}}&\raisebox{1.4ex}[]{$\omega_{pi} \sqrt{1+\frac{{T_F}^4}{128 (\ln 2)^3 T^4}}$}
       &\raisebox{1.4ex}[]{$\gamma_{pi}\left[1-\frac{ (\ln2 )e^2 q }{12 \kappa k_B T }\right]$}
       &\raisebox{1.8ex}[0pt]{\multirow{1}{68pt}{}}\\
      \hline
            \hline
    \multirow{6}{38pt}{Double MLG}&\multirow{2}{38pt}{Two Intrinsic}&{OP}&$\sqrt{2}\omega_{pi}$&$\sqrt{8}\gamma_{pi}$
    &\multirow{2}{68pt}{}\\
      \cline{3-5}{}&{}&{AP}&{$\sqrt{q d} \ \omega_{pi}$}
      &$(q d)^{\frac{3}{2}} \ \gamma_{pi}$&\\
       \cline{2-6}
      &\multirow{2}{38pt}{Intrinsic-Extrinsic}&{OP}&\multicolumn{2}{c||}{\multirow{4}{68pt}{Same  as results for double intrinsic layers}}&\multirow{4}{68pt}{For $T \rightarrow \infty$, chemical potential of MLG $\mu \rightarrow 0$}\\
      \cline{3-3}{}&{}&{AP}&\multicolumn{2}{c||}{}&\\
      \cline{2-3}
      &\multirow{2}{38pt}{Two Extrinsic}&{OP}&\multicolumn{2}{c||}{}&\\
     \cline{3-3} {}&{}&{AP}&\multicolumn{2}{c||}{}&\\
      \hline
      \hline

      \multirow{2}{38pt}{BLG}&\multicolumn{2}{c||}{Intrinsic}&$\sqrt{2}\omega_{pi}$&$\sqrt{8}\gamma_{pi}$&\\
      \cline{2-6}
      {}
      \rule{0pt}{20pt}&\multicolumn{2}{c||}{\raisebox{1.4ex}[0pt]{Extrinsic}}
      &\raisebox{1.4ex}[0pt]{$\sqrt{2 + \frac{T_F^2}{(4 \ln 2)T^2}}\omega_{pi}$}
      &\raisebox{1.4ex}[0pt]{$\sqrt{8} \Big[1-\frac{8 \ln 2-2}{32 \ln2} \frac{T_F^2}{T^2}\Big]\gamma_{pi}$}
      &\\

 \hline
 \hline
          \multirow{6}{38pt}{Double BLG}&\multirow{2}{38pt}{Two Intrinsic}&{OP}&$2 \omega_{pi}$&$8\gamma_{pi}$&\\
      \cline{3-6}{}&{}&{AP}&$\sqrt{2 q d} \omega_{pi}$&$\sqrt{8} (q d)^{\frac{3}{2}} \ \gamma_{pi} $&\\
       \cline{2-6}
      &\multirow{2}{38pt}{Intrinsic-Extrinsic}&{OP}&\multicolumn{2}{c||}{\multirow{4}{68pt}{Same  as results for double intrinsic layers}}&\multirow{4}{68pt}{For $T \rightarrow \infty$, $T_F/T \rightarrow 0$}\\
      \cline{3-3}{}&{}&{AP}&\multicolumn{2}{c||}{}&\\
      \cline{2-3}
      &\multirow{2}{38pt}{Two Extrinsic}&{OP}&\multicolumn{2}{c||}{}&\\
      \cline{3-3}{}&{}&{AP}&\multicolumn{2}{c||}{}&\\
      \hline
      \hline

                  \multirow{3}{38pt}{2DEG}&\multicolumn{2}{c||}{Single layer}
                  &{$\sqrt{\frac{2 \pi n e^2 q}{m \kappa}}[1 + \frac{3}{4}\frac{q \kappa k_B T}{\pi e^2 n}]$}
                  &$\sqrt{\frac{\pi}{k_B T}} \left(\frac{\pi e^2 n}{\kappa q}\right)^{\frac{3}{2}} \text{exp} \left[-\frac{\pi e^2 n}{\kappa k_B T  q}-\frac{3}{2}\right]$&\\
      \cline{2-6}{}&\multirow{2}{38pt}{Double layer$^1$}
     \rule{0pt}{16pt}  &\raisebox{0.8ex}[]{OP}
      &\raisebox{0.8ex}[]{$\sqrt{\frac{2 \pi e^2 q (n_1+n_2)}{m \kappa}}$}
      &\raisebox{0.8ex}[]{$\sqrt{\frac{\pi (n_1+n_2)}{k_B T}} \left(\frac{\pi e^2}{\kappa q}\right)^{\frac{3}{2}} \Big[e^{-\frac{\pi e^2 n_1}{\kappa k_B T  q}-\frac{3}{2}} n_1 + e^{-\frac{\pi e^2 n_2}{\kappa k_B T  q}-\frac{3}{2}} n_2\Big] $}
      &\\
      \cline{3-6}{}&{}
     \rule{0pt}{16pt}  &\raisebox{0.8ex}[]{AP}
      &\raisebox{0.8ex}[]{$\sqrt{\frac{4 \pi e^2 q^2 d}{m \kappa}}\sqrt{\frac{n_1 n_2}{ n_1+n_2}}$}
      &\raisebox{0.8ex}[]{$\sqrt{\frac{\pi}{k_B T}} \left(\frac{2 \pi e^2}{\kappa}\frac{d n_1 n_2}{n_1 + n_2}\right)^{\frac{3}{2}} \frac{\text{exp}[-\frac{\pi e^2 n_2}{\kappa k_B T  q}-\frac{3}{2}] n_1 + \text{exp}[-\frac{\pi e^2 n_1}{\kappa k_B T  q}-\frac{3}{2}] n_2}{n_1+n_2}$}
      &\\

      \hline
   \end{tabular}
            \end{footnotesize}
\caption{Summary of high temperature analytical results. We denote $\omega_{pi}=\sqrt{\frac{4 (\ln2)  (k_{B}T) e^2  q}{\kappa}}$, $\gamma_{pi} = \frac{\pi}{8} \sqrt{\frac{\ln 2}{k_B T}} \Big(\frac{e^2 q}{\kappa}\Big)^{\frac{3}{2}}$. Note 1: only the leading order terms are kept in obtaining results for a double-layer two-dimensional electron gas (2DEG) system.}
\label{tab1}
\end{center}
\end{table*}

\begin{table*}[!ht]
\begin{center}
\renewcommand{\arraystretch}{1.28}
\begin{footnotesize}
    \begin{tabular}{|c|c|c||c|c||c|}
      \cline{1-6} \multirow{2}{*}{Material}&\multicolumn{2}{c||}{\multirow{2}{*}{System}}&\multicolumn{2}{c||}{{Low Temperature $T \rightarrow 0$}}&\\
      \cline{4-5} &\multicolumn{2}{c||}{}&{$\omega_p$}&{$\gamma$}& \raisebox{3.1ex}[0pt]{Comments}\\
        \cline{1-6}
      \cline{2-6} \multirow{3}{36pt}{MLG}
      \rule{0pt}{16pt}
      &\multicolumn{2}{c||}{\raisebox{1.6ex}[0pt]{Intrinsic}}
      &\raisebox{1.6ex}[0pt]{N/A}
      &\raisebox{1.6ex}[0pt]{N/A}
      & \raisebox{2.6ex}[0pt]{\multirow{2}{58pt}{No plasmon mode at $T=0$}}\\
       \cline{2-6}
       \cline{2-6}\rule{0pt}{20pt}
       &\multicolumn{2}{c||}{\raisebox{1.4ex}[]{Extrinsic}}&\raisebox{1.6ex}[]{$\omega_{px1}\sqrt{ \Big[1-\frac{\pi^2}{6}\frac{T^2}{T_F^2}\Big]}$}
       &\raisebox{1.6ex}[]{N/A}
       &\raisebox{2.6ex}[0pt]{\multirow{1}{58pt}{$\gamma$ exponentially suppressed}}\\
      \hline
            \hline
    \multirow{6}{38pt}{Double MLG}&\multirow{2}{38pt}{Two Intrinsic}&{OP}&\multicolumn{2}{c||}{\multirow{2}*{N/A}}
    &\multirow{2}{72pt}{No plasmon mode at $T=0$}\\
      \cline{3-3}{}&{}&{AP}&\multicolumn{2}{c||}{}&\\
       \cline{2-6}
      &\multirow{2}{38pt}{Intrinsic-Extrinsic$^1$}
      \rule{0pt}{20pt} &{OP}
      &{$\omega_{px1} \sqrt{1+ (2 \ln2)\frac{  T}{T_{F1}}-\frac{\pi^2}{6}\frac{T^2}{T^2_{F1}} }$}
      &{$\gamma_{pi} \sqrt{\frac{T_{F1}}{(2 \ln 2 )T} +1-\frac{\pi^2}{12 \ln 2}\frac{T}{T_{F1}} ]}$}
      &\\
      \cline{3-6}{}&{}
      \rule{0pt}{20pt}
      &{AP}
      &{$\omega_{pi} \sqrt{2 q d \Big[ 1 - (2 \ln2)   \frac{T}{T_{F1}} + 4 (\ln2)^2\frac{T^2}{T^2_{F1}}\Big]}  $}
      &
      {$\gamma_{pi} \sqrt{ 8 (q d)^3[1-\frac{(10 \ln2 )T}{T_{F1}}]}$}&\\
      \cline{2-6}
      &\multirow{2}{38pt}{Two Extrinsic}
      \rule{0pt}{20pt}
      &{OP}
      &{$\sqrt{ (\omega_{px1}^2 + \omega_{px2}^2)\Big[1-\frac{\pi^2}{6}\frac{T^2}{T_{F1} T_{F2}}\Big]}$}
      &\raisebox{-1.5ex}[0pt]{\multirow{2}{*}{N/A}}
      &\raisebox{-1.5ex}[0pt]{\multirow{2}{58pt}{$\gamma$ exponentially suppressed}}\\
     \cline{3-4}{}&{}
     \rule{0pt}{20pt}
     &{AP}
     &{$\sqrt{2 q d\omega_{px1}\omega_{px2} \frac{\sqrt{k_{F1}k_{F2}}}{k_{F1}+k_{F2}}}\sqrt{\frac{(1-\frac{\pi^2}{6}\frac{T^2}{T^2_{F1}})(1-\frac{\pi^2}{6}\frac{T^2}{T^2_{F2}})}{1-\frac{\pi^2}{6}\frac{T^2}{T_{F1} T_{F2}}}}$}
     &{}&\\
      \hline
      \hline

      \multirow{2}{38pt}{BLG}
      &\multicolumn{2}{c||}{Intrinsic}&\multicolumn{2}{c||}{N/A}&{No plasmon}\\
      \cline{2-6}\rule{0pt}{20pt}
      {}&\multicolumn{2}{c||}{\raisebox{1.4ex}[]{Extrinsic}}
      &\raisebox{1.4ex}[]{$\sqrt{\frac{2 \pi n e^2 q}{m \kappa}}$}
      &\raisebox{1.4ex}[]{N/A}
      &\raisebox{2.0ex}[]{\multirow{1}{58pt}{$\gamma$ exponentially suppressed}}\\

 \hline
 \hline
          \multirow{6}{38pt}{Double BLG}&\multirow{2}{38pt}{Two Intrinsic}&{OP}
          &\multicolumn{2}{c||}{\raisebox{-1.5ex}[]{N/A}}
          &\raisebox{-1.5ex}[]{No plasmon mode}\\
      \cline{3-3}{}&{}&{AP}&\multicolumn{2}{c||}{}&\\
       \cline{2-6}
      &\multirow{2}{38pt}{Intrinsic-Extrinsic$^1$}
      \rule{0pt}{16pt}
      &\raisebox{0.8ex}[]{OP}
      &\raisebox{0.8ex}[]{$\sqrt{\frac{2 \pi e^2 q n}{m \kappa}}\sqrt{1 + 2 \ln2 \frac{T}{T_{F1}}}$}
      &\raisebox{0.8ex}[]{$\gamma_{pi} \sqrt{\frac{4 T_{F1}}{(\ln 2 )T}}$}
      &\\
      \cline{3-6}{}&{}
       \rule{0pt}{16pt}
      &\raisebox{0.8ex}[]{AP}
      &\raisebox{0.8ex}[]{$\omega_{pi} \sqrt{4 q d \Big[ 1 - (2 \ln2)   \frac{T}{T_{F1}} \Big]}$}
      &\raisebox{0.8ex}[]{$8 (qd)^{\frac{3}{2}}\gamma_{pi}$}
      &\\
      \cline{2-6}
      &\multirow{2}{38pt}{Two Extrinsic}
      \rule{0pt}{16pt}
      &\raisebox{0.8ex}[]{OP}
      &\raisebox{0.8ex}[]{$\sqrt{\frac{2 \pi e^2 q (n_1+n_2)}{m \kappa}}$}
      &\raisebox{0.2ex}[]{\multirow{2}{*}{N/A}}
      &\multirow{2}{58pt}{$\gamma$ exponentially suppressed}\\
      \cline{3-4}
      {}&{}
       \rule{0pt}{16pt}
       &\raisebox{0.8ex}[]{AP}
      &\raisebox{0.8ex}[]{$\sqrt{\frac{4 \pi e^2 q^2 d}{m \kappa}}\sqrt{\frac{n_1 n_2}{ n_1+n_2}}$}
      &&
      \\
      \hline
      \hline

                  \multirow{3}{38pt}{2DEG}
                  \rule{0pt}{18pt}&\multicolumn{2}{c||}{\raisebox{1.4ex}[]{Single layer}}
                  &\raisebox{1.4ex}[]{$\sqrt{\frac{2 \pi n e^2 q}{m \kappa}}$}
                  &\raisebox{1.4ex}[]{N/A}
                  &\raisebox{2.0ex}[]{\multirow{1}{58pt}{$\gamma$ exponentially suppressed}}\\
      \cline{2-6}{}
      &\multirow{2}{38pt}{Double layer}
      \rule{0pt}{16pt}
      &\raisebox{0.8ex}[]{OP}
      &\raisebox{0.8ex}[]{$\sqrt{\frac{2 \pi e^2 q (n_1+n_2)}{m \kappa}}$}
      &\raisebox{0.0ex}[]{\multirow{2}{*}{N/A}}
      &\raisebox{0.0ex}[]{\multirow{2}{58pt}{$\gamma$ exponentially suppressed}}\\
      \cline{3-4}
      \rule{0pt}{16pt} {}&{}
      &\raisebox{0.8ex}[]{AP}
      &\raisebox{0.8ex}[]{$\sqrt{\frac{4 \pi e^2 q^2 d}{m \kappa}}\sqrt{\frac{n_1 n_2}{ n_1+n_2}}$}
      &
      &\\

      \hline
   \end{tabular}
            \end{footnotesize}
\caption{Summary of low temperature analytical results. We denote $\omega_{px1}=\sqrt{2 \frac{e^2 \sqrt{\pi n_1} \hbar v_F q}{\kappa}}$, $\omega_{px2}=\sqrt{2 \frac{e^2 \sqrt{\pi n_2} \hbar v_F q}{\kappa}}$, $\omega_{pi}=\sqrt{\frac{4 (\ln2)  (k_{B}T) e^2  q}{\kappa}}$, and $\gamma_{pi} = \frac{\pi}{8} \sqrt{\frac{\ln 2}{k_B T}} \Big(\frac{e^2 q}{\kappa}\Big)^{\frac{3}{2}}$.  Note 1: results  for a mixed intrinsic-extrinsic graphene double-layer system are obtained in the limit $\hbar v_F q \ll k_B T \ll E_{F1}$. For $k_B T \ll \hbar v_F q $, there is only an optical plamson mode $\omega_{op}$, determined by the extrinsic layer, in a mixed double-layer graphene system. }
\label{tab2}
\end{center}
\end{table*}

Thus, an experimental investigation of intrinsic Dirac plasmons is highly desirable. It may appear to be a straightforward, even a trivial, task to study intrinsic Dirac plasmons experimentally by carrying out plasmon experiments keeping the Fermi level fixed at the charge neutrality point by appropriately tuning the gate voltage in gated graphene. Transport measurements in gated samples routinely enable sitting precisely at the charge neutrality point since the graphene conductivity has a typical ``V" or ``U" shape as a function of gate voltage with the conductivity (resistivity) minimum (maximum) being located at the nominal Dirac point. The serious complication is, however, disorder since it is well-known\cite{dassarma2010,Adam,rossi2008,martin2008,YZhang_NP09,LeRoyPuddle_PRB09,LeRoyPuddle_PRB11} that the charge neutrality point as determined by transport or other experiments is not the theoretical Dirac point because of the existence of electron-hole puddles in the system. In particular, random charged impurities in the graphene environment become qualitatively important in the $n \lesssim n_i$ regime (where $n_i$ is the impurity density and $n$ the carrier density), and drive the system into a highly inhomogeneous state with randomly distributed puddles of electrons and holes dominating the landscape. This puddle-dominated highly inhomogeneous (i.e. carrier density fluctuates spatially) regime has been shown\cite{Adam,Rossi,dassarma2010,qzli_PRB11} to correspond to the so-called minimum conductivity plateau in the graphene transport data (i.e. the bottom of the ``V" or ``U" in the conductivity versus gate voltage plot). Thus, the charge neutrality point in transport corresponds only to the total charge density in the whole sample being zero, and not to the absence of electrons and holes in the system. The puddle-dominated regime around the charge neutrality point should be best thought of as a random spatial variation in the Dirac point with respect to the spatially constant chemical potential (as controlled by the external gate voltage) where at each point in space the Dirac point is either above (``hole regions") or below (``electron regions") the chemical potential. Within the mean field theory\cite{Adam,Rossi,dassarma2010,qzli_PRB11,qzlidbo_PRB12,dashwangsusDirac_PRB13} the system has a finite fluctuation induced carrier density ($n^* \sim n_i$) even at the Dirac point (i.e. the charge neutrality point) because of the Coulomb disorder induced electron-hole puddles.

The existence of electron-hole puddles means that the Dirac point is ill-defined experimentally upto a carrier density of $n^*$ (with $n^*$ being determined by the details of the random charged impurity configuration in the system, but typically $n^* \lesssim n_i$), and $n^*$ can be approximately estimated experimentally by looking at the size of the conductivity minimum plateau region\cite{ChenJ_NPH_2007}. The lack of the precise existence of the Dirac point has obvious implications for the intrinsic plasmon which we discuss below.

Electron-hole puddles make it impossible, as a matter of practice, to explore the precise Dirac point in graphene (or other similar materials), and therefore, our assumption of $n \equiv 0$ at the Dirac point is no longer applicable in our consideration of intrinsic plasmons. This, however, should not be construed as a complete disaster for the observation of the intrinsic plasmon since the Dirac point, being a set of measure zero, would naturally be difficult to approach in any experiment\cite{dashwangsusDirac_PRB13}, and all experimental claims of studying Dirac point phenomena are suspect because of the fragile and unstable nature of this ``measure-zero" fixed point. Any experimental technique requiring the precise placement of the Fermi level at the Dirac point (i.e. $n \equiv 0$ everywhere in the sample) is doomed to fail no matter what.

\begin{figure}[H]
\begin{center}
\includegraphics[width=0.66\columnwidth]{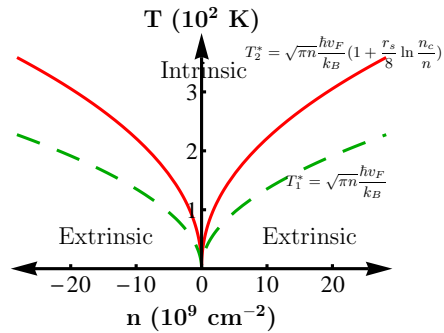}
  \caption{(Color online). The temperature scale for the crossover of the collective mode from being intrinsic to being extrinsic. The dashed curve shows $T_1^*(n) = \sqrt{\pi n} \hbar v_F /k_B$. The solid curve shows $T_2^*(n) = \sqrt{\pi n} \frac{\hbar v_F }{k_B} (1 + \frac{r_s}{8} \ln \frac{n_c}{n})$ including Fermi velocity renormalization due to electron-electron interactions  (see Ref.~[\onlinecite{hwang2007d}] for details), where $r_s = 0.4$ ($\kappa = 5$) and $n_c = 10^{15}$ cm$^{-2}$ is the high density cut off. }
\label{fig23:phase}
\end{center}
\end{figure}

From our presented results in Sec.~\ref{sec2:theory} (see also Tables~\ref{tab1} and \ref{tab2}), it is rather obvious that for the physics of intrinsic plasmons to manifest itself experimentally, there would be a lower cut-off ($T^*$) in the temperature below which (i.e. for $T < T^*$) the electron-hole puddles would inhibit any observation of intrinsic plasmon behavior with the collective mode basically crossing over from being intrinsic for $T \gg T^*$ to being extrinsic for $T \ll T^*$. Clearly, the temperature scale for this crossover is given by
\begin{equation}
T^* = T_F(n^*) = \hbar v_F \sqrt{\pi n^*}/k_B
\label{eq44:criT}
\end{equation}
where $n^*$ is the average puddle-induced carrier density in the system (i.e. $n^*$ is a function of $n_i$). Thus, intrinsic Dirac point behavior is essentially a high-temperature phenomenon with the intrinsic plasmon manifesting itself only for $T \gg T^*$ and then crossing over to extrinsic plasmons of the puddle carriers for $T \ll T^*$. The fact that the intrinsic Dirac point behavior is an effective ``high-temperature" behavior has already been emphasized earlier in the context of approaching the Dirac point through transport measurements\cite{dashwangsusDirac_PRB13}, and in the current work, we establish the same qualitative finding for approaching the Dirac point through collective mode properties. In Fig.~\ref{fig23:phase}, we show two possible crossover behaviors between intrinsic and extrinsic plasmons around the Dirac point with intrinsic plasmon being always the ``high-temperature" mode appearing in the quantum critical ``fan" region and extrinsic plasmon dominating the non-critical low-temperature and high-density region.

In standard graphene on SiO$_2$ substrates, with substantial impurity content in the environment, the typical puddle-induced carrier density $n^* \sim 10^{12}$ cm$^{-2}$, which gives $T^* \sim 1500$ K, and obviously it makes no sense to discuss any experimental study of intrinsic plasmons in such ``impure" graphene samples because the required temperature scale is impractically high. One can, of course, study extrinsic graphene in such samples by creating a doped carrier density $n > n^*$, and induced extrinsic graphene plasmons have been studied in such disordered samples at high carrier density\cite{feizhe_nanoletter11,jianing_nature12,feizhe_nature12}. In suspended graphene\cite{BolotinYacoby} or graphene on h-BN substrates\cite{BNDeanDas}, however, the environmental charged impurity density is very low ($n_i < 10^9$ cm$^{-2}$), and very low $n^*$ ($\sim 10^9 - 10^8$ cm$^{-2}$) has been reported with very sharp and narrow conductivity minimum at the charge neutrality point. Such ultrapure graphene samples (with typical mobility $\mu_m> 100,000$ cm$^{2}$/V$\cdot$S) with very low puddle density are the appropriate samples for studying intrinsic graphene. For $n^* = 10^9 (10^8)$ cm$^{-2}$, $T_F \approx 50$ K ($15$ K), and then the necessary condition ($T \gg T^*$) for the manifestation of intrinsic plasmon collective behavior would necessitate $T \approx 20-100$ K, which is very reasonable from an experimental perspective. We have already emphasized in Section \ref{sec2:theory} (and this is apparent in Tables \ref{tab1} and \ref{tab2}) that the extrinsic plasmon behaves precisely as an intrinsic plasmon (i.e. $\omega_p \propto \sqrt{T}$ and a broadening $\gamma \propto 1/\sqrt{T}$) for $T \gg T_F$, which then crosses over to the temperature-independent plasma frequency and exponentially suppressed broadening, the characteristic features of extrinsic plasmon, at low temperatures $T \ll T_F$. This is what we would expect for high-quality graphene samples ($n^* \sim 10^8 - 10^9$ cm$^{-2}$) at the Dirac point with respect to their plasmon properties. The plasmon properties will behave like those shown in Fig.~\ref{fig9:exhtcom} at higher temperatures ($T \gg T^* \sim 10 - 50$ K) to those shown in Fig.~\ref{fig1:imdis} at lower temperatures ($T \ll T^* \sim 10 - 50$ K). Such an observation will be a direct manifestation of intrinsic Dirac point behavior.

Before concluding this section, we comment on the expected plasmon level broadening (or damping) as manifested by the line width of the experimental plasmon peak. Our calculated damping ($\gamma$) corresponds to the inherent Landau damping induced level broadening which changes the plasmon peak from a pure delta-function like pole in the response function to an approximate broadened Lorentzian shape:
\begin{equation}
\delta(\omega - \omega_p) \rightarrow \frac{\Gamma}{(\omega-\omega_p)^2 + \Gamma^2}
\label{eq45:delta}
\end{equation}
The broadening $\Gamma$ in the Lorentzian plasmon peak in Eq.~\eqref{eq45:delta} would, in general, have two contributions, arising from Landau damping ($\gamma$) and impurity broadening ($\gamma_t$) which can be written as $\gamma_t = \hbar /(2 \tau)$ where $\tau$ is the transport relaxation time for impurity scattering as extracted from the mobility. In the leading order, we can simply use:
\begin{equation}
\Gamma = \gamma + \gamma_t
\label{eq46:gama}
\end{equation}
where $\gamma$ is the Landau damping calculated in Sec.~\ref{sec2:theory} (see e.g., Tables~\ref{tab1} and \ref{tab2}) and $\gamma_t$ is the impurity-scattering induced level broadening which is given by:
\begin{equation}
\gamma_t = \hbar/(2 \tau_t)
\label{eq47:gat}
\end{equation}

For high mobility samples $\tau_t \propto \mu_m$ is very large, and $\gamma_t \propto \mu_m^{-1}$ is small. Such a small impurity induced broadening will also be necessary for an unambiguous identification of the intrinsic plasmon --- it is not enough to just have small values of only the intrinsic Landau damping $\gamma$. Fortunately, the necessary condition for the suppression of $n^*$ and $T^*$ (i.e. the suppression of puddles) also necessitates very large (small) values of $\tau_t (\gamma_t)$ since low impurity density ($n_i \sim n^*$) implies high mobility (since $\mu_m \sim 1/n_i$) and low $\tau_t$. Thus, very high-mobility samples with very low puddle density, ensuring both $\gamma_t$ and $n^*$ to be small, would be necessary for the experimental study of intrinsic plasmons. Fortunately, such high-quality samples already exist in the laboratory\cite{Ponomarenko_arX11}, where the background impurity density ($n_i$) is very low, thus ensuring that both $T^*$ and $\gamma_t$ are low enough for the experimental investigation of intrinsic plasmons.

\section{Conclusion}

We have provided a rather comprehensive theory for the intrinsic collective plasmon modes in graphene associated with the Dirac point within the dynamical finite-temperature random phase approximation. We have considered monolayer graphene, bilayer graphene and double-layer graphene (consisting of two MLG or BLG layers parallel to each other separated by a distance in the third direction). We have obtained within RPA extensive analytical results for both plasmon dispersion and damping in the experimentally relevant long wavelength limit, and have provided detailed numerical results valid for arbitrary wavevectors and temperatures. We have critically discussed the experimental feasibility for observing the graphene intrinsic plasmon modes.

Among our more important qualitative conclusions (arising from the theory developed in this paper) are: (i) although the intrinsic plasmon modes, being inherently finite-temperature excitations, are strongly Landau damped even at low temperatures, they remain well-defined in the sense that the energy of the mode is larger than its damping in a very large regime of temperature, wavevector, and background dielectric constant; (ii) the intrinsic plasmon is inherently a ``high-temperature" phenomenon since the Dirac point has no energy scale (and thus any finite temperature, no matter how low, is inherently a high temperature);  (iii) closely connected with the last item is the corollary that the best experimental approach toward the observation of the intrinsic plasmon is to study low-density extrinsic plasmon at temperatures high enough (i.e. $T \gg T_F$ so that the effect of doping density is minimal); (iv) high-quality currently available graphene samples (either suspended graphene or graphene on h-BN substrates) with very high mobility should manifest clear-cut evidence for the intrinsic plasmon (e.g. plasmon energy increasing as $\sqrt{T}$ and plasmon damping decreasing as $\sqrt{T}$ with increasing temperature) if experiments are carried out at $T \approx 100$ K with the gate voltage tuned to the nominal charge neutrality point; (v) the collective mode dispersion and damping for intrinsic MLG and BLG plasmons is essentially identical (with $\omega_p \propto \sqrt{q T}$ and $\gamma \propto q\sqrt{q T} $ in both) except for numerical factors); (vi) our theoretically calculated analytical formula for plasmon dispersion and damping (both for intrinsic and for extrinsic plasmons) seem to agree with the full RPA numerical results essentially at all wavevectors and temperatures as long as our low-temperature analytical formula is used for $T \lesssim T_F$ and the high-temperature analytical formula is used for $T \gtrsim T_F$ for extrinsic graphene; (vii) for double-layer systems, we establish that it should be experimentally possible to observe both the optical and the acoustic intrinsic plasmon.

Before concluding, we mention that the graphene plasmon frequency is likely to be affected by interaction effects even at long wavelength, unlike the long-wavelength plasma frequency in parabolic band systems which is protected by Galilean invariance and the associated $f$-sum rule so that only the band mass enters the definition of the long-wavelength plasma frequency, since graphene energy dispersion obeys Lorentz invariance.  We believe that RPA is still an excellent approximation for the graphene plasmon properties (since RPA accounts for the long-range Coulomb potential correctly and nonperturbatively) except perhaps that the velocity entering the expression for the graphene plasmon mode should be modified to be the renormalized graphene velocity due to electron-electron interaction as calculated for example in Ref. [\onlinecite{dassarma2007b}].  This simple modification, plus possibly some quantum critical correction arising from the Dirac point which has to be calculated from the renormalization group flow well beyond the scope of our RPA theory, in the spirit of Landau Fermi liquid theory should suffice to incorporate the leading order interaction effect in the plasma frequency since the $r_s$ values characterizing Coulomb interaction strength in graphene are typically not too large.  A more detailed theory for including interaction effects in the graphene plasmon properties is well beyond the scope of our work and would require substantial future theoretical efforts.

We now conclude by discussing one important point which follows directly from our theoretical work with implications for graphene plasmonics. The recent interest in graphene plasmonics arises from the fact that graphene is a nanostructure allowing for very tight size confinement\cite{weihuawang_PRB11} and that the graphene (extrinsic) plasmon has energy tunable by gate voltage (i.e. $\omega_p \sim q^{1/2} n^{1/4}$) through the carrier density. In the case of intrinsic plasmon, however, the plasma energy goes as $\omega_p \sim q^{1/2} T^{1/2}$ since basically the doping density $n$ in the extrinsic formula gets replaced by the thermal electron-hole excitation density (i.e. $n \sim T^2$). This implies that it should be relatively easy to tune the intrinsic plasmon energy simply by changing temperature while sitting at the charge neutrality point,. Since the temperature dependence ($\sim \sqrt{T}$) of the intrinsic plasmon frequency is much stronger than the doping density dependence (i.e. $n^{1/4}$) of the extrinsic plasmon frequency, it may be more convenient to use the thermal tuning of the plasma frequency than the gate voltage tuning considered so far. In addition, the intrinsic plasmon should be a relatively strong and well-defined mode at room temperatures ($T \sim 300$ K) in high-mobility graphene samples, making it an interesting candidate for possible plasmonic applications.

\begin{acknowledgments}

This work is supported by US-ONR and LPS-NSA-CMTC.

\end{acknowledgments}


\begin{thebibliography}{123}
\expandafter\ifx\csname natexlab\endcsname\relax\def\natexlab#1{#1}\fi
\expandafter\ifx\csname bibnamefont\endcsname\relax
  \def\bibnamefont#1{#1}\fi
\expandafter\ifx\csname bibfnamefont\endcsname\relax
  \def\bibfnamefont#1{#1}\fi
\expandafter\ifx\csname citenamefont\endcsname\relax
  \def\citenamefont#1{#1}\fi
\expandafter\ifx\csname url\endcsname\relax
  \def\url#1{\texttt{#1}}\fi
\expandafter\ifx\csname urlprefix\endcsname\relax\def\urlprefix{URL }\fi
\providecommand{\bibinfo}[2]{#2}
\providecommand{\eprint}[2][]{\url{#2}}

\bibitem[{\citenamefont{Hwang and {Das Sarma}}(2007)}]{HwangDas_PRB07}
\bibinfo{author}{\bibfnamefont{E.~H.} \bibnamefont{Hwang}} \bibnamefont{and}
  \bibinfo{author}{\bibfnamefont{S.}~\bibnamefont{{Das Sarma}}},
  \bibinfo{journal}{Phys. Rev. B} \textbf{\bibinfo{volume}{75}},
  \bibinfo{pages}{205418} (\bibinfo{year}{2007}).

\bibitem[{\citenamefont{Vafek}(2006)}]{vafek_PRL06}
\bibinfo{author}{\bibfnamefont{O.}~\bibnamefont{Vafek}},
  \bibinfo{journal}{Phys. Rev. Lett.} \textbf{\bibinfo{volume}{97}},
  \bibinfo{pages}{266406} (\bibinfo{year}{2006}).

\bibitem[{\citenamefont{Ryzhii et~al.}(2007)\citenamefont{Ryzhii, Satou, and
  Otsuji}}]{ryzhii2007}
\bibinfo{author}{\bibfnamefont{V.}~\bibnamefont{Ryzhii}},
  \bibinfo{author}{\bibfnamefont{A.}~\bibnamefont{Satou}}, \bibnamefont{and}
  \bibinfo{author}{\bibfnamefont{T.}~\bibnamefont{Otsuji}},
  \bibinfo{journal}{J. Appl. Phys.} \textbf{\bibinfo{volume}{101}},
  \bibinfo{pages}{024509} (\bibinfo{year}{2007}).

\bibitem[{\citenamefont{Gangadharaiah et~al.}(2008)\citenamefont{Gangadharaiah,
  Farid, and Mishchenko}}]{gangadharaiah_PRL08}
\bibinfo{author}{\bibfnamefont{S.}~\bibnamefont{Gangadharaiah}},
  \bibinfo{author}{\bibfnamefont{A.~M.} \bibnamefont{Farid}}, \bibnamefont{and}
  \bibinfo{author}{\bibfnamefont{E.~G.} \bibnamefont{Mishchenko}},
  \bibinfo{journal}{Phys. Rev. Lett.} \textbf{\bibinfo{volume}{100}},
  \bibinfo{pages}{166802} (\bibinfo{year}{2008}).

\bibitem[{\citenamefont{Polini et~al.}(2008)\citenamefont{Polini, Asgari,
  Borghi, Barlas, Pereg-Barnea, and MacDonald}}]{polini_PRB08}
\bibinfo{author}{\bibfnamefont{M.}~\bibnamefont{Polini}},
  \bibinfo{author}{\bibfnamefont{R.}~\bibnamefont{Asgari}},
  \bibinfo{author}{\bibfnamefont{G.}~\bibnamefont{Borghi}},
  \bibinfo{author}{\bibfnamefont{Y.}~\bibnamefont{Barlas}},
  \bibinfo{author}{\bibfnamefont{T.}~\bibnamefont{Pereg-Barnea}},
  \bibnamefont{and} \bibinfo{author}{\bibfnamefont{A.~H.}
  \bibnamefont{MacDonald}}, \bibinfo{journal}{Phys. Rev. B}
  \textbf{\bibinfo{volume}{77}}, \bibinfo{pages}{081411}
  (\bibinfo{year}{2008}).

\bibitem[{\citenamefont{Liu et~al.}(2008)\citenamefont{Liu, Willis, Emtsev, and
  Seyller}}]{yuliu_PRB08}
\bibinfo{author}{\bibfnamefont{Y.}~\bibnamefont{Liu}},
  \bibinfo{author}{\bibfnamefont{R.~F.} \bibnamefont{Willis}},
  \bibinfo{author}{\bibfnamefont{K.~V.} \bibnamefont{Emtsev}},
  \bibnamefont{and} \bibinfo{author}{\bibfnamefont{T.}~\bibnamefont{Seyller}},
  \bibinfo{journal}{Phys. Rev. B} \textbf{\bibinfo{volume}{78}},
  \bibinfo{pages}{201403} (\bibinfo{year}{2008}).

\bibitem[{\citenamefont{Kramberger et~al.}(2008)\citenamefont{Kramberger,
  Hambach, Giorgetti, R\"ummeli, Knupfer, Fink, B\"uchner, Reining, Einarsson,
  Maruyama et~al.}}]{kramberger_PRL08}
\bibinfo{author}{\bibfnamefont{C.}~\bibnamefont{Kramberger}},
  \bibinfo{author}{\bibfnamefont{R.}~\bibnamefont{Hambach}},
  \bibinfo{author}{\bibfnamefont{C.}~\bibnamefont{Giorgetti}},
  \bibinfo{author}{\bibfnamefont{M.~H.} \bibnamefont{R\"ummeli}},
  \bibinfo{author}{\bibfnamefont{M.}~\bibnamefont{Knupfer}},
  \bibinfo{author}{\bibfnamefont{J.}~\bibnamefont{Fink}},
  \bibinfo{author}{\bibfnamefont{B.}~\bibnamefont{B\"uchner}},
  \bibinfo{author}{\bibfnamefont{L.}~\bibnamefont{Reining}},
  \bibinfo{author}{\bibfnamefont{E.}~\bibnamefont{Einarsson}},
  \bibinfo{author}{\bibfnamefont{S.}~\bibnamefont{Maruyama}},
  \bibnamefont{et~al.}, \bibinfo{journal}{Phys. Rev. Lett.}
  \textbf{\bibinfo{volume}{100}}, \bibinfo{pages}{196803}
  (\bibinfo{year}{2008}).

\bibitem[{\citenamefont{Lu et~al.}(2009)\citenamefont{Lu, Loh, Huang, Chen, and
  Wee}}]{lujiong_PRB09}
\bibinfo{author}{\bibfnamefont{J.}~\bibnamefont{Lu}},
  \bibinfo{author}{\bibfnamefont{K.~P.} \bibnamefont{Loh}},
  \bibinfo{author}{\bibfnamefont{H.}~\bibnamefont{Huang}},
  \bibinfo{author}{\bibfnamefont{W.}~\bibnamefont{Chen}}, \bibnamefont{and}
  \bibinfo{author}{\bibfnamefont{A.~T.~S.} \bibnamefont{Wee}},
  \bibinfo{journal}{Phys. Rev. B} \textbf{\bibinfo{volume}{80}},
  \bibinfo{pages}{113410} (\bibinfo{year}{2009}).

\bibitem[{\citenamefont{Jablan et~al.}(2009)\citenamefont{Jablan, Buljan, and
  Soljacic}}]{jablanmarinko_PRB09}
\bibinfo{author}{\bibfnamefont{M.}~\bibnamefont{Jablan}},
  \bibinfo{author}{\bibfnamefont{H.}~\bibnamefont{Buljan}}, \bibnamefont{and}
  \bibinfo{author}{\bibfnamefont{M.}~\bibnamefont{Soljacic}},
  \bibinfo{journal}{Phys. Rev. B} \textbf{\bibinfo{volume}{80}},
  \bibinfo{pages}{245435} (\bibinfo{year}{2009}).

\bibitem[{\citenamefont{{Das Sarma} and Hwang}(2009)}]{dassarma2009}
\bibinfo{author}{\bibfnamefont{S.}~\bibnamefont{{Das Sarma}}} \bibnamefont{and}
  \bibinfo{author}{\bibfnamefont{E.~H.} \bibnamefont{Hwang}},
  \bibinfo{journal}{Phys. Rev. Lett.} \textbf{\bibinfo{volume}{102}},
  \bibinfo{pages}{206412} (\bibinfo{year}{2009}).

\bibitem[{\citenamefont{Stauber et~al.}(2010)\citenamefont{Stauber, Schliemann,
  and Peres}}]{stauberperes_PRB10}
\bibinfo{author}{\bibfnamefont{T.}~\bibnamefont{Stauber}},
  \bibinfo{author}{\bibfnamefont{J.}~\bibnamefont{Schliemann}},
  \bibnamefont{and} \bibinfo{author}{\bibfnamefont{N.~M.~R.}
  \bibnamefont{Peres}}, \bibinfo{journal}{Phys. Rev. B}
  \textbf{\bibinfo{volume}{81}}, \bibinfo{pages}{085409}
  (\bibinfo{year}{2010}).

\bibitem[{\citenamefont{Langer et~al.}(2010)\citenamefont{Langer, Baringhaus,
  Pfnur, Schumacher, and Tegenkamp}}]{langer_njp10}
\bibinfo{author}{\bibfnamefont{T.}~\bibnamefont{Langer}},
  \bibinfo{author}{\bibfnamefont{J.}~\bibnamefont{Baringhaus}},
  \bibinfo{author}{\bibfnamefont{H.}~\bibnamefont{Pfnur}},
  \bibinfo{author}{\bibfnamefont{H.~W.} \bibnamefont{Schumacher}},
  \bibnamefont{and}
  \bibinfo{author}{\bibfnamefont{C.}~\bibnamefont{Tegenkamp}},
  \bibinfo{journal}{New Journal of Physics} \textbf{\bibinfo{volume}{12}},
  \bibinfo{pages}{033017} (\bibinfo{year}{2010}).

\bibitem[{\citenamefont{Liu and Willis}(2010)}]{liuyu_PRB10}
\bibinfo{author}{\bibfnamefont{Y.}~\bibnamefont{Liu}} \bibnamefont{and}
  \bibinfo{author}{\bibfnamefont{R.~F.} \bibnamefont{Willis}},
  \bibinfo{journal}{Phys. Rev. B} \textbf{\bibinfo{volume}{81}},
  \bibinfo{pages}{081406} (\bibinfo{year}{2010}).

\bibitem[{\citenamefont{Tudorovskiy and Mikhailov}(2010)}]{tudorovskiy_PRB10}
\bibinfo{author}{\bibfnamefont{T.}~\bibnamefont{Tudorovskiy}} \bibnamefont{and}
  \bibinfo{author}{\bibfnamefont{S.~A.} \bibnamefont{Mikhailov}},
  \bibinfo{journal}{Phys. Rev. B} \textbf{\bibinfo{volume}{82}},
  \bibinfo{pages}{073411} (\bibinfo{year}{2010}).

\bibitem[{\citenamefont{Hwang et~al.}(2010)\citenamefont{Hwang, Sensarma, and
  Das~Sarma}}]{hwangsensardas_PRB10}
\bibinfo{author}{\bibfnamefont{E.~H.} \bibnamefont{Hwang}},
  \bibinfo{author}{\bibfnamefont{R.}~\bibnamefont{Sensarma}}, \bibnamefont{and}
  \bibinfo{author}{\bibfnamefont{S.}~\bibnamefont{Das~Sarma}},
  \bibinfo{journal}{Phys. Rev. B} \textbf{\bibinfo{volume}{82}},
  \bibinfo{pages}{195406} (\bibinfo{year}{2010}).

\bibitem[{\citenamefont{Mishchenko et~al.}(2010)\citenamefont{Mishchenko,
  Shytov, and Silvestrov}}]{mishchenko_prl10}
\bibinfo{author}{\bibfnamefont{E.~G.} \bibnamefont{Mishchenko}},
  \bibinfo{author}{\bibfnamefont{A.~V.} \bibnamefont{Shytov}},
  \bibnamefont{and} \bibinfo{author}{\bibfnamefont{P.~G.}
  \bibnamefont{Silvestrov}}, \bibinfo{journal}{Phys. Rev. Lett.}
  \textbf{\bibinfo{volume}{104}}, \bibinfo{pages}{156806}
  (\bibinfo{year}{2010}).

\bibitem[{\citenamefont{Muniz et~al.}(2010)\citenamefont{Muniz, Dahal,
  Balatsky, and Haas}}]{muniz_PRB10}
\bibinfo{author}{\bibfnamefont{R.~A.} \bibnamefont{Muniz}},
  \bibinfo{author}{\bibfnamefont{H.~P.} \bibnamefont{Dahal}},
  \bibinfo{author}{\bibfnamefont{A.~V.} \bibnamefont{Balatsky}},
  \bibnamefont{and} \bibinfo{author}{\bibfnamefont{S.}~\bibnamefont{Haas}},
  \bibinfo{journal}{Phys. Rev. B} \textbf{\bibinfo{volume}{82}},
  \bibinfo{pages}{081411} (\bibinfo{year}{2010}).

\bibitem[{\citenamefont{Koch et~al.}(2010)\citenamefont{Koch, Seyller, and
  Schaefer}}]{koch_PRB10}
\bibinfo{author}{\bibfnamefont{R.~J.} \bibnamefont{Koch}},
  \bibinfo{author}{\bibfnamefont{T.}~\bibnamefont{Seyller}}, \bibnamefont{and}
  \bibinfo{author}{\bibfnamefont{J.~A.} \bibnamefont{Schaefer}},
  \bibinfo{journal}{Phys. Rev. B} \textbf{\bibinfo{volume}{82}},
  \bibinfo{pages}{201413} (\bibinfo{year}{2010}).

\bibitem[{\citenamefont{Sch\"utt et~al.}(2011)\citenamefont{Sch\"utt,
  Ostrovsky, Gornyi, and Mirlin}}]{schutt_PRB11}
\bibinfo{author}{\bibfnamefont{M.}~\bibnamefont{Sch\"utt}},
  \bibinfo{author}{\bibfnamefont{P.~M.} \bibnamefont{Ostrovsky}},
  \bibinfo{author}{\bibfnamefont{I.~V.} \bibnamefont{Gornyi}},
  \bibnamefont{and} \bibinfo{author}{\bibfnamefont{A.~D.}
  \bibnamefont{Mirlin}}, \bibinfo{journal}{Phys. Rev. B}
  \textbf{\bibinfo{volume}{83}}, \bibinfo{pages}{155441}
  (\bibinfo{year}{2011}).

\bibitem[{\citenamefont{Politano et~al.}(2011)\citenamefont{Politano, Marino,
  Formoso, Farias, Miranda, and Chiarello}}]{politano_PRB11}
\bibinfo{author}{\bibfnamefont{A.}~\bibnamefont{Politano}},
  \bibinfo{author}{\bibfnamefont{A.~R.} \bibnamefont{Marino}},
  \bibinfo{author}{\bibfnamefont{V.}~\bibnamefont{Formoso}},
  \bibinfo{author}{\bibfnamefont{D.}~\bibnamefont{Farias}},
  \bibinfo{author}{\bibfnamefont{R.}~\bibnamefont{Miranda}}, \bibnamefont{and}
  \bibinfo{author}{\bibfnamefont{G.}~\bibnamefont{Chiarello}},
  \bibinfo{journal}{Phys. Rev. B} \textbf{\bibinfo{volume}{84}},
  \bibinfo{pages}{033401} (\bibinfo{year}{2011}).

\bibitem[{\citenamefont{Villegas and Tavares}(2011)}]{Cesar_diamond11}
\bibinfo{author}{\bibfnamefont{C.~E.} \bibnamefont{Villegas}} \bibnamefont{and}
  \bibinfo{author}{\bibfnamefont{M.~R.} \bibnamefont{Tavares}},
  \bibinfo{journal}{Diamond and Related Materials}
  \textbf{\bibinfo{volume}{20}}, \bibinfo{pages}{170 } (\bibinfo{year}{2011}).

\bibitem[{\citenamefont{Abedinpour et~al.}(2011)\citenamefont{Abedinpour,
  Vignale, Principi, Polini, Tse, and MacDonald}}]{abedinpur_PRB11}
\bibinfo{author}{\bibfnamefont{S.~H.} \bibnamefont{Abedinpour}},
  \bibinfo{author}{\bibfnamefont{G.}~\bibnamefont{Vignale}},
  \bibinfo{author}{\bibfnamefont{A.}~\bibnamefont{Principi}},
  \bibinfo{author}{\bibfnamefont{M.}~\bibnamefont{Polini}},
  \bibinfo{author}{\bibfnamefont{W.-K.} \bibnamefont{Tse}}, \bibnamefont{and}
  \bibinfo{author}{\bibfnamefont{A.~H.} \bibnamefont{MacDonald}},
  \bibinfo{journal}{Phys. Rev. B} \textbf{\bibinfo{volume}{84}},
  \bibinfo{pages}{045429} (\bibinfo{year}{2011}).

\bibitem[{\citenamefont{Walter et~al.}(2011)\citenamefont{Walter, Bostwick,
  Jeon, Speck, Ostler, Seyller, Moreschini, Chang, Polini, Asgari
  et~al.}}]{walter_PRB11}
\bibinfo{author}{\bibfnamefont{A.~L.} \bibnamefont{Walter}},
  \bibinfo{author}{\bibfnamefont{A.}~\bibnamefont{Bostwick}},
  \bibinfo{author}{\bibfnamefont{K.-J.} \bibnamefont{Jeon}},
  \bibinfo{author}{\bibfnamefont{F.}~\bibnamefont{Speck}},
  \bibinfo{author}{\bibfnamefont{M.}~\bibnamefont{Ostler}},
  \bibinfo{author}{\bibfnamefont{T.}~\bibnamefont{Seyller}},
  \bibinfo{author}{\bibfnamefont{L.}~\bibnamefont{Moreschini}},
  \bibinfo{author}{\bibfnamefont{Y.~J.} \bibnamefont{Chang}},
  \bibinfo{author}{\bibfnamefont{M.}~\bibnamefont{Polini}},
  \bibinfo{author}{\bibfnamefont{R.}~\bibnamefont{Asgari}},
  \bibnamefont{et~al.}, \bibinfo{journal}{Phys. Rev. B}
  \textbf{\bibinfo{volume}{84}}, \bibinfo{pages}{085410}
  (\bibinfo{year}{2011}).

\bibitem[{\citenamefont{Tegenkamp et~al.}(2011)\citenamefont{Tegenkamp, Pfnur,
  Langer, Baringhaus, and Schumacher}}]{tegenkamp_jph11}
\bibinfo{author}{\bibfnamefont{C.}~\bibnamefont{Tegenkamp}},
  \bibinfo{author}{\bibfnamefont{H.}~\bibnamefont{Pfnur}},
  \bibinfo{author}{\bibfnamefont{T.}~\bibnamefont{Langer}},
  \bibinfo{author}{\bibfnamefont{J.}~\bibnamefont{Baringhaus}},
  \bibnamefont{and} \bibinfo{author}{\bibfnamefont{H.~W.}
  \bibnamefont{Schumacher}}, \bibinfo{journal}{Journal of Physics: Condensed
  Matter} \textbf{\bibinfo{volume}{23}}, \bibinfo{pages}{012001}
  (\bibinfo{year}{2011}).

\bibitem[{\citenamefont{Gan et~al.}(2012)\citenamefont{Gan, Chu, and
  Li}}]{ganchoon_PRB12}
\bibinfo{author}{\bibfnamefont{C.~H.} \bibnamefont{Gan}},
  \bibinfo{author}{\bibfnamefont{H.~S.} \bibnamefont{Chu}}, \bibnamefont{and}
  \bibinfo{author}{\bibfnamefont{E.~P.} \bibnamefont{Li}},
  \bibinfo{journal}{Phys. Rev. B} \textbf{\bibinfo{volume}{85}},
  \bibinfo{pages}{125431} (\bibinfo{year}{2012}).

\bibitem[{\citenamefont{G\'{o}mez-Santos and Stauber}(2012)}]{gomezsanto_EPL12}
\bibinfo{author}{\bibfnamefont{G.}~\bibnamefont{G\'{o}mez-Santos}}
  \bibnamefont{and} \bibinfo{author}{\bibfnamefont{T.}~\bibnamefont{Stauber}},
  \bibinfo{journal}{Europhysics Letters} \textbf{\bibinfo{volume}{99}},
  \bibinfo{pages}{27006} (\bibinfo{year}{2012}).

\bibitem[{\citenamefont{Dong et~al.}(2012)\citenamefont{Dong, Li, Wang, Zhang,
  Zhao, and Xu}}]{Dong20121889}
\bibinfo{author}{\bibfnamefont{H.}~\bibnamefont{Dong}},
  \bibinfo{author}{\bibfnamefont{L.}~\bibnamefont{Li}},
  \bibinfo{author}{\bibfnamefont{W.}~\bibnamefont{Wang}},
  \bibinfo{author}{\bibfnamefont{S.}~\bibnamefont{Zhang}},
  \bibinfo{author}{\bibfnamefont{C.}~\bibnamefont{Zhao}}, \bibnamefont{and}
  \bibinfo{author}{\bibfnamefont{W.}~\bibnamefont{Xu}},
  \bibinfo{journal}{Physica E: Low-dimensional Systems and Nanostructures}
  \textbf{\bibinfo{volume}{44}}, \bibinfo{pages}{1889 } (\bibinfo{year}{2012}).

\bibitem[{\citenamefont{Krstaji\ifmmode~\acute{c}\else \'{c}\fi{} and
  Peeters}(2012{\natexlab{a}})}]{krstaji_RPB12}
\bibinfo{author}{\bibfnamefont{P.~M.}
  \bibnamefont{Krstaji\ifmmode~\acute{c}\else \'{c}\fi{}}} \bibnamefont{and}
  \bibinfo{author}{\bibfnamefont{F.~M.} \bibnamefont{Peeters}},
  \bibinfo{journal}{Phys. Rev. B} \textbf{\bibinfo{volume}{85}},
  \bibinfo{pages}{205454} (\bibinfo{year}{2012}{\natexlab{a}}).

\bibitem[{\citenamefont{Krstaji\ifmmode~\acute{c}\else \'{c}\fi{} and
  Peeters}(2012{\natexlab{b}})}]{krstajipee_PRB12}
\bibinfo{author}{\bibfnamefont{P.~M.}
  \bibnamefont{Krstaji\ifmmode~\acute{c}\else \'{c}\fi{}}} \bibnamefont{and}
  \bibinfo{author}{\bibfnamefont{F.~M.} \bibnamefont{Peeters}},
  \bibinfo{journal}{Phys. Rev. B} \textbf{\bibinfo{volume}{85}},
  \bibinfo{pages}{085436} (\bibinfo{year}{2012}{\natexlab{b}}).

\bibitem[{\citenamefont{Langer et~al.}(2012)\citenamefont{Langer, Pfnur,
  Tegenkamp, Forti, Emtsev, and Starke}}]{thomas_NJP12}
\bibinfo{author}{\bibfnamefont{T.}~\bibnamefont{Langer}},
  \bibinfo{author}{\bibfnamefont{H.}~\bibnamefont{Pfnur}},
  \bibinfo{author}{\bibfnamefont{C.}~\bibnamefont{Tegenkamp}},
  \bibinfo{author}{\bibfnamefont{S.}~\bibnamefont{Forti}},
  \bibinfo{author}{\bibfnamefont{K.}~\bibnamefont{Emtsev}}, \bibnamefont{and}
  \bibinfo{author}{\bibfnamefont{U.}~\bibnamefont{Starke}},
  \bibinfo{journal}{New Journal of Physics} \textbf{\bibinfo{volume}{14}},
  \bibinfo{pages}{103045} (\bibinfo{year}{2012}).

\bibitem[{\citenamefont{Politano et~al.}(2012)\citenamefont{Politano, Marino,
  and Chiarello}}]{politano_PRB12}
\bibinfo{author}{\bibfnamefont{A.}~\bibnamefont{Politano}},
  \bibinfo{author}{\bibfnamefont{A.~R.} \bibnamefont{Marino}},
  \bibnamefont{and}
  \bibinfo{author}{\bibfnamefont{G.}~\bibnamefont{Chiarello}},
  \bibinfo{journal}{Phys. Rev. B} \textbf{\bibinfo{volume}{86}},
  \bibinfo{pages}{085420} (\bibinfo{year}{2012}).

\bibitem[{\citenamefont{Gorbach}(2013)}]{gorbach_PRA13}
\bibinfo{author}{\bibfnamefont{A.~V.} \bibnamefont{Gorbach}},
  \bibinfo{journal}{Phys. Rev. A} \textbf{\bibinfo{volume}{87}},
  \bibinfo{pages}{013830} (\bibinfo{year}{2013}).

\bibitem[{\citenamefont{Zhu et~al.}(2013)\citenamefont{Zhu, Badalyan, and
  Peeters}}]{zhujj_PRB13}
\bibinfo{author}{\bibfnamefont{J.-J.} \bibnamefont{Zhu}},
  \bibinfo{author}{\bibfnamefont{S.~M.} \bibnamefont{Badalyan}},
  \bibnamefont{and} \bibinfo{author}{\bibfnamefont{F.~M.}
  \bibnamefont{Peeters}}, \bibinfo{journal}{Phys. Rev. B}
  \textbf{\bibinfo{volume}{87}}, \bibinfo{pages}{085401}
  (\bibinfo{year}{2013}).

\bibitem[{\citenamefont{Das~Sarma and Hwang}(1998)}]{dashwang_PRL98}
\bibinfo{author}{\bibfnamefont{S.}~\bibnamefont{Das~Sarma}} \bibnamefont{and}
  \bibinfo{author}{\bibfnamefont{E.~H.} \bibnamefont{Hwang}},
  \bibinfo{journal}{Phys. Rev. Lett.} \textbf{\bibinfo{volume}{81}},
  \bibinfo{pages}{4216} (\bibinfo{year}{1998}).

\bibitem[{\citenamefont{Hwang and Das~Sarma}(2001)}]{hwangdas_PRB01}
\bibinfo{author}{\bibfnamefont{E.~H.} \bibnamefont{Hwang}} \bibnamefont{and}
  \bibinfo{author}{\bibfnamefont{S.}~\bibnamefont{Das~Sarma}},
  \bibinfo{journal}{Phys. Rev. B} \textbf{\bibinfo{volume}{64}},
  \bibinfo{pages}{165409} (\bibinfo{year}{2001}).

\bibitem[{\citenamefont{Koppens et~al.}(2011)\citenamefont{Koppens, Chang, and
  Garcia~de Abajo}}]{koppens_nano11}
\bibinfo{author}{\bibfnamefont{F.~H.~L.} \bibnamefont{Koppens}},
  \bibinfo{author}{\bibfnamefont{D.~E.} \bibnamefont{Chang}}, \bibnamefont{and}
  \bibinfo{author}{\bibfnamefont{F.~J.} \bibnamefont{Garcia~de Abajo}},
  \bibinfo{journal}{Nano Letters} \textbf{\bibinfo{volume}{11}},
  \bibinfo{pages}{3370} (\bibinfo{year}{2011}).

\bibitem[{\citenamefont{Nikitin et~al.}(2011)\citenamefont{Nikitin, Guinea,
  Garcia-Vidal, and Martin-Moreno}}]{nikitin_PRB11}
\bibinfo{author}{\bibfnamefont{A.~Y.} \bibnamefont{Nikitin}},
  \bibinfo{author}{\bibfnamefont{F.}~\bibnamefont{Guinea}},
  \bibinfo{author}{\bibfnamefont{F.~J.} \bibnamefont{Garcia-Vidal}},
  \bibnamefont{and}
  \bibinfo{author}{\bibfnamefont{L.}~\bibnamefont{Martin-Moreno}},
  \bibinfo{journal}{Phys. Rev. B} \textbf{\bibinfo{volume}{84}},
  \bibinfo{pages}{195446} (\bibinfo{year}{2011}).

\bibitem[{\citenamefont{Ju et~al.}(2011)\citenamefont{Ju, Geng, Horng, Girit,
  Martin, Hao, Bechtel, Liang, Zettl, Shen et~al.}}]{jufeng_NN11}
\bibinfo{author}{\bibfnamefont{L.}~\bibnamefont{Ju}},
  \bibinfo{author}{\bibfnamefont{B.}~\bibnamefont{Geng}},
  \bibinfo{author}{\bibfnamefont{J.}~\bibnamefont{Horng}},
  \bibinfo{author}{\bibfnamefont{C.}~\bibnamefont{Girit}},
  \bibinfo{author}{\bibfnamefont{M.}~\bibnamefont{Martin}},
  \bibinfo{author}{\bibfnamefont{Z.}~\bibnamefont{Hao}},
  \bibinfo{author}{\bibfnamefont{H.~A.} \bibnamefont{Bechtel}},
  \bibinfo{author}{\bibfnamefont{X.}~\bibnamefont{Liang}},
  \bibinfo{author}{\bibfnamefont{A.}~\bibnamefont{Zettl}},
  \bibinfo{author}{\bibfnamefont{Y.~R.} \bibnamefont{Shen}},
  \bibnamefont{et~al.}, \bibinfo{journal}{Nat Nano}
  \textbf{\bibinfo{volume}{6}}, \bibinfo{pages}{630} (\bibinfo{year}{2011}).

\bibitem[{\citenamefont{Echtermeyer et~al.}(2011)\citenamefont{Echtermeyer,
  Britnell, Jasnos, Lombardo, Gorbachev, Grigorenko, Geim, Ferrari, and
  Novoselov}}]{Echtermeyer_natcommun11}
\bibinfo{author}{\bibfnamefont{T.~J.} \bibnamefont{Echtermeyer}},
  \bibinfo{author}{\bibfnamefont{L.}~\bibnamefont{Britnell}},
  \bibinfo{author}{\bibfnamefont{P.~K.} \bibnamefont{Jasnos}},
  \bibinfo{author}{\bibfnamefont{A.}~\bibnamefont{Lombardo}},
  \bibinfo{author}{\bibfnamefont{R.~V.} \bibnamefont{Gorbachev}},
  \bibinfo{author}{\bibfnamefont{A.~N.} \bibnamefont{Grigorenko}},
  \bibinfo{author}{\bibfnamefont{A.~K.} \bibnamefont{Geim}},
  \bibinfo{author}{\bibfnamefont{A.~C.} \bibnamefont{Ferrari}},
  \bibnamefont{and} \bibinfo{author}{\bibfnamefont{K.~S.}
  \bibnamefont{Novoselov}}, \bibinfo{journal}{Nat. Commun.}
  \textbf{\bibinfo{volume}{2}}, \bibinfo{pages}{458} (\bibinfo{year}{2011}).

\bibitem[{\citenamefont{Fei et~al.}(2011)\citenamefont{Fei, Andreev, Bao,
  Zhang, S.~McLeod, Wang, Stewart, Zhao, Dominguez, Thiemens
  et~al.}}]{feizhe_nanoletter11}
\bibinfo{author}{\bibfnamefont{Z.}~\bibnamefont{Fei}},
  \bibinfo{author}{\bibfnamefont{G.~O.} \bibnamefont{Andreev}},
  \bibinfo{author}{\bibfnamefont{W.}~\bibnamefont{Bao}},
  \bibinfo{author}{\bibfnamefont{L.~M.} \bibnamefont{Zhang}},
  \bibinfo{author}{\bibfnamefont{A.}~\bibnamefont{S.~McLeod}},
  \bibinfo{author}{\bibfnamefont{C.}~\bibnamefont{Wang}},
  \bibinfo{author}{\bibfnamefont{M.~K.} \bibnamefont{Stewart}},
  \bibinfo{author}{\bibfnamefont{Z.}~\bibnamefont{Zhao}},
  \bibinfo{author}{\bibfnamefont{G.}~\bibnamefont{Dominguez}},
  \bibinfo{author}{\bibfnamefont{M.}~\bibnamefont{Thiemens}},
  \bibnamefont{et~al.}, \bibinfo{journal}{Nano Letters}
  \textbf{\bibinfo{volume}{11}}, \bibinfo{pages}{4701} (\bibinfo{year}{2011}).

\bibitem[{\citenamefont{Yan et~al.}(2012)\citenamefont{Yan, Li, Chandra,
  Tulevski, Wu, Freitag, Zhu, Avouris, and Xia}}]{yanhugen_nanaotech12}
\bibinfo{author}{\bibfnamefont{H.}~\bibnamefont{Yan}},
  \bibinfo{author}{\bibfnamefont{X.}~\bibnamefont{Li}},
  \bibinfo{author}{\bibfnamefont{B.}~\bibnamefont{Chandra}},
  \bibinfo{author}{\bibfnamefont{G.}~\bibnamefont{Tulevski}},
  \bibinfo{author}{\bibfnamefont{Y.}~\bibnamefont{Wu}},
  \bibinfo{author}{\bibfnamefont{M.}~\bibnamefont{Freitag}},
  \bibinfo{author}{\bibfnamefont{W.}~\bibnamefont{Zhu}},
  \bibinfo{author}{\bibfnamefont{P.}~\bibnamefont{Avouris}}, \bibnamefont{and}
  \bibinfo{author}{\bibfnamefont{F.}~\bibnamefont{Xia}},
  \bibinfo{journal}{Nature Nanotechnology} \textbf{\bibinfo{volume}{7}},
  \bibinfo{pages}{330} (\bibinfo{year}{2012}).

\bibitem[{\citenamefont{Grigorenko et~al.}(2012)\citenamefont{Grigorenko,
  Polini, and Novoselov}}]{grigorenko_natpho12}
\bibinfo{author}{\bibfnamefont{A.~N.} \bibnamefont{Grigorenko}},
  \bibinfo{author}{\bibfnamefont{M.}~\bibnamefont{Polini}}, \bibnamefont{and}
  \bibinfo{author}{\bibfnamefont{K.~S.} \bibnamefont{Novoselov}},
  \bibinfo{journal}{Nature Photon.} \textbf{\bibinfo{volume}{6}},
  \bibinfo{pages}{749} (\bibinfo{year}{2012}).

\bibitem[{\citenamefont{Fei et~al.}(2012)\citenamefont{Fei, Rodin, Andreev,
  Bao, S.~McLeod, Wagner, Zhang, Zhao, Thiemens, Dominguez
  et~al.}}]{feizhe_nature12}
\bibinfo{author}{\bibfnamefont{Z.}~\bibnamefont{Fei}},
  \bibinfo{author}{\bibfnamefont{A.~S.} \bibnamefont{Rodin}},
  \bibinfo{author}{\bibfnamefont{G.~O.} \bibnamefont{Andreev}},
  \bibinfo{author}{\bibfnamefont{W.}~\bibnamefont{Bao}},
  \bibinfo{author}{\bibfnamefont{A.}~\bibnamefont{S.~McLeod}},
  \bibinfo{author}{\bibfnamefont{M.}~\bibnamefont{Wagner}},
  \bibinfo{author}{\bibfnamefont{L.~M.} \bibnamefont{Zhang}},
  \bibinfo{author}{\bibfnamefont{Z.}~\bibnamefont{Zhao}},
  \bibinfo{author}{\bibfnamefont{M.}~\bibnamefont{Thiemens}},
  \bibinfo{author}{\bibfnamefont{G.}~\bibnamefont{Dominguez}},
  \bibnamefont{et~al.}, \bibinfo{journal}{Nature}
  \textbf{\bibinfo{volume}{487}}, \bibinfo{pages}{82} (\bibinfo{year}{2012}).

\bibitem[{\citenamefont{Chen et~al.}(2012)\citenamefont{Chen, Badioli,
  Alonso-Gonzalez, Thongrattanasiri, Huth, Osmond, Spasenovic, Centeno,
  Pesquera, Godignon et~al.}}]{jianing_nature12}
\bibinfo{author}{\bibfnamefont{J.}~\bibnamefont{Chen}},
  \bibinfo{author}{\bibfnamefont{M.}~\bibnamefont{Badioli}},
  \bibinfo{author}{\bibfnamefont{P.}~\bibnamefont{Alonso-Gonzalez}},
  \bibinfo{author}{\bibfnamefont{S.}~\bibnamefont{Thongrattanasiri}},
  \bibinfo{author}{\bibfnamefont{F.}~\bibnamefont{Huth}},
  \bibinfo{author}{\bibfnamefont{J.}~\bibnamefont{Osmond}},
  \bibinfo{author}{\bibfnamefont{M.}~\bibnamefont{Spasenovic}},
  \bibinfo{author}{\bibfnamefont{A.}~\bibnamefont{Centeno}},
  \bibinfo{author}{\bibfnamefont{A.}~\bibnamefont{Pesquera}},
  \bibinfo{author}{\bibfnamefont{P.}~\bibnamefont{Godignon}},
  \bibnamefont{et~al.}, \bibinfo{journal}{Nature}
  \textbf{\bibinfo{volume}{487}}, \bibinfo{pages}{77} (\bibinfo{year}{2012}).

\bibitem[{\citenamefont{Thongrattanasiri
  et~al.}(2012)\citenamefont{Thongrattanasiri, Manjavacas, and Garcia~de
  Abajo}}]{thongra_ACSnano12}
\bibinfo{author}{\bibfnamefont{S.}~\bibnamefont{Thongrattanasiri}},
  \bibinfo{author}{\bibfnamefont{A.}~\bibnamefont{Manjavacas}},
  \bibnamefont{and} \bibinfo{author}{\bibfnamefont{F.~J.}
  \bibnamefont{Garcia~de Abajo}}, \bibinfo{journal}{ACS Nano}
  \textbf{\bibinfo{volume}{6}}, \bibinfo{pages}{1766} (\bibinfo{year}{2012}).

\bibitem[{\citenamefont{Davoyan et~al.}(2012)\citenamefont{Davoyan, Popov, and
  Nikitov}}]{davoyan_PRL12}
\bibinfo{author}{\bibfnamefont{A.~R.} \bibnamefont{Davoyan}},
  \bibinfo{author}{\bibfnamefont{V.~V.} \bibnamefont{Popov}}, \bibnamefont{and}
  \bibinfo{author}{\bibfnamefont{S.~A.} \bibnamefont{Nikitov}},
  \bibinfo{journal}{Phys. Rev. Lett.} \textbf{\bibinfo{volume}{108}},
  \bibinfo{pages}{127401} (\bibinfo{year}{2012}).

\bibitem[{\citenamefont{Zhan et~al.}(2012)\citenamefont{Zhan, Zhao, Hu, Liu,
  and Zi}}]{zhantr_PRB12}
\bibinfo{author}{\bibfnamefont{T.~R.} \bibnamefont{Zhan}},
  \bibinfo{author}{\bibfnamefont{F.~Y.} \bibnamefont{Zhao}},
  \bibinfo{author}{\bibfnamefont{X.~H.} \bibnamefont{Hu}},
  \bibinfo{author}{\bibfnamefont{X.~H.} \bibnamefont{Liu}}, \bibnamefont{and}
  \bibinfo{author}{\bibfnamefont{J.}~\bibnamefont{Zi}}, \bibinfo{journal}{Phys.
  Rev. B} \textbf{\bibinfo{volume}{86}}, \bibinfo{pages}{165416}
  (\bibinfo{year}{2012}).

\bibitem[{\citenamefont{Carbotte et~al.}(2012)\citenamefont{Carbotte, LeBlanc,
  and Nicol}}]{carbotte_PRB12}
\bibinfo{author}{\bibfnamefont{J.~P.} \bibnamefont{Carbotte}},
  \bibinfo{author}{\bibfnamefont{J.~P.~F.} \bibnamefont{LeBlanc}},
  \bibnamefont{and} \bibinfo{author}{\bibfnamefont{E.~J.} \bibnamefont{Nicol}},
  \bibinfo{journal}{Phys. Rev. B} \textbf{\bibinfo{volume}{85}},
  \bibinfo{pages}{201411} (\bibinfo{year}{2012}).

\bibitem[{\citenamefont{Ilic et~al.}(2012)\citenamefont{Ilic, Jablan,
  Joannopoulos, Celanovic, and Solja\v{c}i\'{c}}}]{Ilic_opletter12}
\bibinfo{author}{\bibfnamefont{O.}~\bibnamefont{Ilic}},
  \bibinfo{author}{\bibfnamefont{M.}~\bibnamefont{Jablan}},
  \bibinfo{author}{\bibfnamefont{J.~D.} \bibnamefont{Joannopoulos}},
  \bibinfo{author}{\bibfnamefont{I.}~\bibnamefont{Celanovic}},
  \bibnamefont{and}
  \bibinfo{author}{\bibfnamefont{M.}~\bibnamefont{Solja\v{c}i\'{c}}},
  \bibinfo{journal}{Opt. Express} \textbf{\bibinfo{volume}{20}},
  \bibinfo{pages}{A366} (\bibinfo{year}{2012}).

\bibitem[{\citenamefont{Rast et~al.}(2013)\citenamefont{Rast, Sullivan, and
  Tewary}}]{rast_PRB13}
\bibinfo{author}{\bibfnamefont{L.}~\bibnamefont{Rast}},
  \bibinfo{author}{\bibfnamefont{T.~J.} \bibnamefont{Sullivan}},
  \bibnamefont{and} \bibinfo{author}{\bibfnamefont{V.~K.}
  \bibnamefont{Tewary}}, \bibinfo{journal}{Phys. Rev. B}
  \textbf{\bibinfo{volume}{87}}, \bibinfo{pages}{045428}
  (\bibinfo{year}{2013}).

\bibitem[{\citenamefont{Xia et~al.}(2009)\citenamefont{Xia, Mueller, ming Lin,
  Valdes-Garcia, and Avouris}}]{fengnianxia_natnano09}
\bibinfo{author}{\bibfnamefont{F.}~\bibnamefont{Xia}},
  \bibinfo{author}{\bibfnamefont{T.}~\bibnamefont{Mueller}},
  \bibinfo{author}{\bibfnamefont{Y.}~\bibnamefont{ming Lin}},
  \bibinfo{author}{\bibfnamefont{A.}~\bibnamefont{Valdes-Garcia}},
  \bibnamefont{and} \bibinfo{author}{\bibfnamefont{P.}~\bibnamefont{Avouris}},
  \bibinfo{journal}{Nature Nanotechnology} \textbf{\bibinfo{volume}{4}},
  \bibinfo{pages}{839} (\bibinfo{year}{2009}).

\bibitem[{\citenamefont{Xu et~al.}(2010)\citenamefont{Xu, Gabor, Alden, van~der
  Zande, and McEuen}}]{xiaodongxu_Nanoletter10}
\bibinfo{author}{\bibfnamefont{X.}~\bibnamefont{Xu}},
  \bibinfo{author}{\bibfnamefont{N.~M.} \bibnamefont{Gabor}},
  \bibinfo{author}{\bibfnamefont{J.~S.} \bibnamefont{Alden}},
  \bibinfo{author}{\bibfnamefont{A.~M.} \bibnamefont{van~der Zande}},
  \bibnamefont{and} \bibinfo{author}{\bibfnamefont{P.~L.}
  \bibnamefont{McEuen}}, \bibinfo{journal}{Nano Letters}
  \textbf{\bibinfo{volume}{10}}, \bibinfo{pages}{562} (\bibinfo{year}{2010}).

\bibitem[{\citenamefont{Mueller et~al.}(2010)\citenamefont{Mueller, Xia, and
  Avouris}}]{mueller_natphoto10}
\bibinfo{author}{\bibfnamefont{T.}~\bibnamefont{Mueller}},
  \bibinfo{author}{\bibfnamefont{F.}~\bibnamefont{Xia}}, \bibnamefont{and}
  \bibinfo{author}{\bibfnamefont{P.}~\bibnamefont{Avouris}},
  \bibinfo{journal}{Nat. Photonics} \textbf{\bibinfo{volume}{4}},
  \bibinfo{pages}{297} (\bibinfo{year}{2010}).

\bibitem[{\citenamefont{Rana}(2011)}]{farhan_natnano11}
\bibinfo{author}{\bibfnamefont{F.}~\bibnamefont{Rana}},
  \bibinfo{journal}{Nature Nanotechnology} \textbf{\bibinfo{volume}{6}},
  \bibinfo{pages}{611} (\bibinfo{year}{2011}).

\bibitem[{\citenamefont{Rana et~al.}(2011)\citenamefont{Rana, Strait, Wang, and
  Manolatou}}]{Farhan_PRB11}
\bibinfo{author}{\bibfnamefont{F.}~\bibnamefont{Rana}},
  \bibinfo{author}{\bibfnamefont{J.~H.} \bibnamefont{Strait}},
  \bibinfo{author}{\bibfnamefont{H.}~\bibnamefont{Wang}}, \bibnamefont{and}
  \bibinfo{author}{\bibfnamefont{C.}~\bibnamefont{Manolatou}},
  \bibinfo{journal}{Phys. Rev. B} \textbf{\bibinfo{volume}{84}},
  \bibinfo{pages}{045437} (\bibinfo{year}{2011}).

\bibitem[{\citenamefont{Liu et~al.}(2011)\citenamefont{Liu, Yin, Ulin-Avila,
  Geng, Zentgraf, Ju, Wang, and Zhang}}]{liu_Nature11}
\bibinfo{author}{\bibfnamefont{M.}~\bibnamefont{Liu}},
  \bibinfo{author}{\bibfnamefont{X.}~\bibnamefont{Yin}},
  \bibinfo{author}{\bibfnamefont{E.}~\bibnamefont{Ulin-Avila}},
  \bibinfo{author}{\bibfnamefont{B.}~\bibnamefont{Geng}},
  \bibinfo{author}{\bibfnamefont{T.}~\bibnamefont{Zentgraf}},
  \bibinfo{author}{\bibfnamefont{L.}~\bibnamefont{Ju}},
  \bibinfo{author}{\bibfnamefont{F.}~\bibnamefont{Wang}}, \bibnamefont{and}
  \bibinfo{author}{\bibfnamefont{X.}~\bibnamefont{Zhang}},
  \bibinfo{journal}{Nature} \textbf{\bibinfo{volume}{474}}, \bibinfo{pages}{64}
  (\bibinfo{year}{2011}).

\bibitem[{\citenamefont{Sun et~al.}(2010)\citenamefont{Sun, Hasan, Torrisi,
  Popa, Privitera, Wang, Bonaccorso, Basko, and Ferrari}}]{sun_ACSnano10}
\bibinfo{author}{\bibfnamefont{Z.}~\bibnamefont{Sun}},
  \bibinfo{author}{\bibfnamefont{T.}~\bibnamefont{Hasan}},
  \bibinfo{author}{\bibfnamefont{F.}~\bibnamefont{Torrisi}},
  \bibinfo{author}{\bibfnamefont{D.}~\bibnamefont{Popa}},
  \bibinfo{author}{\bibfnamefont{G.}~\bibnamefont{Privitera}},
  \bibinfo{author}{\bibfnamefont{F.}~\bibnamefont{Wang}},
  \bibinfo{author}{\bibfnamefont{F.}~\bibnamefont{Bonaccorso}},
  \bibinfo{author}{\bibfnamefont{D.~M.} \bibnamefont{Basko}}, \bibnamefont{and}
  \bibinfo{author}{\bibfnamefont{A.~C.} \bibnamefont{Ferrari}},
  \bibinfo{journal}{ACS Nano} \textbf{\bibinfo{volume}{4}},
  \bibinfo{pages}{803} (\bibinfo{year}{2010}).

\bibitem[{\citenamefont{Bonaccorso et~al.}(2010)\citenamefont{Bonaccorso, Sun,
  Hasan, and Ferrari}}]{bonaccorso_natpho10}
\bibinfo{author}{\bibfnamefont{F.}~\bibnamefont{Bonaccorso}},
  \bibinfo{author}{\bibfnamefont{Z.}~\bibnamefont{Sun}},
  \bibinfo{author}{\bibfnamefont{T.}~\bibnamefont{Hasan}}, \bibnamefont{and}
  \bibinfo{author}{\bibfnamefont{A.~C.} \bibnamefont{Ferrari}},
  \bibinfo{journal}{Nature Photonics} \textbf{\bibinfo{volume}{4}},
  \bibinfo{pages}{611} (\bibinfo{year}{2010}).

\bibitem[{\citenamefont{Vakil and Engheta}(2011)}]{ashkan_science11}
\bibinfo{author}{\bibfnamefont{A.}~\bibnamefont{Vakil}} \bibnamefont{and}
  \bibinfo{author}{\bibfnamefont{N.}~\bibnamefont{Engheta}},
  \bibinfo{journal}{Science} \textbf{\bibinfo{volume}{332}},
  \bibinfo{pages}{1291} (\bibinfo{year}{2011}).

\bibitem[{\citenamefont{Bao and Loh}(2012)}]{qiaobao_ACSNano12}
\bibinfo{author}{\bibfnamefont{Q.}~\bibnamefont{Bao}} \bibnamefont{and}
  \bibinfo{author}{\bibfnamefont{K.~P.} \bibnamefont{Loh}},
  \bibinfo{journal}{ACS Nano} \textbf{\bibinfo{volume}{6}},
  \bibinfo{pages}{3677} (\bibinfo{year}{2012}).

\bibitem[{\citenamefont{Eberlein et~al.}(2008)\citenamefont{Eberlein, Bangert,
  Nair, Jones, Gass, Bleloch, Novoselov, Geim, and Briddon}}]{eberlein_PRB08}
\bibinfo{author}{\bibfnamefont{T.}~\bibnamefont{Eberlein}},
  \bibinfo{author}{\bibfnamefont{U.}~\bibnamefont{Bangert}},
  \bibinfo{author}{\bibfnamefont{R.~R.} \bibnamefont{Nair}},
  \bibinfo{author}{\bibfnamefont{R.}~\bibnamefont{Jones}},
  \bibinfo{author}{\bibfnamefont{M.}~\bibnamefont{Gass}},
  \bibinfo{author}{\bibfnamefont{A.~L.} \bibnamefont{Bleloch}},
  \bibinfo{author}{\bibfnamefont{K.~S.} \bibnamefont{Novoselov}},
  \bibinfo{author}{\bibfnamefont{A.}~\bibnamefont{Geim}}, \bibnamefont{and}
  \bibinfo{author}{\bibfnamefont{P.~R.} \bibnamefont{Briddon}},
  \bibinfo{journal}{Phys. Rev. B} \textbf{\bibinfo{volume}{77}},
  \bibinfo{pages}{233406} (\bibinfo{year}{2008}).

\bibitem[{\citenamefont{Yuan et~al.}(2011)\citenamefont{Yuan, Rold\'an, and
  Katsnelson}}]{yuan_PRB11}
\bibinfo{author}{\bibfnamefont{S.}~\bibnamefont{Yuan}},
  \bibinfo{author}{\bibfnamefont{R.}~\bibnamefont{Rold\'an}}, \bibnamefont{and}
  \bibinfo{author}{\bibfnamefont{M.~I.} \bibnamefont{Katsnelson}},
  \bibinfo{journal}{Phys. Rev. B} \textbf{\bibinfo{volume}{84}},
  \bibinfo{pages}{035439} (\bibinfo{year}{2011}).

\bibitem[{\citenamefont{Shin et~al.}(2011)\citenamefont{Shin, Hwang, Sung, Kim,
  Kim, and Chung}}]{shin_prb11}
\bibinfo{author}{\bibfnamefont{S.~Y.} \bibnamefont{Shin}},
  \bibinfo{author}{\bibfnamefont{C.~G.} \bibnamefont{Hwang}},
  \bibinfo{author}{\bibfnamefont{S.~J.} \bibnamefont{Sung}},
  \bibinfo{author}{\bibfnamefont{N.~D.} \bibnamefont{Kim}},
  \bibinfo{author}{\bibfnamefont{H.~S.} \bibnamefont{Kim}}, \bibnamefont{and}
  \bibinfo{author}{\bibfnamefont{J.~W.} \bibnamefont{Chung}},
  \bibinfo{journal}{Phys. Rev. B} \textbf{\bibinfo{volume}{83}},
  \bibinfo{pages}{161403} (\bibinfo{year}{2011}).

\bibitem[{\citenamefont{Yan et~al.}(2011)\citenamefont{Yan, Thygesen, and
  Jacobsen}}]{yanjunFP_prl11}
\bibinfo{author}{\bibfnamefont{J.}~\bibnamefont{Yan}},
  \bibinfo{author}{\bibfnamefont{K.~S.} \bibnamefont{Thygesen}},
  \bibnamefont{and} \bibinfo{author}{\bibfnamefont{K.~W.}
  \bibnamefont{Jacobsen}}, \bibinfo{journal}{Phys. Rev. Lett.}
  \textbf{\bibinfo{volume}{106}}, \bibinfo{pages}{146803}
  (\bibinfo{year}{2011}).

\bibitem[{\citenamefont{Kinyanjui et~al.}(2012)\citenamefont{Kinyanjui,
  Kramberger, Pichler, Meyer, Wachsmuth, Benner, and Kaiser}}]{kinyanjui_EPL12}
\bibinfo{author}{\bibfnamefont{M.~K.} \bibnamefont{Kinyanjui}},
  \bibinfo{author}{\bibfnamefont{C.}~\bibnamefont{Kramberger}},
  \bibinfo{author}{\bibfnamefont{T.}~\bibnamefont{Pichler}},
  \bibinfo{author}{\bibfnamefont{J.~C.} \bibnamefont{Meyer}},
  \bibinfo{author}{\bibfnamefont{P.}~\bibnamefont{Wachsmuth}},
  \bibinfo{author}{\bibfnamefont{G.}~\bibnamefont{Benner}}, \bibnamefont{and}
  \bibinfo{author}{\bibfnamefont{U.}~\bibnamefont{Kaiser}},
  \bibinfo{journal}{Europhysics Letters} \textbf{\bibinfo{volume}{97}},
  \bibinfo{pages}{57005} (\bibinfo{year}{2012}).

\bibitem[{\citenamefont{Despoja et~al.}(2013)\citenamefont{Despoja, Novko,
  Dekanic, Sunjic, and Marusic}}]{despoja_PRB13}
\bibinfo{author}{\bibfnamefont{V.}~\bibnamefont{Despoja}},
  \bibinfo{author}{\bibfnamefont{D.}~\bibnamefont{Novko}},
  \bibinfo{author}{\bibfnamefont{K.}~\bibnamefont{Dekanic}},
  \bibinfo{author}{\bibfnamefont{M.}~\bibnamefont{Sunjic}}, \bibnamefont{and}
  \bibinfo{author}{\bibfnamefont{L.}~\bibnamefont{Marusic}},
  \bibinfo{journal}{Phys. Rev. B} \textbf{\bibinfo{volume}{87}},
  \bibinfo{pages}{075447} (\bibinfo{year}{2013}).

\bibitem[{\citenamefont{Hwang and Das~Sarma}(2009)}]{HwangScreen_PRB09}
\bibinfo{author}{\bibfnamefont{E.~H.} \bibnamefont{Hwang}} \bibnamefont{and}
  \bibinfo{author}{\bibfnamefont{S.}~\bibnamefont{Das~Sarma}},
  \bibinfo{journal}{Phys. Rev. B} \textbf{\bibinfo{volume}{79}},
  \bibinfo{pages}{165404} (\bibinfo{year}{2009}).

\bibitem[{\citenamefont{Wunsch et~al.}(2006)\citenamefont{Wunsch, Stauber,
  Sols, and Guinea}}]{wunsch2006}
\bibinfo{author}{\bibfnamefont{B.}~\bibnamefont{Wunsch}},
  \bibinfo{author}{\bibfnamefont{T.}~\bibnamefont{Stauber}},
  \bibinfo{author}{\bibfnamefont{F.}~\bibnamefont{Sols}}, \bibnamefont{and}
  \bibinfo{author}{\bibfnamefont{F.}~\bibnamefont{Guinea}},
  \bibinfo{journal}{New J. Phys.} \textbf{\bibinfo{volume}{8}},
  \bibinfo{pages}{318} (\bibinfo{year}{2006}).

\bibitem[{\citenamefont{{Sensarma} et~al.}(2010)\citenamefont{{Sensarma},
  {Hwang}, and {Das Sarma}}}]{sensarma2010}
\bibinfo{author}{\bibfnamefont{R.}~\bibnamefont{{Sensarma}}},
  \bibinfo{author}{\bibfnamefont{E.~H.} \bibnamefont{{Hwang}}},
  \bibnamefont{and} \bibinfo{author}{\bibfnamefont{S.}~\bibnamefont{{Das
  Sarma}}}, \bibinfo{journal}{Phys. Rev. B} \textbf{\bibinfo{volume}{82}},
  \bibinfo{pages}{195428} (\bibinfo{year}{2010}).

\bibitem[{\citenamefont{Hwang and {Das Sarma}}(2009)}]{hwang2009b}
\bibinfo{author}{\bibfnamefont{E.~H.} \bibnamefont{Hwang}} \bibnamefont{and}
  \bibinfo{author}{\bibfnamefont{S.}~\bibnamefont{{Das Sarma}}},
  \bibinfo{journal}{Phys. Rev. B} \textbf{\bibinfo{volume}{80}},
  \bibinfo{pages}{205405} (\bibinfo{year}{2009}).

\bibitem[{\citenamefont{Vazifehshenas et~al.}(2010)\citenamefont{Vazifehshenas,
  Amlaki, Farmanbar, and Parhizgar}}]{varifeh_PLA10}
\bibinfo{author}{\bibfnamefont{T.}~\bibnamefont{Vazifehshenas}},
  \bibinfo{author}{\bibfnamefont{T.}~\bibnamefont{Amlaki}},
  \bibinfo{author}{\bibfnamefont{M.}~\bibnamefont{Farmanbar}},
  \bibnamefont{and}
  \bibinfo{author}{\bibfnamefont{F.}~\bibnamefont{Parhizgar}},
  \bibinfo{journal}{Physics Letters A} \textbf{\bibinfo{volume}{374}},
  \bibinfo{pages}{4899} (\bibinfo{year}{2010}).

\bibitem[{\citenamefont{Bahrami and Vazifehshenas}(2012)}]{Bahrami20123518}
\bibinfo{author}{\bibfnamefont{B.}~\bibnamefont{Bahrami}} \bibnamefont{and}
  \bibinfo{author}{\bibfnamefont{T.}~\bibnamefont{Vazifehshenas}},
  \bibinfo{journal}{Physics Letters A} \textbf{\bibinfo{volume}{376}},
  \bibinfo{pages}{3518 } (\bibinfo{year}{2012}).

\bibitem[{\citenamefont{Stauber and
  G\'omez-Santos}(2012)}]{staubersantos_PRB12}
\bibinfo{author}{\bibfnamefont{T.}~\bibnamefont{Stauber}} \bibnamefont{and}
  \bibinfo{author}{\bibfnamefont{G.}~\bibnamefont{G\'omez-Santos}},
  \bibinfo{journal}{Phys. Rev. B} \textbf{\bibinfo{volume}{85}},
  \bibinfo{pages}{075410} (\bibinfo{year}{2012}).

\bibitem[{\citenamefont{Stauber and Gomez-Santos}(2012)}]{stauber_NJP12}
\bibinfo{author}{\bibfnamefont{T.}~\bibnamefont{Stauber}} \bibnamefont{and}
  \bibinfo{author}{\bibfnamefont{G.}~\bibnamefont{Gomez-Santos}},
  \bibinfo{journal}{New J. Phys.} \textbf{\bibinfo{volume}{14}},
  \bibinfo{pages}{105018} (\bibinfo{year}{2012}).

\bibitem[{\citenamefont{Tuan and Khanh}(2012)}]{tuan_commphys12}
\bibinfo{author}{\bibfnamefont{D.~V.} \bibnamefont{Tuan}} \bibnamefont{and}
  \bibinfo{author}{\bibfnamefont{N.~Q.} \bibnamefont{Khanh}},
  \bibinfo{journal}{Communications in Physics} \textbf{\bibinfo{volume}{22}},
  \bibinfo{pages}{45} (\bibinfo{year}{2012}).

\bibitem[{\citenamefont{Profumo et~al.}(2012)\citenamefont{Profumo, Asgari,
  Polini, and MacDonald}}]{profumo_PRB12}
\bibinfo{author}{\bibfnamefont{R.~E.~V.} \bibnamefont{Profumo}},
  \bibinfo{author}{\bibfnamefont{R.}~\bibnamefont{Asgari}},
  \bibinfo{author}{\bibfnamefont{M.}~\bibnamefont{Polini}}, \bibnamefont{and}
  \bibinfo{author}{\bibfnamefont{A.~H.} \bibnamefont{MacDonald}},
  \bibinfo{journal}{Phys. Rev. B} \textbf{\bibinfo{volume}{85}},
  \bibinfo{pages}{085443} (\bibinfo{year}{2012}).

\bibitem[{\citenamefont{Badalyan and
  Peeters}(2012{\natexlab{a}})}]{badalyan_PRB12}
\bibinfo{author}{\bibfnamefont{S.~M.} \bibnamefont{Badalyan}} \bibnamefont{and}
  \bibinfo{author}{\bibfnamefont{F.~M.} \bibnamefont{Peeters}},
  \bibinfo{journal}{Phys. Rev. B} \textbf{\bibinfo{volume}{85}},
  \bibinfo{pages}{195444} (\bibinfo{year}{2012}{\natexlab{a}}).

\bibitem[{\citenamefont{Badalyan and
  Peeters}(2012{\natexlab{b}})}]{badalyanpeeter_PRB12}
\bibinfo{author}{\bibfnamefont{S.~M.} \bibnamefont{Badalyan}} \bibnamefont{and}
  \bibinfo{author}{\bibfnamefont{F.~M.} \bibnamefont{Peeters}},
  \bibinfo{journal}{Phys. Rev. B} \textbf{\bibinfo{volume}{86}},
  \bibinfo{pages}{121405} (\bibinfo{year}{2012}{\natexlab{b}}).

\bibitem[{\citenamefont{Triola and Rossi}(2012)}]{Triolacenrico_PRB12}
\bibinfo{author}{\bibfnamefont{C.}~\bibnamefont{Triola}} \bibnamefont{and}
  \bibinfo{author}{\bibfnamefont{E.}~\bibnamefont{Rossi}},
  \bibinfo{journal}{Phys. Rev. B} \textbf{\bibinfo{volume}{86}},
  \bibinfo{pages}{161408} (\bibinfo{year}{2012}).

\bibitem[{\citenamefont{Das~Sarma and Madhukar}(1981)}]{dasmadhukar_PRB82}
\bibinfo{author}{\bibfnamefont{S.}~\bibnamefont{Das~Sarma}} \bibnamefont{and}
  \bibinfo{author}{\bibfnamefont{A.}~\bibnamefont{Madhukar}},
  \bibinfo{journal}{Phys. Rev. B} \textbf{\bibinfo{volume}{23}},
  \bibinfo{pages}{805} (\bibinfo{year}{1981}).

\bibitem[{\citenamefont{Wang and Chakraborty}(2007)}]{wang2007}
\bibinfo{author}{\bibfnamefont{X.}~\bibnamefont{Wang}} \bibnamefont{and}
  \bibinfo{author}{\bibfnamefont{T.}~\bibnamefont{Chakraborty}},
  \bibinfo{journal}{Phys. Rev. B} \textbf{\bibinfo{volume}{75}},
  \bibinfo{pages}{041404} (\bibinfo{year}{2007}).

\bibitem[{\citenamefont{Borghi et~al.}(2009)\citenamefont{Borghi, Polini,
  Asgari, and MacDonald}}]{borghi2009}
\bibinfo{author}{\bibfnamefont{G.}~\bibnamefont{Borghi}},
  \bibinfo{author}{\bibfnamefont{M.}~\bibnamefont{Polini}},
  \bibinfo{author}{\bibfnamefont{R.}~\bibnamefont{Asgari}}, \bibnamefont{and}
  \bibinfo{author}{\bibfnamefont{A.~H.} \bibnamefont{MacDonald}},
  \bibinfo{journal}{Phys. Rev. B} \textbf{\bibinfo{volume}{80}},
  \bibinfo{pages}{241402} (\bibinfo{year}{2009}).

\bibitem[{\citenamefont{Wang and Chakraborty}(2010)}]{wang2010}
\bibinfo{author}{\bibfnamefont{X.-F.} \bibnamefont{Wang}} \bibnamefont{and}
  \bibinfo{author}{\bibfnamefont{T.}~\bibnamefont{Chakraborty}},
  \bibinfo{journal}{Phys. Rev. B} \textbf{\bibinfo{volume}{81}},
  \bibinfo{pages}{081402} (\bibinfo{year}{2010}).

\bibitem[{\citenamefont{{Gamayun}}(2011)}]{gamayun2011}
\bibinfo{author}{\bibfnamefont{O.~V.} \bibnamefont{{Gamayun}}},
  \bibinfo{journal}{Phys. Rev. B} \textbf{\bibinfo{volume}{84}},
  \bibinfo{pages}{085112} (\bibinfo{year}{2011}).

\bibitem[{\citenamefont{Weitz et~al.}(2010)\citenamefont{Weitz, Allen, Feldman,
  Martin, and Yacoby}}]{weitz2010}
\bibinfo{author}{\bibfnamefont{R.~T.} \bibnamefont{Weitz}},
  \bibinfo{author}{\bibfnamefont{M.~T.} \bibnamefont{Allen}},
  \bibinfo{author}{\bibfnamefont{B.~E.} \bibnamefont{Feldman}},
  \bibinfo{author}{\bibfnamefont{J.}~\bibnamefont{Martin}}, \bibnamefont{and}
  \bibinfo{author}{\bibfnamefont{A.}~\bibnamefont{Yacoby}},
  \bibinfo{journal}{Science} \textbf{\bibinfo{volume}{330}},
  \bibinfo{pages}{812} (\bibinfo{year}{2010}).

\bibitem[{\citenamefont{Ponomarenko et~al.}(2011)\citenamefont{Ponomarenko,
  Geim, Zhukov, Jalil, Morozov, Novoselov, Grigorieva, Hill, Cheianov, Fal'ko
  et~al.}}]{Ponomarenko_arX11}
\bibinfo{author}{\bibfnamefont{L.~A.} \bibnamefont{Ponomarenko}},
  \bibinfo{author}{\bibfnamefont{A.~K.} \bibnamefont{Geim}},
  \bibinfo{author}{\bibfnamefont{A.~A.} \bibnamefont{Zhukov}},
  \bibinfo{author}{\bibfnamefont{R.}~\bibnamefont{Jalil}},
  \bibinfo{author}{\bibfnamefont{S.~V.} \bibnamefont{Morozov}},
  \bibinfo{author}{\bibfnamefont{K.~S.} \bibnamefont{Novoselov}},
  \bibinfo{author}{\bibfnamefont{I.~V.} \bibnamefont{Grigorieva}},
  \bibinfo{author}{\bibfnamefont{E.~H.} \bibnamefont{Hill}},
  \bibinfo{author}{\bibfnamefont{V.~V.} \bibnamefont{Cheianov}},
  \bibinfo{author}{\bibfnamefont{V.~I.} \bibnamefont{Fal'ko}},
  \bibnamefont{et~al.}, \bibinfo{journal}{Nat. Phys.}
  \textbf{\bibinfo{volume}{7}}, \bibinfo{pages}{958} (\bibinfo{year}{2011}).

\bibitem[{\citenamefont{Jr et~al.}(2012)\citenamefont{Jr, Jing, Bao, Lee,
  Kratz, Aji, Bockrath, Lau, Varma, Stillwell et~al.}}]{velasco_natnano12}
\bibinfo{author}{\bibfnamefont{J.~V.} \bibnamefont{Jr}},
  \bibinfo{author}{\bibfnamefont{L.}~\bibnamefont{Jing}},
  \bibinfo{author}{\bibfnamefont{W.}~\bibnamefont{Bao}},
  \bibinfo{author}{\bibfnamefont{Y.}~\bibnamefont{Lee}},
  \bibinfo{author}{\bibfnamefont{P.}~\bibnamefont{Kratz}},
  \bibinfo{author}{\bibfnamefont{V.}~\bibnamefont{Aji}},
  \bibinfo{author}{\bibfnamefont{M.}~\bibnamefont{Bockrath}},
  \bibinfo{author}{\bibfnamefont{C.~N.} \bibnamefont{Lau}},
  \bibinfo{author}{\bibfnamefont{C.}~\bibnamefont{Varma}},
  \bibinfo{author}{\bibfnamefont{R.}~\bibnamefont{Stillwell}},
  \bibnamefont{et~al.}, \bibinfo{journal}{Nature Nanotechnology}
  \textbf{\bibinfo{volume}{7}}, \bibinfo{pages}{156} (\bibinfo{year}{2012}).

\bibitem[{\citenamefont{Bao et~al.}(2012)\citenamefont{Bao, Velasco, Zhang,
  Jing, Standley, Smirnov, Bockrath, MacDonald, and Lau}}]{Bao03072012}
\bibinfo{author}{\bibfnamefont{W.}~\bibnamefont{Bao}},
  \bibinfo{author}{\bibfnamefont{J.}~\bibnamefont{Velasco}},
  \bibinfo{author}{\bibfnamefont{F.}~\bibnamefont{Zhang}},
  \bibinfo{author}{\bibfnamefont{L.}~\bibnamefont{Jing}},
  \bibinfo{author}{\bibfnamefont{B.}~\bibnamefont{Standley}},
  \bibinfo{author}{\bibfnamefont{D.}~\bibnamefont{Smirnov}},
  \bibinfo{author}{\bibfnamefont{M.}~\bibnamefont{Bockrath}},
  \bibinfo{author}{\bibfnamefont{A.~H.} \bibnamefont{MacDonald}},
  \bibnamefont{and} \bibinfo{author}{\bibfnamefont{C.~N.} \bibnamefont{Lau}},
  \bibinfo{journal}{Proceedings of the National Academy of Sciences}
  \textbf{\bibinfo{volume}{109}}, \bibinfo{pages}{10802}
  (\bibinfo{year}{2012}).

\bibitem[{\citenamefont{Min et~al.}(2008)\citenamefont{Min, Borghi, Polini, and
  MacDonald}}]{min2008b}
\bibinfo{author}{\bibfnamefont{H.}~\bibnamefont{Min}},
  \bibinfo{author}{\bibfnamefont{G.}~\bibnamefont{Borghi}},
  \bibinfo{author}{\bibfnamefont{M.}~\bibnamefont{Polini}}, \bibnamefont{and}
  \bibinfo{author}{\bibfnamefont{A.~H.} \bibnamefont{MacDonald}},
  \bibinfo{journal}{Phys. Rev. B} \textbf{\bibinfo{volume}{77}},
  \bibinfo{pages}{041407} (\bibinfo{year}{2008}).

\bibitem[{\citenamefont{Barlas and Yang}(2009)}]{yafis_PRB09}
\bibinfo{author}{\bibfnamefont{Y.}~\bibnamefont{Barlas}} \bibnamefont{and}
  \bibinfo{author}{\bibfnamefont{K.}~\bibnamefont{Yang}},
  \bibinfo{journal}{Phys. Rev. B} \textbf{\bibinfo{volume}{80}},
  \bibinfo{pages}{161408} (\bibinfo{year}{2009}).

\bibitem[{\citenamefont{Zhang et~al.}(2010)\citenamefont{Zhang, Min, Polini,
  and MacDonald}}]{zhang2010}
\bibinfo{author}{\bibfnamefont{F.}~\bibnamefont{Zhang}},
  \bibinfo{author}{\bibfnamefont{H.}~\bibnamefont{Min}},
  \bibinfo{author}{\bibfnamefont{M.}~\bibnamefont{Polini}}, \bibnamefont{and}
  \bibinfo{author}{\bibfnamefont{A.~H.} \bibnamefont{MacDonald}},
  \bibinfo{journal}{Phys. Rev. B} \textbf{\bibinfo{volume}{81}},
  \bibinfo{pages}{041402} (\bibinfo{year}{2010}).

\bibitem[{\citenamefont{Lemonik et~al.}(2010)\citenamefont{Lemonik, Aleiner,
  Toke, and Fal'ko}}]{lemonik2010}
\bibinfo{author}{\bibfnamefont{Y.}~\bibnamefont{Lemonik}},
  \bibinfo{author}{\bibfnamefont{I.~L.} \bibnamefont{Aleiner}},
  \bibinfo{author}{\bibfnamefont{C.}~\bibnamefont{Toke}}, \bibnamefont{and}
  \bibinfo{author}{\bibfnamefont{V.~I.} \bibnamefont{Fal'ko}},
  \bibinfo{journal}{Phys. Rev. B} \textbf{\bibinfo{volume}{82}},
  \bibinfo{pages}{201408} (\bibinfo{year}{2010}).

\bibitem[{\citenamefont{Nandkishore and
  Levitov}(2010{\natexlab{a}})}]{nandkishore2010}
\bibinfo{author}{\bibfnamefont{R.}~\bibnamefont{Nandkishore}} \bibnamefont{and}
  \bibinfo{author}{\bibfnamefont{L.}~\bibnamefont{Levitov}},
  \bibinfo{journal}{Phys. Rev. Lett.} \textbf{\bibinfo{volume}{104}},
  \bibinfo{pages}{156803} (\bibinfo{year}{2010}{\natexlab{a}}).

\bibitem[{\citenamefont{Nandkishore and
  Levitov}(2010{\natexlab{b}})}]{nandkishore2010b}
\bibinfo{author}{\bibfnamefont{R.}~\bibnamefont{Nandkishore}} \bibnamefont{and}
  \bibinfo{author}{\bibfnamefont{L.}~\bibnamefont{Levitov}},
  \bibinfo{journal}{Phys. Rev. B} \textbf{\bibinfo{volume}{82}},
  \bibinfo{pages}{115431} (\bibinfo{year}{2010}{\natexlab{b}}).

\bibitem[{\citenamefont{Nandkishore and
  Levitov}(2010{\natexlab{c}})}]{nadkishorelevi_PRB10}
\bibinfo{author}{\bibfnamefont{R.}~\bibnamefont{Nandkishore}} \bibnamefont{and}
  \bibinfo{author}{\bibfnamefont{L.}~\bibnamefont{Levitov}},
  \bibinfo{journal}{Phys. Rev. B} \textbf{\bibinfo{volume}{82}},
  \bibinfo{pages}{115124} (\bibinfo{year}{2010}{\natexlab{c}}).

\bibitem[{\citenamefont{Vafek and Yang}(2010)}]{vafek2010}
\bibinfo{author}{\bibfnamefont{O.}~\bibnamefont{Vafek}} \bibnamefont{and}
  \bibinfo{author}{\bibfnamefont{K.}~\bibnamefont{Yang}},
  \bibinfo{journal}{Phys. Rev. B} \textbf{\bibinfo{volume}{81}},
  \bibinfo{pages}{041401} (\bibinfo{year}{2010}).

\bibitem[{\citenamefont{Jung et~al.}(2011)\citenamefont{Jung, Zhang, and
  MacDonald}}]{jjmac_PRB11}
\bibinfo{author}{\bibfnamefont{J.}~\bibnamefont{Jung}},
  \bibinfo{author}{\bibfnamefont{F.}~\bibnamefont{Zhang}}, \bibnamefont{and}
  \bibinfo{author}{\bibfnamefont{A.~H.} \bibnamefont{MacDonald}},
  \bibinfo{journal}{Phys. Rev. B} \textbf{\bibinfo{volume}{83}},
  \bibinfo{pages}{115408} (\bibinfo{year}{2011}).

\bibitem[{\citenamefont{{Das Sarma} et~al.}(2007)\citenamefont{{Das Sarma},
  Hwang, and Tse}}]{dassarma2007b}
\bibinfo{author}{\bibfnamefont{S.}~\bibnamefont{{Das Sarma}}},
  \bibinfo{author}{\bibfnamefont{E.~H.} \bibnamefont{Hwang}}, \bibnamefont{and}
  \bibinfo{author}{\bibfnamefont{W.-K.} \bibnamefont{Tse}},
  \bibinfo{journal}{Phys. Rev. B} \textbf{\bibinfo{volume}{75}},
  \bibinfo{pages}{121406} (\bibinfo{year}{2007}).

\bibitem[{\citenamefont{Herbut}(2006)}]{igor_PRL06}
\bibinfo{author}{\bibfnamefont{I.~F.} \bibnamefont{Herbut}},
  \bibinfo{journal}{Phys. Rev. Lett.} \textbf{\bibinfo{volume}{97}},
  \bibinfo{pages}{146401} (\bibinfo{year}{2006}).

\bibitem[{\citenamefont{Nomura and MacDonald}(2006)}]{nomura2006}
\bibinfo{author}{\bibfnamefont{K.}~\bibnamefont{Nomura}} \bibnamefont{and}
  \bibinfo{author}{\bibfnamefont{A.~H.} \bibnamefont{MacDonald}},
  \bibinfo{journal}{Phys. Rev. Lett.} \textbf{\bibinfo{volume}{96}},
  \bibinfo{pages}{256602} (\bibinfo{year}{2006}).

\bibitem[{\citenamefont{Yang et~al.}(2006)\citenamefont{Yang, {Das Sarma}, and
  MacDonald}}]{yang2006}
\bibinfo{author}{\bibfnamefont{K.}~\bibnamefont{Yang}},
  \bibinfo{author}{\bibfnamefont{S.}~\bibnamefont{{Das Sarma}}},
  \bibnamefont{and} \bibinfo{author}{\bibfnamefont{A.~H.}
  \bibnamefont{MacDonald}}, \bibinfo{journal}{Phys. Rev. B}
  \textbf{\bibinfo{volume}{74}}, \bibinfo{pages}{075423}
  (\bibinfo{year}{2006}).

\bibitem[{\citenamefont{Drut and
  L\"ahde}(2009{\natexlab{a}})}]{drut_criticalPRB09}
\bibinfo{author}{\bibfnamefont{J.~E.} \bibnamefont{Drut}} \bibnamefont{and}
  \bibinfo{author}{\bibfnamefont{T.~A.} \bibnamefont{L\"ahde}},
  \bibinfo{journal}{Phys. Rev. B} \textbf{\bibinfo{volume}{79}},
  \bibinfo{pages}{241405} (\bibinfo{year}{2009}{\natexlab{a}}).

\bibitem[{\citenamefont{Neto et~al.}(2009)\citenamefont{Neto, Guinea, Peres,
  Novoselov, and Geim}}]{neto2009}
\bibinfo{author}{\bibfnamefont{A.~H.~C.} \bibnamefont{Neto}},
  \bibinfo{author}{\bibfnamefont{F.}~\bibnamefont{Guinea}},
  \bibinfo{author}{\bibfnamefont{N.~M.~R.} \bibnamefont{Peres}},
  \bibinfo{author}{\bibfnamefont{K.~S.} \bibnamefont{Novoselov}},
  \bibnamefont{and} \bibinfo{author}{\bibfnamefont{A.~K.} \bibnamefont{Geim}},
  \bibinfo{journal}{Rev. Mod. Phys.} \textbf{\bibinfo{volume}{81}},
  \bibinfo{pages}{109} (\bibinfo{year}{2009}).

\bibitem[{\citenamefont{Lemonik et~al.}(2012)\citenamefont{Lemonik, Aleiner,
  and Fal'ko}}]{lemonik_PRB12}
\bibinfo{author}{\bibfnamefont{Y.}~\bibnamefont{Lemonik}},
  \bibinfo{author}{\bibfnamefont{I.}~\bibnamefont{Aleiner}}, \bibnamefont{and}
  \bibinfo{author}{\bibfnamefont{V.~I.} \bibnamefont{Fal'ko}},
  \bibinfo{journal}{Phys. Rev. B} \textbf{\bibinfo{volume}{85}},
  \bibinfo{pages}{245451} (\bibinfo{year}{2012}).

\bibitem[{\citenamefont{Kotov et~al.}(2012)\citenamefont{Kotov, Uchoa, Pereira,
  Guinea, and Castro~Neto}}]{kotov_RMP12}
\bibinfo{author}{\bibfnamefont{V.~N.} \bibnamefont{Kotov}},
  \bibinfo{author}{\bibfnamefont{B.}~\bibnamefont{Uchoa}},
  \bibinfo{author}{\bibfnamefont{V.~M.} \bibnamefont{Pereira}},
  \bibinfo{author}{\bibfnamefont{F.}~\bibnamefont{Guinea}}, \bibnamefont{and}
  \bibinfo{author}{\bibfnamefont{A.~H.} \bibnamefont{Castro~Neto}},
  \bibinfo{journal}{Rev. Mod. Phys.} \textbf{\bibinfo{volume}{84}},
  \bibinfo{pages}{1067} (\bibinfo{year}{2012}).

\bibitem[{\citenamefont{Drut and L\"ahde}(2009{\natexlab{b}})}]{drut_PRL09}
\bibinfo{author}{\bibfnamefont{J.~E.} \bibnamefont{Drut}} \bibnamefont{and}
  \bibinfo{author}{\bibfnamefont{T.~A.} \bibnamefont{L\"ahde}},
  \bibinfo{journal}{Phys. Rev. Lett.} \textbf{\bibinfo{volume}{102}},
  \bibinfo{pages}{026802} (\bibinfo{year}{2009}{\natexlab{b}}).

\bibitem[{\citenamefont{Drut and
  L\"ahde}(2009{\natexlab{c}})}]{drut_latticePRB09}
\bibinfo{author}{\bibfnamefont{J.~E.} \bibnamefont{Drut}} \bibnamefont{and}
  \bibinfo{author}{\bibfnamefont{T.~A.} \bibnamefont{L\"ahde}},
  \bibinfo{journal}{Phys. Rev. B} \textbf{\bibinfo{volume}{79}},
  \bibinfo{pages}{165425} (\bibinfo{year}{2009}{\natexlab{c}}).

\bibitem[{\citenamefont{{Das Sarma} et~al.}(2011)\citenamefont{{Das Sarma},
  Adam, Hwang, and Rossi}}]{dassarma2010}
\bibinfo{author}{\bibfnamefont{S.}~\bibnamefont{{Das Sarma}}},
  \bibinfo{author}{\bibfnamefont{S.}~\bibnamefont{Adam}},
  \bibinfo{author}{\bibfnamefont{E.~H.} \bibnamefont{Hwang}}, \bibnamefont{and}
  \bibinfo{author}{\bibfnamefont{E.}~\bibnamefont{Rossi}},
  \bibinfo{journal}{Rev. Mod. Phys.} \textbf{\bibinfo{volume}{83}},
  \bibinfo{pages}{407} (\bibinfo{year}{2011}).

\bibitem[{\citenamefont{Adam et~al.}(2007)\citenamefont{Adam, Hwang, Galitski,
  and {Das Sarma}}}]{Adam}
\bibinfo{author}{\bibfnamefont{S.}~\bibnamefont{Adam}},
  \bibinfo{author}{\bibfnamefont{E.~H.} \bibnamefont{Hwang}},
  \bibinfo{author}{\bibfnamefont{V.~M.} \bibnamefont{Galitski}},
  \bibnamefont{and} \bibinfo{author}{\bibfnamefont{S.}~\bibnamefont{{Das
  Sarma}}}, \bibinfo{journal}{Proc.\ Natl.\ Acad.\ Sci.\ USA}
  \textbf{\bibinfo{volume}{104}}, \bibinfo{pages}{18392}
  (\bibinfo{year}{2007}).

\bibitem[{\citenamefont{Rossi and {Das Sarma}}(2008)}]{rossi2008}
\bibinfo{author}{\bibfnamefont{E.}~\bibnamefont{Rossi}} \bibnamefont{and}
  \bibinfo{author}{\bibfnamefont{S.}~\bibnamefont{{Das Sarma}}},
  \bibinfo{journal}{Phys. Rev. Lett.} \textbf{\bibinfo{volume}{101}},
  \bibinfo{pages}{166803} (\bibinfo{year}{2008}).

\bibitem[{\citenamefont{Martin et~al.}(2008)\citenamefont{Martin, Akerman,
  Ulbricht, Lohmann, Smet, \mbox{von Klitzing}, and Yacobi}}]{martin2008}
\bibinfo{author}{\bibfnamefont{J.}~\bibnamefont{Martin}},
  \bibinfo{author}{\bibfnamefont{N.}~\bibnamefont{Akerman}},
  \bibinfo{author}{\bibfnamefont{G.}~\bibnamefont{Ulbricht}},
  \bibinfo{author}{\bibfnamefont{T.}~\bibnamefont{Lohmann}},
  \bibinfo{author}{\bibfnamefont{J.~H.} \bibnamefont{Smet}},
  \bibinfo{author}{\bibfnamefont{K.}~\bibnamefont{\mbox{von Klitzing}}},
  \bibnamefont{and} \bibinfo{author}{\bibfnamefont{A.}~\bibnamefont{Yacobi}},
  \bibinfo{journal}{Nature Physics} \textbf{\bibinfo{volume}{4}},
  \bibinfo{pages}{144} (\bibinfo{year}{2008}).

\bibitem[{\citenamefont{Zhang et~al.}(2009)\citenamefont{Zhang, Brar, Girit,
  Zettl, and Crommie}}]{YZhang_NP09}
\bibinfo{author}{\bibfnamefont{Y.}~\bibnamefont{Zhang}},
  \bibinfo{author}{\bibfnamefont{V.~W.} \bibnamefont{Brar}},
  \bibinfo{author}{\bibfnamefont{C.}~\bibnamefont{Girit}},
  \bibinfo{author}{\bibfnamefont{A.}~\bibnamefont{Zettl}}, \bibnamefont{and}
  \bibinfo{author}{\bibfnamefont{M.~F.} \bibnamefont{Crommie}},
  \bibinfo{journal}{Nat. Phys.} \textbf{\bibinfo{volume}{5}},
  \bibinfo{pages}{722} (\bibinfo{year}{2009}).

\bibitem[{\citenamefont{Deshpande et~al.}(2009)\citenamefont{Deshpande, Bao,
  Miao, Lau, and LeRoy}}]{LeRoyPuddle_PRB09}
\bibinfo{author}{\bibfnamefont{A.}~\bibnamefont{Deshpande}},
  \bibinfo{author}{\bibfnamefont{W.}~\bibnamefont{Bao}},
  \bibinfo{author}{\bibfnamefont{F.}~\bibnamefont{Miao}},
  \bibinfo{author}{\bibfnamefont{C.~N.} \bibnamefont{Lau}}, \bibnamefont{and}
  \bibinfo{author}{\bibfnamefont{B.~J.} \bibnamefont{LeRoy}},
  \bibinfo{journal}{Phys. Rev. B} \textbf{\bibinfo{volume}{79}},
  \bibinfo{pages}{205411} (\bibinfo{year}{2009}).

\bibitem[{\citenamefont{Deshpande et~al.}(2011)\citenamefont{Deshpande, Bao,
  Zhao, Lau, and LeRoy}}]{LeRoyPuddle_PRB11}
\bibinfo{author}{\bibfnamefont{A.}~\bibnamefont{Deshpande}},
  \bibinfo{author}{\bibfnamefont{W.}~\bibnamefont{Bao}},
  \bibinfo{author}{\bibfnamefont{Z.}~\bibnamefont{Zhao}},
  \bibinfo{author}{\bibfnamefont{C.~N.} \bibnamefont{Lau}}, \bibnamefont{and}
  \bibinfo{author}{\bibfnamefont{B.~J.} \bibnamefont{LeRoy}},
  \bibinfo{journal}{Phys. Rev. B} \textbf{\bibinfo{volume}{83}},
  \bibinfo{pages}{155409} (\bibinfo{year}{2011}).

\bibitem[{\citenamefont{Rossi et~al.}(2009)\citenamefont{Rossi, Adam, and {Das
  Sarma}}}]{Rossi}
\bibinfo{author}{\bibfnamefont{E.}~\bibnamefont{Rossi}},
  \bibinfo{author}{\bibfnamefont{S.}~\bibnamefont{Adam}}, \bibnamefont{and}
  \bibinfo{author}{\bibfnamefont{S.}~\bibnamefont{{Das Sarma}}},
  \bibinfo{journal}{Phys. Rev. B} \textbf{\bibinfo{volume}{79}},
  \bibinfo{pages}{245423} (\bibinfo{year}{2009}).

\bibitem[{\citenamefont{Li et~al.}(2011)\citenamefont{Li, Hwang, and {Das
  Sarma}}}]{qzli_PRB11}
\bibinfo{author}{\bibfnamefont{Q.}~\bibnamefont{Li}},
  \bibinfo{author}{\bibfnamefont{E.~H.} \bibnamefont{Hwang}}, \bibnamefont{and}
  \bibinfo{author}{\bibfnamefont{S.}~\bibnamefont{{Das Sarma}}},
  \bibinfo{journal}{Phys. Rev. B} \textbf{\bibinfo{volume}{84}},
  \bibinfo{pages}{115442} (\bibinfo{year}{2011}).

\bibitem[{\citenamefont{{Das Sarma} et~al.}(2012)\citenamefont{{Das Sarma},
  Hwang, and Li}}]{qzlidbo_PRB12}
\bibinfo{author}{\bibfnamefont{S.}~\bibnamefont{{Das Sarma}}},
  \bibinfo{author}{\bibfnamefont{E.~H.} \bibnamefont{Hwang}}, \bibnamefont{and}
  \bibinfo{author}{\bibfnamefont{Q.}~\bibnamefont{Li}}, \bibinfo{journal}{Phys.
  Rev. B} \textbf{\bibinfo{volume}{85}}, \bibinfo{pages}{195451}
  (\bibinfo{year}{2012}).

\bibitem[{\citenamefont{Das~Sarma and Hwang}(2013)}]{dashwangsusDirac_PRB13}
\bibinfo{author}{\bibfnamefont{S.}~\bibnamefont{Das~Sarma}} \bibnamefont{and}
  \bibinfo{author}{\bibfnamefont{E.~H.} \bibnamefont{Hwang}},
  \bibinfo{journal}{Phys. Rev. B} \textbf{\bibinfo{volume}{87}},
  \bibinfo{pages}{035415} (\bibinfo{year}{2013}).

\bibitem[{\citenamefont{Chen et~al.}(2008)\citenamefont{Chen, Jang, Adam,
  Fuhrer, Williams, and Ishigami}}]{ChenJ_NPH_2007}
\bibinfo{author}{\bibfnamefont{J.-H.} \bibnamefont{Chen}},
  \bibinfo{author}{\bibfnamefont{C.}~\bibnamefont{Jang}},
  \bibinfo{author}{\bibfnamefont{S.}~\bibnamefont{Adam}},
  \bibinfo{author}{\bibfnamefont{M.~S.} \bibnamefont{Fuhrer}},
  \bibinfo{author}{\bibfnamefont{E.~D.} \bibnamefont{Williams}},
  \bibnamefont{and} \bibinfo{author}{\bibfnamefont{M.}~\bibnamefont{Ishigami}},
  \bibinfo{journal}{Nature Phys.} \textbf{\bibinfo{volume}{4}},
  \bibinfo{pages}{377} (\bibinfo{year}{2008}).

\bibitem[{\citenamefont{Hwang et~al.}(2007)\citenamefont{Hwang, Hu, and
  Sarma}}]{hwang2007d}
\bibinfo{author}{\bibfnamefont{E.~H.} \bibnamefont{Hwang}},
  \bibinfo{author}{\bibfnamefont{B.~Y.-K.} \bibnamefont{Hu}}, \bibnamefont{and}
  \bibinfo{author}{\bibfnamefont{S.~D.} \bibnamefont{Sarma}},
  \bibinfo{journal}{Phys. Rev. Lett.} \textbf{\bibinfo{volume}{99}},
  \bibinfo{eid}{226801} (\bibinfo{year}{2007}).

\bibitem[{\citenamefont{Bolotin et~al.}(2008)\citenamefont{Bolotin, Sikes,
  Jiang, Klima, Fudenberg, Hone, Kim, and Stormer}}]{BolotinYacoby}
\bibinfo{author}{\bibfnamefont{K.}~\bibnamefont{Bolotin}},
  \bibinfo{author}{\bibfnamefont{K.}~\bibnamefont{Sikes}},
  \bibinfo{author}{\bibfnamefont{Z.}~\bibnamefont{Jiang}},
  \bibinfo{author}{\bibfnamefont{M.}~\bibnamefont{Klima}},
  \bibinfo{author}{\bibfnamefont{G.}~\bibnamefont{Fudenberg}},
  \bibinfo{author}{\bibfnamefont{J.}~\bibnamefont{Hone}},
  \bibinfo{author}{\bibfnamefont{P.}~\bibnamefont{Kim}}, \bibnamefont{and}
  \bibinfo{author}{\bibfnamefont{H.}~\bibnamefont{Stormer}},
  \bibinfo{journal}{Solid State Commun.} \textbf{\bibinfo{volume}{146}},
  \bibinfo{pages}{351} (\bibinfo{year}{2008}).

\bibitem[{\citenamefont{Dean et~al.}(2010)\citenamefont{Dean, Young, Meric,
  Lee, Wang, Sorgenfrei, Watanabe, Taniguchi, Kim, Shepard et~al.}}]{BNDeanDas}
\bibinfo{author}{\bibfnamefont{C.~R.} \bibnamefont{Dean}},
  \bibinfo{author}{\bibfnamefont{A.~F.} \bibnamefont{Young}},
  \bibinfo{author}{\bibfnamefont{I.}~\bibnamefont{Meric}},
  \bibinfo{author}{\bibfnamefont{C.}~\bibnamefont{Lee}},
  \bibinfo{author}{\bibfnamefont{L.}~\bibnamefont{Wang}},
  \bibinfo{author}{\bibfnamefont{S.}~\bibnamefont{Sorgenfrei}},
  \bibinfo{author}{\bibfnamefont{K.}~\bibnamefont{Watanabe}},
  \bibinfo{author}{\bibfnamefont{T.}~\bibnamefont{Taniguchi}},
  \bibinfo{author}{\bibfnamefont{P.}~\bibnamefont{Kim}},
  \bibinfo{author}{\bibfnamefont{K.~L.} \bibnamefont{Shepard}},
  \bibnamefont{et~al.}, \bibinfo{journal}{Nat. Nanotechnol.}
  \textbf{\bibinfo{volume}{5}}, \bibinfo{pages}{722} (\bibinfo{year}{2010}).

\bibitem[{\citenamefont{Wang et~al.}(2011)\citenamefont{Wang, Apell, and
  Kinaret}}]{weihuawang_PRB11}
\bibinfo{author}{\bibfnamefont{W.}~\bibnamefont{Wang}},
  \bibinfo{author}{\bibfnamefont{P.}~\bibnamefont{Apell}}, \bibnamefont{and}
  \bibinfo{author}{\bibfnamefont{J.}~\bibnamefont{Kinaret}},
  \bibinfo{journal}{Phys. Rev. B} \textbf{\bibinfo{volume}{84}},
  \bibinfo{pages}{085423} (\bibinfo{year}{2011}).

\end{thebibliography}
\end{document}